\documentclass[12pt, draftclsnofoot, onecolumn]{IEEEtran}
\usepackage{geometry}
\ifCLASSINFOpdf

\else

\fi
\usepackage[cmex10]{amsmath}
\usepackage{amssymb}
\usepackage{cite}
\usepackage{graphicx}
\usepackage{array,color}
\usepackage{amsmath}
\usepackage{stfloats}
\usepackage{graphicx}
\usepackage{subfigure}
\usepackage{tabularx}
\usepackage{epsfig,epsf,color,balance,cite}
\usepackage{algorithmic}
\usepackage{algorithm}
\usepackage{bm}
\usepackage{textcomp}

\usepackage{epstopdf}
\usepackage{graphicx}
\usepackage{epstopdf}
\usepackage{hyperref}

\allowdisplaybreaks[4]

\usepackage{bm}
\usepackage{multirow}
\usepackage{amsmath}
\usepackage{amssymb}
\usepackage{graphicx}

\newtheorem{remark}{Remark}

\makeatletter

\begin{document}
\title{Position and Orientation Estimation in RIS-aided MIMO-OFDM Systems with Practical Scatterers}
	
\author{Sheng Hong, Cunhua Pan, Wenchao Tang, Marco Di Renzo, \IEEEmembership{Fellow, IEEE}, Wei Zhang, \IEEEmembership{Fellow, IEEE}

\thanks{Sheng Hong is with the School of Information Engineering, Nanchang University, Nanchang 330031, China (email: shenghong@ncu.edu.cn). Cunhua Pan is with the National Mobile Communications Research Laboratory, Southeast University, Nanjing, 210096, China (e-mail:cpan@seu.edu.cn). Wenchao Tang is with the Institute of Space Science and Technology, Nanchang university (email: tangwc@ncu.edu.cn). M. Di Renzo is with Universit\'e Paris-Saclay, CNRS, CentraleSup\'elec, Laboratoire des Signaux et Syst\`emes, 3 Rue Joliot-Curie, 91192 Gif-sur-Yvette, France (marco.di-renzo@universite-paris-saclay.fr). Wei Zhang is with the School of Electrical Engineering and Telecommunications, University of New South Wales, Sydney, Australia (e-mail: w.zhang@unsw.edu.au). (Corresponding author: Cunhua Pan)
}
}
\vspace{-1.9cm}
\maketitle
\vspace{-1.9cm}
	\begin{abstract}
		In this paper, we investigate the problem of estimating the position and the angle of rotation of a mobile station (MS) in a millimeter wave (mmWave)
		multiple-input-multiple-output (MIMO) system aided by a reconfigurable intelligent surface (RIS).
		The virtual line-of-sight (VLoS) link created by the RIS and the non-line-of-sight (NLoS) links that originate from scatterers in the considered environment are utilized
		to facilitate the estimation. A two-step positioning scheme is exploited, where the channel parameters are first acquired, and the position-related parameters are then estimated. The channel parameters are obtained through a coarse and a subsequent fine estimation processes. As for the coarse estimation, the distributed compressed
		sensing orthogonal simultaneous matching pursuit (DCS-SOMP) algorithm, the maximum likelihood (ML) algorithm, and the discrete Fourier transform (DFT) are utilized to separately estimate the channel parameters. The obtained channel parameters are then jointly refined by using the space-alternating generalized expectation
		maximization (SAGE) algorithm, which circumvents the high-dimensional
		optimization issue of ML estimation. Departing from the estimated channel parameters, the positioning-related parameters
		are estimated. The performance of estimating the channel-related and position-related parameters is theoretically quantified by using the Cramer-Rao lower bound (CRLB). Simulation results demonstrate the
		superior performance of the proposed positioning algorithms.
	\end{abstract}
	
	
	
	\section{Introduction}
Reconfigurable intelligent surface (RIS) has recently emerged as
a promising technology to realize reconfigurable wireless propagation environments for future communication networks \cite{wu2021intelligent}. RIS can be integrated into existing radio localization systems
at low cost and high energy efficiency, and can enhance the positioning reliability and accuracy \cite{pan2022overview}. If a given target is out of coverage, e.g., the line-of-sight (LoS) link is obstructed by obstacles, RIS can provide an alternative virtual LoS (VLoS) link that may offer high-precision positioning performance. Furthermore, the
large physical aperture of an RIS favors high-precise estimates
of the angle of arrival (AOA) and angle of departure (AOD) at the RIS, which is beneficial for increasing the positioning accuracy.

To evaluate the potential benefits of RIS in the context of radio
localization, early research has derived and analyzed the the theoretic position error bound, i.e., Cramer-Rao lower bound (CRLB), position error bound (PEB) and orientation error bound (OEB), to investigate the fundamental limits of RIS-aided terminal positioning \cite{paper7,paper9,emenonye_fundamentals_2024,emenonye_ris-aided_2023,wang_location_2022,win_location_2022}. The works in \cite{paper7,paper9} derived and analyzed the CRLB, PEB, and OEB, demonstrating the potential of RIS to improve positioning performance by comparing the performance limits with and without RIS. The work in \cite{emenonye_fundamentals_2024} examined the structural properties of the CRLB, offering key insights into far-field RIS-aided localization. The impact of RIS uncertainty of location and orientation offsets on the terminal positioning performance was further investigated in \cite{emenonye_ris-aided_2023}. The works in \cite{wang_location_2022,win_location_2022} established the performance limits of RIS-aided network localization, showing that RISs can improve network localization accuracy.

To harness the potential of RIS, algorithm design for RIS-aided terminal positioning has been widely investigated, which falls into two main categories of direct positioning and two-step positioning. Direct positioning is primarily proposed for near-field conditions, where channel parameters are difficult to estimate efficiently. In \cite{paper10,paper14}, near-field direct positioning algorithms based on maximum likelihood estimate (MLE) in RIS-aided
	single-input-single-output (SISO) systems were proposed. Two-step positioning is mostly developed for far-field conditions, where channel parameters are first estimated, followed by the recovery of position parameters from them. For example, in \cite{paper12}, a two-step positioning algorithm for joint three-dimensional (3D) localization and synchronization problem was proposed in a RIS-enabled SISO system. The same localization problem under user mobility and spatial-wideband effects was further solved by a low-complexity two-step positioning algorithm in a RIS-enabled SISO system \cite{keykhosravi_ris-enabled_2022}. In \cite{paper13}, the MLE of AOA and AOD was obtained first, then the 3D coordinates of a single-antenna mobile station (MS) were estimated in a multiple-RIS aided
	multiple-input-single-output (MISO) system. In \cite{fascista_ris-aided_2022}, a reduced-complexity MLE algorithm was devised to jointly recover the user's two-dimensional (2D) position and the synchronization offset directly in RIS-aided MISO systems. Moreover, the terminal positioning was extended to the user tracking problem, where channel paramters and user positions were estimated in multi-RIS-aided MISO systems \cite{teng_bayesian_2022,teng_variational_2023}. 
	
	It is worth noting that the algorithm development in above works is based on the multipath-free model, which either ignore the non-line-of-sight (NLoS) links caused by scatterers \cite{paper12, keykhosravi_ris-enabled_2022, fascista_ris-aided_2022} or treat them as complex Gaussian noise \cite{paper13,teng_bayesian_2022,teng_variational_2023}. The considered systems are also relatively simple, e.g., assuming a single-antenna user instead of multiple antennas, an RIS with a uniform linear array (ULA) instead of a uniform planar array (UPA), and non-orthogonal frequency division multiplexing (OFDM) signals. Consequently, only a few parameters need to be estimated, and low-dimensional MLE problems were solved.

However, in practical scenarios, the presence of NLoS links caused by scatterers is inevitable. When scatterers exist, neglecting to estimate scatterers' positions or treating their contributions as Gaussian noise can lead to model mismatch, thereby resulting in estimation bias and degraded MS positioning accuracy. Moreover, as stated in \cite{wymeersch2022radio}, multipath exploitation is a generalization of both localization and sensing, benefiting from the NLoS part of the channel, rather than considering it as a disturbance \cite{witrisal2016high,fascista2021downlink}. The position-related information in multipath components can be used for designing positioning algorithms \cite{leitinger2015evaluation} and provide valuable information for localization \cite{abu2018error,mendrzik2018harnessing}. Actually, the NLoS links have already been exploited to aid localization in wireless communication systems in \cite{shahmansoori2017position,li_joint_2022,gong2022multipath}, where the parameters of NLoS links were estimated to retrieve the position-related information. However, above works contained no RIS. In RIS-aided MIMO-OFDM systems, there remains a significant gap in leveraging NLoS links for positioning. It is more challenging to estimate these parameters since RIS operates without any radio frequency (RF) links. Especially for a planar RIS-enabled MIMO-OFDM systems in environments with scatterers, the number of parameters to be estimated greatly increases, leading to a challenging high-dimensional parameter estimation problem. Our work deals with these challenges and fills this gap.

Motivated by these considerations, we study positioning
algorithms conceived for RIS-aided MIMO-OFDM systems in the presence of
practical scatterers, where an uplink MIMO-OFDM system
assisted by a single planar RIS is considered, and all
available channel parameters, including those on the VLoS
link made available by the RIS and the NLoS links due to the
presence of scatterers, are exploited to locate an MS in a 3D
space. A two-step positioning scheme is utilized,
according to which the channel parameters are first estimated,
and the location coordinates and rotation angle of the MS
are then obtained. For estimating the channel parameters,
we formulate an MLE problem, which results in two main
challenges: good initial values are required by the
algorithms and the problem has a high dimensionality.

To glean good initial values, we design an RIS phase
shift profile for enabling us to decouple the channel parameters,
and for estimating these parameters sequentially. Specifically, the
AODs at the MS are estimated by using the distributed
compressed sensing simultaneous orthogonal matching pursuit
(DCS-SOMP) algorithm \cite{tropp2005simultaneous} and the maximum likelihood
(ML) algorithm. The AOAs at the RIS are estimated by using
the DCS-SOMP algorithm as well, and the times of arrival
(TOAs) are estimated by using the discrete Fourier transform
(DFT) approach. To tackle the high dimensionality of the
MLE problem, the classic methods for low-complexity MLE under multipath scenarios mainly include the Expectation-Maximization (EM) algorithm \cite{moon1996expectation} and the Space-Alternating Generalized Expectation-Maximization (SAGE) algorithm \cite{fessler1994space}. While retaining the stability of the EM algorithm, the SAGE algorithm has the advantages of faster convergence and lower complexity over the EM algorithm \cite{fleury1999channel}. Therefore, we adopt the SAGE algorithm \footnote{Since the considered channel environment and the objective of parameter estimation differ from SAGE variants, such as the SAGE-MAP \cite{senol2012nondata} and Bayesian SAGE \cite{shutin2011sparse}, adopting SAGE variants is unnecessary.} for refined channel parameter estimation, where parameters associated with distinct paths are updated alternately, enabling near-optimal estimation performance. Once the channel parameters
are estimated, the location coordinates and the rotation angle
of the MS are calculated in a closed-form expression by using
geometric relationships. By taking these closed-form solutions
as initial values, the positioning problem is formulated as
an equivalent MLE problem, and it is solved by using the
Levenberg-Marquardt algorithm (LMA) \cite{marquardt1963algorithm}.

Specifically, the main contributions are
the following.
\begin{enumerate}
	\item While non-RIS systems have exploited NLoS links for user positioning \cite{shahmansoori2017position,li_joint_2022,gong2022multipath}, RIS-aided systems face challenges in extracting scatterer-related parameters, leading relevant works \cite{paper10,paper14,paper12,keykhosravi_ris-enabled_2022,paper13,fascista_ris-aided_2022,teng_bayesian_2022,teng_variational_2023} to neglect NLoS links or treat them as noise. Differently, we beneficially harness NLoS links to retrieve valuable information for designing 3D MS positioning algorithms in RIS-aided MIMO-OFDM systems.
	\item In existing works \cite{paper10,paper14,paper12,keykhosravi_ris-enabled_2022,paper13,fascista_ris-aided_2022,teng_bayesian_2022,teng_variational_2023}, scatterers are often ignored, and only the position (without orientation) of a single-antenna user is estimated in MISO/SISO systems with a ULA-based RIS. Thus, the parameter estimation problem therein has a low dimension. Different from these works, we consider a more general MIMO-OFDM system enabled by a UPA-based RIS with practical scatterers, and jointly estimate the position and orientation of the MS as well as the coordinates of the scatterers. Thus, a very high-dimensional parameter estimation problem is solved by our proposed algorithm.
	\item We propose a two-step positioning algorithm to first estimate the channel parameters and then, based on them, the positions and orientation. The channel parameters are first decoupled and coarsely estimated using DCS-OMP, MLE, DFT, and least squares (LS) methods, which are then refined by the SAGE algorithm. The positioning parameters are first obtained in closed-form by solving geometric equations, then refined by solving an equivalent MLE problem. This two-prong strategy of coarse estimation complemented by refined estimation ensures high estimation accuracy with low complexity.
	\item The CRLB for the estimates of the channel parameters,
	and the PEB/OEB for the position and orientation estimation are
	derived as performance benchmarks for the proposed
	algorithms. Simulation results show that the estimation
	errors of proposed algorithms approach the theoretic error
	bounds at relatively high transmit power.
\end{enumerate}
The remainder of this paper is organized as follows. Section \ref{System and Channel Model} introduces the system models. In Section \ref{Problem Formulation}, the estimation problems are formulated. The coarse estimation of the channel parameters is discussed in Section \ref{Coarse Estimation of Channel Parameters},
and the joint refinement is described in Section \ref{Fine Estimation of Channel Parameters}. Estimating coordinates and orientation angles of the MS is introduced in Section \ref{Position and Rotation Angle Estimation}. In Section \ref{Position and Orientation Estimation Fundermental Bounds}, the CRLB, PEB and OEB are derived. Section \ref{Complexity} analyzes the complexity. In Section \ref{Simulation},
simulation results are illustrated. Section
\ref{Conclusion} concludes this work.

\subsubsection*{Notations}
The symbols $[\cdot]^{*}$ , $[\cdot]^{T}$ and $[\cdot]^{H}$ denote
the conjugate, transpose and Hermitian transpose, respectively.
$[\mathbf{X}]_{u,v}$ denotes the element in the $u$-th row and $v$-th
column of matrix $\mathbf{X}$. $[\mathrm{\mathbf{x}}]_{u}$ denotes
the $u$-th element of vector $\mathbf{x}$. ${\mathbf{I}}_{M}$ represents
an $M\times M$ identity matrix. $\mathrm{diag}(\mathbf{x})$ represents
a diagonal matrix whose main diagonal is given by the elements of vector $\mathbf{x}$. The Hadamard and Kronecker products are denoted
by $\circ$ and $\otimes$, respectively. The symbols $\mathrm{Tr}(\cdot)$,
$\mathfrak{Re}(\cdot)$, $\mathfrak{Im}(\cdot)$, $\mathrm{det}(\cdot)$
and $\Vert\cdot\Vert$ denote the trace, the real part, the imaginary part, the determinant
and the norm, respectively. $\mathbb{E}[\cdot]$ denotes the expectation. $\left\lceil \cdot\right\rceil $ is
the ceiling function. $\ensuremath{\propto}$ denotes equality up to some irrelevant constants.
	
\section{System and Channel Model}
\label{System and Channel Model}
\subsection{System Model}
\label{System Model}
\begin{figure}[b]
	\centering \includegraphics[width=4.5in]{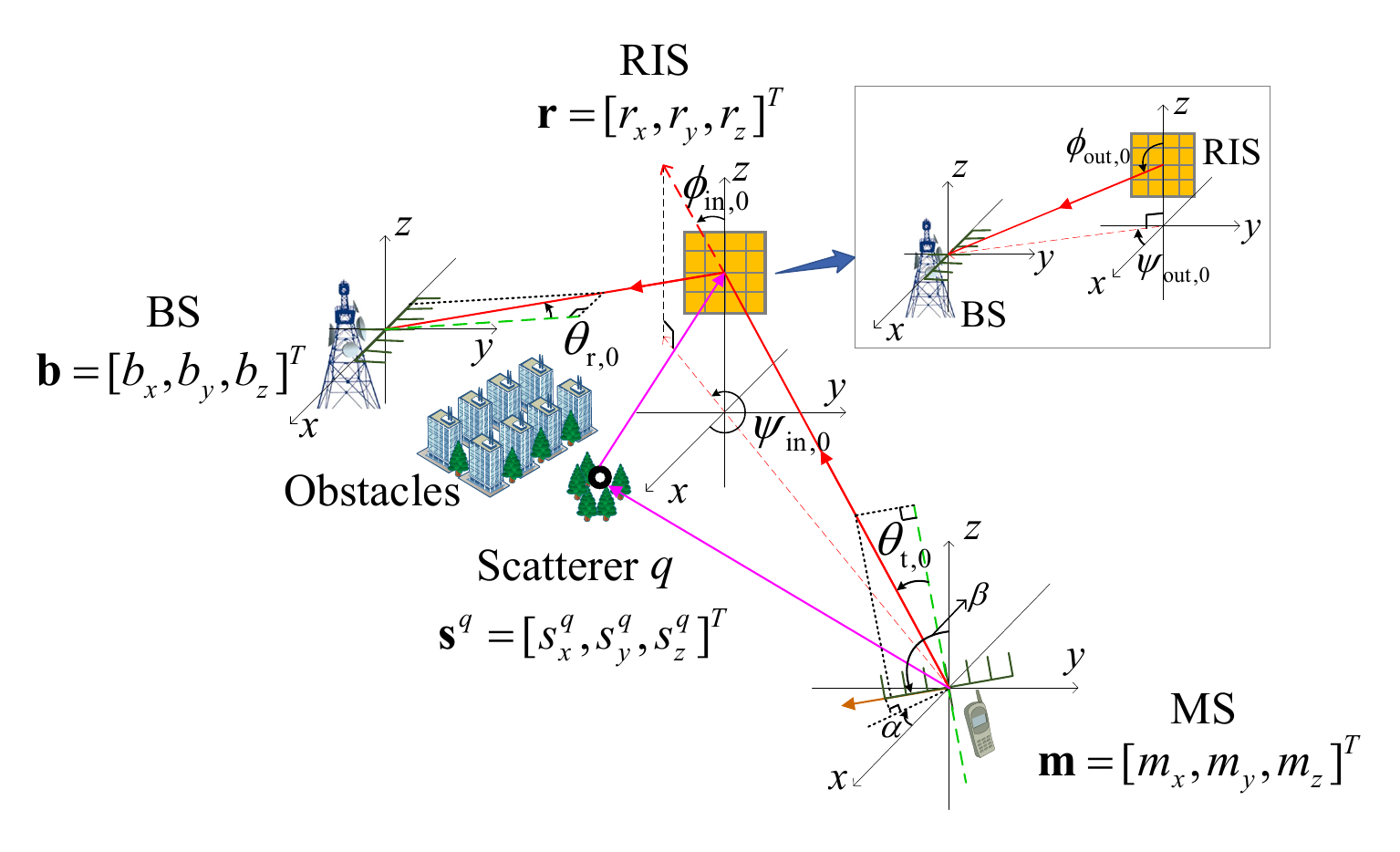}\caption{The RIS-aided positioning system.}
	\vspace{-0.3cm}
	\label{figpositioning}
\end{figure}

As shown in Fig. \ref{figpositioning}, we consider an uplink RIS-aided
mmWave MIMO system operating at a carrier frequency of $f_{c}$ (corresponding
to wavelength of $\lambda=c/f_{c}$) and bandwidth of $B$, in which
an RIS is adopted to locate the MS. The BS is equipped with a uniform
linear array (ULA) of $N_{{\rm b}}$ antennas, and the MS is equipped
with a ULA of $N_{{\rm m}}$ antennas. The reflecting elements on the
RIS constitute an UPA of $N_{{\rm r}}=N_{{\rm {a}}}N_{{\rm {e}}}$ antennas, where
$N_{{\rm {a}}}$ passive reflecting elements are on the horizontal direction
of the UPA, and $N_{{\rm {e}}}$ passive reflecting elements are on the vertical
direction of the UPA.

The ULA of the BS is placed parallel to the $x$ axis with the center
located at ${\bf b}=[b_{x},b_{y},b_{z}]^{T}\in\mathbb{R}^{3}$. The
UPA of the RIS is placed parallel to the ${y-o-z}$ plane with their centers
located at ${\bf r}=[r_{x},r_{y},r_{z}]^{T}$. Dividing the ULA at the MS into left and right halves about ${\bf m}=[m_{x},m_{y},m_{z}]^{T}$, the part nearer the positive $x$-axis is defined as the positive array direction (orange arrow in Fig. \ref{figpositioning}). The azimuth rotation angle of the MS defines the counter-clockwise rotation around the $x$-axis of the vector on the ${x-o-y}$ plane projected from positive direction of the ULA, which is $-\alpha$, $\alpha\in[0,\pi)$. The elevation rotation angle $\beta$ depicts the counter-clockwise rotation of the positive direction of the ULA around the $z$-axis, where $\beta\in[0,\pi)$. Then the unit direction
	vector of the positive direction of ULA at MS can be written as
	$\mathbf{e}_{\textrm{rot}}\triangleq[\sin\beta\cos\alpha,-\sin\beta\sin\alpha,\cos\beta]^{T}$.

We assume that the RIS and the BS are relatively close and thus there is only one path between them. On the other hand, since the MS is generally deployed on the ground with rich scatterers, we assume that there are $Q$ scatterers between the MS and the RIS. The
location of the $q$-th scatterer is denoted by ${\bf s}^{q}=[s_{x}^{q},s_{y}^{q},s_{z}^{q}]^{T},q=1,2,\cdots,Q$, which
is unknown. Generally, once the RIS and the BS are deployed, the
coordinates of $\mathbf{r}$ and ${\bf b}$ are known and invariant.

\subsection{Transmitter Model}

The OFDM signals are exploited for an uplink position estimation in
the RIS-aided mmWave MIMO system. At the MS, $T$ OFDM symbols with
$N$ subcarriers are transmitted sequentially. The transmitted signal
over subcarrier $n$ at the $t$-th time slot can be expressed as $\mathbf{x}_{t}[n]=\sqrt{P_{\mathrm{m}}}\mathbf{\bar{x}}_{t}[n]\in\mathbb{C}^{N_{m}\times1}$,
where $n=1,2,\cdots,N$ and $t=1,2,\cdots,T$. $P_{\mathrm{m}}$ is
the transmit power of the MS, and $\mathbf{\bar{x}}_{t}[n]$ is the
power-normalized pilot signal. We assume bandwidth
$B\ll f_{c}$ for a narrow-band model.

\subsection{Wireless Channel Model}

The mmWave propagation channels can be described by the array steering
vectors. Generally, the steering vector of an ULA can be represented
as
\begin{equation}
	\mathbf{f}(u_{{\rm sf}},N_{{\rm ant}})=[1,e^{-j2\pi u_{{\rm sf}}},\cdots,e^{-j2\pi(N_{{\rm ant}}-1)u_{{\rm sf}}}]^{T}\in\mathbb{C^{\mathit{N_{{\rm ant}}\times\mathrm{1}}}},\label{ULAsteeringVctr}
\end{equation}
where $u_{{\rm sf}}=\frac{d}{\lambda}\sin\theta$ is the spatial frequency
at the ULA corresponding to the physical angle $\theta$, $d$ is
the distance between adjacent antennas, and $N_{{\rm ant}}$ is the
number of antennas. We assume that $d\leq\frac{\lambda}{2}$ so that
there is a one-to-one relationship between the spatial frequencies
and the physical angles on either the ULA or UPA.

As shown in Fig. \ref{figpositioning}, there is only an LoS path
from the RIS to the BS. There are ($Q+1$) paths from the MS to the
RIS, in which the ($q=0$)-th path represents the LoS path, and the
others are the NLoS paths. We term the cascaded
path of the MS-RIS link and the RIS-BS link as the VLoS path. Then,
the channel of the RIS-BS link ${\bf H}_{{\rm RB}}[n]\in\mathbb{C}^{N_{{\rm b}}\times N_{\mathrm{r}}}$
and the channel of the MS-RIS link ${\bf H}_{{\rm MR}}[n]\in\mathbb{C}^{N_{\mathrm{r}}\times N_{\mathrm{m}}}$
can be respectively modeled as
\begin{align}
	{\bf H}_{{\rm RB}}[n] & =\delta_{{\rm RB},0}e^{-j2\pi\tau_{{\rm RB},0}\frac{(n-1)B}{N}}{\bf a}_{{\rm B}}(\vartheta_{{\rm r},0}){\bf a}_{{\rm R}}^{H}(\omega_{{\rm out},0}^{{\rm a}},\omega_{{\rm out},0}^{{\rm e}}),\label{SVmodlChnnlRIS2BS}\\
	{\bf H}_{{\rm MR}}[n] & =\sum\limits _{q=0}^{Q}\delta_{{\rm MR},q}e^{-j2\pi\tau_{{\rm MR},q}\frac{(n-1)B}{N}}{\bf a}_{{\rm R}}(\omega_{{\rm in},q}^{{\rm a}},\omega_{{\rm in},q}^{{\rm e}}){\bf a}_{{\rm M}}^{H}(\vartheta_{{\rm t},q}),\label{SVmodlChnnlMS2RIS}
\end{align}
where $\delta_{{\rm RB},0}$ and $\tau_{{\rm RB},0}$ are respectively
the channel gain and TOA of the direct RIS-BS path,
while $\delta_{{\rm MR},q}$ and $\tau_{{\rm MR},q}$ are respectively
the channel gain and TOA of the $q$-th path from the MS to the RIS.
By considering a narrow-band model, the signal wavelength at the $n$-th
subcarrier satisfies $\lambda_{n}\approx\lambda$, and the steering
vectors ${\bf a}_{{\rm B}}(\vartheta_{{\rm r},0})\in\mathbb{C^{\mathit{N_{{\rm b}}\times\mathrm{1}}}}$
at the BS, ${\bf a}_{{\rm M}}(\vartheta_{{\rm t},q})\in\mathbb{C^{\mathit{N_{{\rm m}}\times\mathrm{1}}}}$
at the MS, and ${\bf a}_{{\rm R}}(\omega_{x,y}^{{\rm {a}}},\omega_{x,y}^{{\rm {e}}})\in\mathbb{C^{\mathit{N_{{\rm r}}\times\mathrm{1}}}}$
at the RIS are the same for all $N$ subcarriers. These array steering
vectors can be represented by ${\bf a}_{{\rm B}}(\vartheta_{{\rm r},0})=\mathbf{f}(\vartheta_{{\rm r},0},N_{\mathrm{b}})$, ${\bf a}_{{\rm M}}(\vartheta_{{\rm t},q})=\mathbf{f}(\vartheta_{{\rm t},q},N_{\mathrm{m}})$ and ${\bf a}_{{\rm R}}(\omega_{x,y}^{{\rm {a}}},\omega_{x,y}^{{\rm {e}}})=\mathbf{f}(\omega_{x,y}^{{\rm {e}}},N_{{\rm {e}}})\otimes\mathbf{f}(\omega_{x,y}^{{\rm {a}}},N_{{\rm {a}}})$, where $\ensuremath{\vartheta_{{\rm r},0}=\frac{d_{\textrm{{\rm {BS}}}}}{\lambda}\sin\theta_{{\rm r},0}}$,
$\ensuremath{\vartheta_{{\rm t},q}=\frac{d_{\textrm{{\rm {MS}}}}}{\lambda}\sin\theta_{{\rm t},q}}$,
$\ensuremath{\omega_{x,y}^{{\rm {a}}}=\frac{d_{\textrm{{\rm {RIS,a}}}}}{\lambda}\sin\psi_{x,y}\sin\phi_{x,y}}$,
$\ensuremath{\omega_{x,y}^{{\rm {e}}}=\frac{d_{\textrm{{\rm {RIS,e}}}}}{\lambda}\cos\phi_{x,y}}$,
$\ensuremath{x\in\{{\rm out,in}\}}$, $y\in\{0,q\}$ are the spatial
frequencies.

The physical angles are defined as follows. The angles at ULA departed
from MS or arrived at BS are defined as the counter-clockwise
rotation angle of the wave vector with respect to the normal direction
of the ULA , which is on the plane formed by the line collinear with
the ULA and the line collinear with the wave vector, and is denoted
by the green dotted line in Fig. \ref{figpositioning}. Then, the
positive AODs $\theta_{\mathrm{t},0}$ ,$\theta_{\mathrm{t},q}$ at
MS and the positive AOA $\theta_{{\rm r,}0}$ at BS are obtained by
using wave vectors ${\bf r}-{\bf m}$, ${\bf s}_{q}-{\bf m}$ and
${\bf b}-{\bf r}$, respectively. The positive elevation angle at
UPA of RIS is defined as the counter-clockwise rotation
angle of the wave vector with respect to the positive $z$-axis, while
the positive azimuth angle at UPA of RIS is defined as the counter-clockwise
rotation angle of the vector on the $x-o-y$ plane projected from
the wave vector with respect to the positive $x$-axis. Then, the
positive elevation/azimuth AOA $\phi_{\mathrm{in},0}$/$\psi_{\mathrm{in},0}$,
elevation/azimuth AOA $\phi_{\mathrm{in},q}$/$\psi_{\mathrm{in},q}$,
and elevation/azimuth AOD $\phi_{{\rm out,}0}$/$\psi_{{\rm out,}0}$
at RIS are obtained by using wave vectors ${\bf r}-{\bf m}$, ${\bf s}_{q}-{\bf m}$
and ${\bf b}-{\bf r}$, respectively. These angles become negative
when the counter-clockwise rotation is changed to clockwise rotation.

By using \eqref{SVmodlChnnlRIS2BS} and \eqref{SVmodlChnnlMS2RIS},
the complete channel from the MS to the BS via the RIS is expressed
as \begin{align}
	{\bf H}_{t}[n] & ={\bf H}_{{\rm RB}}[n]\text{\ensuremath{\bm{{\Omega}}}}_{t}{\bf H}_{{\rm MR}}[n]\nonumber \\
	& =\sum\limits _{q=0}^{Q}\delta_{{\rm RB},0}\delta_{{\rm MR},q}e^{-j2\pi(\tau_{{\rm RB},0}+\tau_{{\rm MR},q})\frac{(n-1)B}{N}}{\bf a}_{{\rm B}}(\vartheta_{{\rm r},0}){\bf a}_{{\rm R}}^{H}(\omega_{{\rm out},0}^{{\rm a}},\omega_{{\rm out},0}^{{\rm e}})\text{\ensuremath{\bm{\Omega}}}_{t}{\bf a}_{{\rm R}}(\omega_{{\rm in},q}^{{\rm a}},\omega_{{\rm in},q}^{{\rm e}}){\bf a}_{{\rm M}}^{H}(\vartheta_{{\rm t},q})\nonumber \\
	& \triangleq\sum\limits _{q=0}^{Q}\delta_{q}e^{-j2\pi\tau_{q}\frac{(n-1)B}{N}}{\bf a}_{{\rm B}}(\vartheta_{{\rm r},0}){\bf a}_{{\rm R}}^{H}(\omega_{{\rm out},0}^{{\rm a}},\omega_{{\rm out},0}^{{\rm e}})\text{\ensuremath{\bm{\Omega}}}_{t}{\bf a}_{{\rm R}}(\omega_{{\rm in},q}^{{\rm a}},\omega_{{\rm in},q}^{{\rm e}}){\bf a}_{{\rm M}}^{H}(\vartheta_{{\rm t},q}),\label{TotalChnnl}
\end{align}
where $\ensuremath{\delta_{q}\triangleq\delta_{{\rm RB},0}\delta_{{\rm MR},q}}$
and $\ensuremath{\tau_{q}\triangleq\tau_{{\rm RB},0}+\tau_{{\rm MR},q}}$.
For later use, we introduce $\tilde{\delta}_{q}[n]\triangleq\delta_{q}e^{-j2\pi\tau_{q}\frac{(n-1)B}{N}}$.
$\text{\ensuremath{\bm{{\Omega}}}}_{t}=\textrm{diag}\{{\bf g}_{t}\}$
is a diagonal matrix with ${\bf g}_{t}$ on the diagonal. ${\bf g}_{t}$ contains the reflection coefficients of the
RIS reflecting elements at the $t$-th time slot, and can be represented by ${\bf g}_{t}=[e^{j\varphi_{t}^{(1)}},e^{j\varphi_{t}^{(2)}},\cdots,e^{j\varphi_{t}^{(N_{\textrm{r}})}}]^{T}$,
where $\varphi_{t}^{(n)},n=1,2,\cdots,N_{\textrm{r}}$
are the phase shifts of reflection coefficients. In \eqref{TotalChnnl}, there are $(Q+1)$ paths, and the $q$-th
path can be characterized by the channel
parameters $[\tau_{q},\delta_{q},\theta_{{\rm r},0},\phi_{{\rm out},0},\psi_{{\rm out},0},\phi_{{\rm in},q},\psi_{{\rm in},q},\theta_{{\rm t},q}]^{T},q=0,1,\cdots,Q$.

\subsection{Sparse Representation of mmWave Channel}

Based on \eqref{SVmodlChnnlRIS2BS} and \eqref{SVmodlChnnlMS2RIS},
the two channel matrices can be re-expressed in an angular domain
representation as \begin{subequations}
	\begin{align}
		{\bf H}_{{\rm RB}}[n] & ={\bf A}_{{\rm B}}\bm{{\varLambda}}_{\textrm{RB}}[n]\mathbf{A}_{{\rm R}}^{H},\label{SprsChnnlHRB}\\
		{\bf H}_{{\rm MR}}[n] & ={\bf A}_{{\rm R}}\bm{{\varLambda}}_{\textrm{MR}}[n]{\bf A}_{{\rm M}}^{H},\label{SprsChnnlHMR}
	\end{align}
\end{subequations} where the three matrices of ${\bf A}_{{\rm B}}\in\mathbb{C}^{N_{{\rm b}}\times G_{\mathrm{b}}}$($N_{{\rm b}}< G_{\mathrm{b}}$), ${\bf A}_{{\rm R}}\in\mathbb{C}^{N_{{\rm r}}\times G_{\mathrm{r}}}$($N_{{\rm r}}< G_{\mathrm{r}}$), and
${\bf A}_{{\rm M}}\in\mathbb{C}^{N_{{\rm m}}\times G_{\mathrm{m}}}$($N_{{\rm m}}< G_{\mathrm{m}}$)
are over-complete dictionary matrices in the angle domain. The columns
of ${\bf A}_{{\rm B}}$, ${\bf A}_{{\rm R}}$ and ${\bf A}_{{\rm M}}$
consist of the array steering vectors ${\bf a}_{{\rm B}}(\vartheta)$,
${\bf a}_{{\rm R}}(\omega^{{\rm a}},\omega^{{\rm e}})$ and ${\bf a}_{{\rm M}}(\vartheta)$
respectively at the specific angles. The $G_{\mathrm{b}}$, $G_{\mathrm{r}}$
and $G_{\mathrm{m}}$ are the numbers of the per-discretized grids.
Since there is only one path between the RIS and the BS, the $\bm{{\varLambda}}_{\textrm{RB}}[n]\in\mathbb{C}^{G_{\mathrm{b}}\times G_{\mathrm{r}}}$
is a sparse matrix containing only one non-zero element given by $\tilde{\delta}_{{\rm RB,}0}[n]\triangleq\delta_{{\rm RB,}0}e^{-j2\pi\tau_{{\rm RB,}0}\frac{(n-1)B}{N}}$.
The $\bm{{\varLambda}}_{\textrm{MR}}[n]\in\mathbb{C}^{G_{\mathrm{r}}\times G_{\mathrm{m}}}$
is a sparse matrix containing ($Q+1$) non-zero elements given by
$\{\tilde{\delta}_{{\rm MR,}q}[n]\triangleq\delta_{{\rm MR,}q}e^{-j2\pi\tau_{{\rm MR,}q}\frac{(n-1)B}{N}}\}_{q=0,1,\cdots,Q}$.
By using \eqref{ULAsteeringVctr}, ${\bf A}_{{\rm B}}$ and ${\bf A}_{{\rm M}}$
can be respectively written as \begin{subequations}
	\begin{align}
		{\bf A}_{{\rm B}} & =[\mathbf{f}(-1\frac{d_{\textrm{{\rm {BS}}}}}{\lambda},N_{\mathrm{b}}),\mathbf{f}((-1+\frac{2}{G_{\mathrm{b}}})\frac{d_{\textrm{{\rm {BS}}}}}{\lambda},N_{\mathrm{b}}),\cdots,\mathbf{f}((1-\frac{2}{G_{\mathrm{b}}})\frac{d_{\textrm{{\rm {BS}}}}}{\lambda},N_{\mathrm{b}})],\\
		{\bf A}_{{\rm M}} & =[\mathbf{f}(-1\frac{d_{\textrm{{\rm {MS}}}}}{\lambda},N_{\mathrm{m}}),\mathbf{f}((-1+\frac{2}{G_{\mathrm{m}}})\frac{d_{\textrm{{\rm {MS}}}}}{\lambda},N_{\mathrm{m}}),\cdots,\mathbf{f}((1-\frac{2}{G_{\mathrm{m}}})\frac{d_{\textrm{{\rm {MS}}}}}{\lambda},N_{\mathrm{m}})].
	\end{align}
\end{subequations}

Since the UPA is utilized at the RIS, ${\bf A}_{{\rm R}}$ can be respectively
written as
\begin{align}
	{\bf A}_{{\rm R}} & ={\bf A}_{{\rm R}}^{{\rm e}}\otimes{\bf A}_{{\rm R}}^{{\rm a}},\label{OverCompleteMatrxRIS}
\end{align}
where ${\bf A}_{{\rm R}}^{{\rm e}}\in\mathbb{C}^{N_{{\rm {e}}}\times G_{\mathrm{e}}}$
and ${\bf A}_{{\rm R}}^{{\rm a}}\in\mathbb{C}^{N_{{\rm {a}}}\times G_{\mathrm{a}}}$
can be written as\begin{subequations}
	\begin{align}
		{\bf A}_{{\rm R}}^{{\rm a}} & =[\mathbf{f}(-1\frac{d_{\textrm{{\rm {RIS,a}}}}}{\lambda},N_{{\rm {a}}}),\mathbf{f}((-1+\frac{2}{G_{\mathrm{a}}})\frac{d_{\textrm{{\rm {RIS,a}}}}}{\lambda},N_{{\rm {a}}}),\cdots,\mathbf{f}((1-\frac{2}{G_{\mathrm{a}}})\frac{d_{\textrm{{\rm {RIS,a}}}}}{\lambda},N_{{\rm {a}}})],\\
		{\bf A}_{{\rm R}}^{{\rm e}} & =[\mathbf{f}(-1\frac{d_{\textrm{{\rm {RIS,e}}}}}{\lambda},N_{{\rm {e}}}),\mathbf{f}((-1+\frac{2}{G_{\mathrm{e}}})\frac{d_{\textrm{{\rm {RIS,e}}}}}{\lambda},N_{{\rm {e}}}),\cdots,\mathbf{f}((1-\frac{2}{G_{\mathrm{e}}})\frac{d_{\textrm{{\rm {RIS,e}}}}}{\lambda},N_{{\rm {e}}})],
	\end{align}
\end{subequations} and $G_{\mathrm{r}}=G_{\mathrm{a}}\times G_{\mathrm{e}}$.

\subsection{Received Signal Model}

We assume that the direct path from the MS to the BS is obstructed
by obstacles. Then, the received signal at the $t$-th time slot of the $n$-th subcarrier can be written as
\begin{align}
	\mathbf{y}_{t}[n] & ={\bf H}_{t}[n]\mathbf{x}_{t}[n]+\mathbf{z}_{t}[n]\nonumber \\
	& ={\bf H}_{{\rm RB}}[n]\text{\ensuremath{\bm{{\Omega}}}}_{t}{\bf H}_{{\rm MR}}[n]\mathbf{x}_{t}[n]+\mathbf{z}_{t}[n],\label{ReceivdYt}
\end{align}
where $t=1,2,\cdots,T$. $\mathbf{x}_{t}[n]$ is the pilot signal
transmitted at the $t$-th time slot, which is known at the BS.
The transmit power of the pilot signal is given by $P_{\mathrm{m}}=\left\Vert \mathbf{x}_{t}[n]\right\Vert ^{2}$.
$\mathbf{z}_{t}[n] \sim \mathcal{CN}(0,\sigma^{2}\mathbf{I}_{N_{{\rm b}}})$ is the noise vector, where $\sigma^{2}$ denotes the noise power.

\section{Estimation Problem Formulation}
\label{Problem Formulation}

\subsection{Parameters to be Estimated}

From the received signal $\mathbf{y}_{t}[n]$ in \eqref{ReceivdYt},
we should first estimate all the unknown channel parameters, which
are collected in the vector $\bm{{\eta}}\in\mathbb{C}^{6(Q+1)}$ as
\begin{align}
	\bm{{\eta}} & =[\bm{{\eta}}_{0}^{T},\bm{{\eta}}_{1}^{T},\cdots,\bm{{\eta}}_{q}^{T},\cdots,\bm{{\eta}}_{Q}^{T}]^{T},\label{Channel_Parameters}
\end{align}
where $\bm{{\eta}}_{q}=[\tau_{q},\bm{{h}}_{q}^{T},\bm{{\theta}}_{q}^{T}]^{T}$, $\bm{{h}}_{q}=[\delta_{q,\textrm{R}},\delta_{q,\textrm{I}}]^{T}$, $\delta_{q,\textrm{R}}=\mathfrak{Re}(\delta_{q})$, $\delta_{q,\textrm{I}}=\mathfrak{Im}(\delta_{q})$ and $\bm{{\theta}}_{q}=[\theta_{{\rm t},q},\phi_{{\rm in},q},\psi_{{\rm in},q}]^{T}$. We term these
channel parameters as the intermediate parameters.

Then, from these intermediate parameters, we estimate the location
coordinates of the MS, rotation angles of the MS, location coordinates
of the scatterer, etc., which are termed as the final parameters. We
collect the final parameters in a vector $\tilde{\bm{{\eta}}}\in\mathbb{C}^{2(Q+1)+3Q+5}$
, which can be represented as
\begin{align}
	\tilde{\bm{{\eta}}} & =[\bm{{h}}_{0}^{T},\cdots,\bm{{h}}_{q}^{T},\cdots,\bm{{h}}_{Q}^{T},\tilde {\bf {m}}^{T},({\bf s}^{1})^{T},\cdots,({\bf s}^{q})^{T},\cdots,({\bf s}^{Q})^{T}]^{T},
\end{align}
where $\tilde{\bf {m}}=[{\bf m}^{T},\alpha,\beta]^{T}$. The final parameters
in $\tilde{\bm{{\eta}}}$ are related with the intermediate
parameters $\boldsymbol{{\bf \eta}}$ according to the geometric relationship
as follows.

\subsection{The Geometric Relationship}

\label{The Geometric Relationship}
Obviously, the channel parameters $\{\theta_{{\rm r},0},\phi_{{\rm out},0},\psi_{{\rm out},0}\}$
in the RIS-BS link can be calculated directly from the coordinates
of $\mathbf{b}$ and $\mathbf{r}$. Other channel parameters $\{\tau_{q},\delta_{q},\phi_{{\rm in},q},\psi_{{\rm in},q},\theta_{{\rm t},q}\}_{q=0,1,\cdots,Q}$
are determined by the location coordinates ${\bf m}$ of the MS, rotation angles $\alpha,\beta$ of the MS, and location coordinates ${\bf s}^{q}$
of the scatterer according to the geometric relationship described in
Fig. \ref{figpositioning}, which is illustrated by equations
as follows.

\subsubsection{The Angles in the MS-RIS Link}

The angles of the LoS path from the MS to the RIS are $\theta_{{\rm t},0}$,
$\psi_{{\rm in},0}$ and $\phi_{{\rm in},0}$. These angles are related
to $\tilde{\bf {m}}$ as \begin{subequations} \label{AnglesMS-RIS-VLOS}
	\begin{align}
		\theta_{{\rm t},0} & =\arcsin\frac{({\bf r}-{\bf m})^{T}\mathbf{e}_{\textrm{rot}}}{\left\Vert {\bf r}-{\bf m}\right\Vert _{2}},\label{thetat0_revised}\\
		\psi_{{\rm in},0} & =\pi-\arcsin\frac{r_{y}-m_{y}}{\sqrt{(r_{x}-m_{x})^{2}+(r_{y}-m_{y})^{2}}},\label{psi_in0_revised}\\
		\phi_{{\rm in},0} & =\arccos\frac{r_{z}-m_{z}}{\left\Vert {\bf r}-{\bf m}\right\Vert _{2}}.
	\end{align}
\end{subequations}

The angles of the NLoS path from the MS to the RIS via scatterers are $\theta_{{\rm t},q}$,
$\psi_{{\rm in},q}$ and $\phi_{{\rm in},q}$. These angles are related
to $\tilde{\bf {m}}$ and ${\bf s}^{q}$ as \begin{subequations}
	\label{AnglesMS-RIS-NLOS}
	\begin{align}
		\theta_{{\rm t},q} & =\arcsin\frac{({\bf s}^{q}-{\bf m})^{T}\mathbf{e}_{\textrm{rot}}}{\left\Vert {\bf s}^{q}-{\bf m}\right\Vert _{2}},\label{thetatq_revised}\\
		\psi_{{\rm in},q} & =\pi-\arcsin\frac{r_{y}-s_{y}^{q}}{\sqrt{(r_{x}-s_{x}^{q})^{2}+(r_{y}-s_{y}^{q})^{2}}},\label{psi_inq_revised}\\
		\phi_{{\rm in},q} & =\arccos\frac{r_{z}-s_{z}^{q}}{\left\Vert {\bf r}-{\bf s}^{q}\right\Vert _{2}},
	\end{align}
\end{subequations} where $q=1,2,\cdots,Q$.

\subsubsection{The Angles in the RIS-BS Link}

The angles of the LoS path from the RIS to the BS are $\theta_{{\rm r},0}$,
$\psi_{{\rm out},0}$ and $\phi_{{\rm out},0}$. These angles are
related to $\mathbf{b}$ and $\mathbf{r}$ as \begin{subequations}
	\label{AnlgesRIS-BS-VLOS}
	\begin{align}
		\theta_{{\rm r},0} & =\arcsin\frac{b_{x}-r_{x}}{\left\Vert {\bf b}-{\bf r}\right\Vert _{2}},\\
		\psi_{{\rm out},0} & =\arcsin\frac{b_{y}-r_{y}}{\sqrt{(b_{x}-r_{x})^{2}+(b_{y}-r_{y})^{2}}},\label{psi_out0_revised}\\
		\phi_{{\rm out},0} & =\arccos\frac{b_{z}-r_{z}}{\left\Vert {\bf b}-{\bf r}\right\Vert _{2}}. \label{phi_out0_revised}
	\end{align}
\end{subequations}
It is noted that $\theta_{{\rm r},0}$, $\psi_{{\rm out},0}$
and $\phi_{{\rm out},0}$ can be calculated directly from $\mathbf{b}$
and $\mathbf{r}$, and can be taken as known values.

\subsubsection{The TOAs in All Paths}

The TOA is equivalent to the distance. Thus, the TOAs are related to the
coordinates of all nodes as \begin{subequations} \label{Distances}
	\begin{align}
		\tau_{0} & =\left\Vert {\bf r}-{\bf b}\right\Vert _{2}/c+\left\Vert {\bf m}-{\bf r}\right\Vert _{2}/c,\label{DistVlos}\\
		\tau_{q} & =\left\Vert {\bf r}-{\bf b}\right\Vert _{2}/c+\left\Vert {\bf s}^{q}-{\bf r}\right\Vert _{2}/c+\left\Vert {\bf m}-{\bf s}^{q}\right\Vert _{2}/c,{\ }q=1,\cdots,Q.\label{DistScttr0q}
	\end{align}
\end{subequations}

\begin{remark} According to the definition of physical angles above,
	the positive azimuth angle at the RIS is defined as the anticlockwise
	rotation angle of the projection vector with respect to the positive
	$x$ axis, while the clockwise rotation angle of the projection vector
	with respect to the positive $x$ axis is the negative azimuth angle.
	Thus, from the system model described in Fig. \ref{figpositioning},
	the ranges of azimuth angles are $\psi_{{\rm in},0}\in[\pi,3\pi/2]$,
	$\psi_{{\rm in},q}\in[\pi/2,\pi]$ and $\psi_{{\rm out},0}\in[-\pi/2,0]$,
	respectively. Since the function of $\arcsin(\cdotp)$ can only generate
	angles within $[-\pi/2,\pi/2]$, the expressions in \eqref{psi_in0_revised}
	and \eqref{psi_inq_revised} are utilized to ensure the correct angle
	range based on the periodicity of $\sin(\cdotp)$ function. \end{remark}

\subsection{Estimation Framework}
The classic two-step positioning strategy is utilized in this paper.
Our task is to first estimate the intermediate parameters in $\boldsymbol{{\bf \eta}}$
from the received signal, then estimate the final parameters in $\tilde{\bm{{\eta}}}$
from the estimated ${\boldsymbol{{\bf \eta}}}$. Thus, two parameter
estimation problems should be formulated and solved. 	\vspace{-0.8cm}
\begin{figure}[h]
	\centering \includegraphics[width=4.5in]{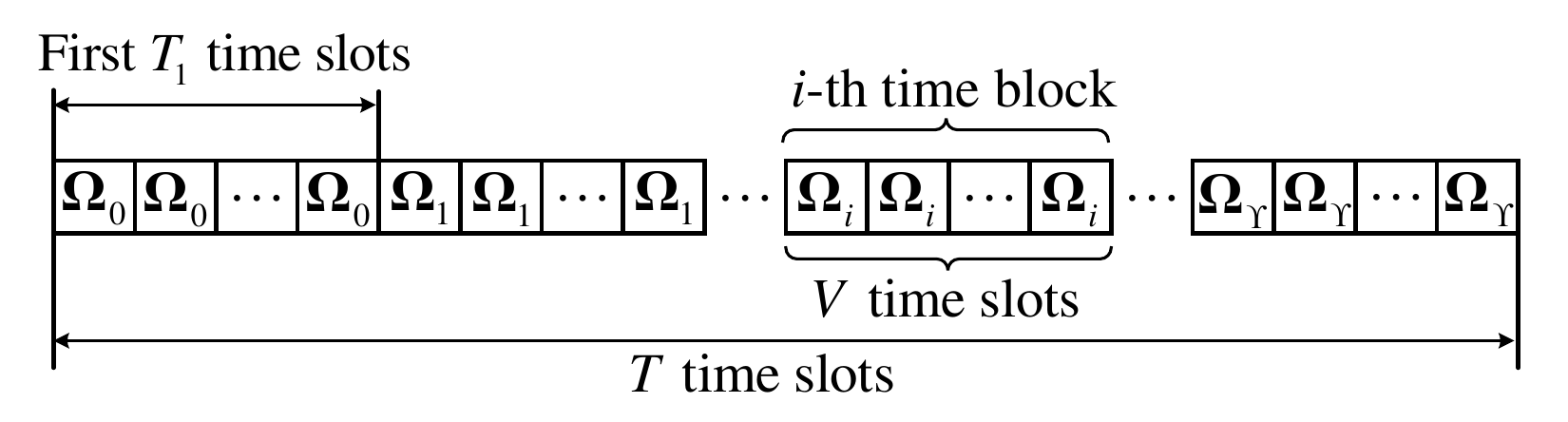}\caption{The configuration for the RIS phase shifts.}
	\vspace{-0.6cm}
	\label{phaseProtocol}
\end{figure}
\begin{figure}[b!]
	\centering \includegraphics[width=6.5in]{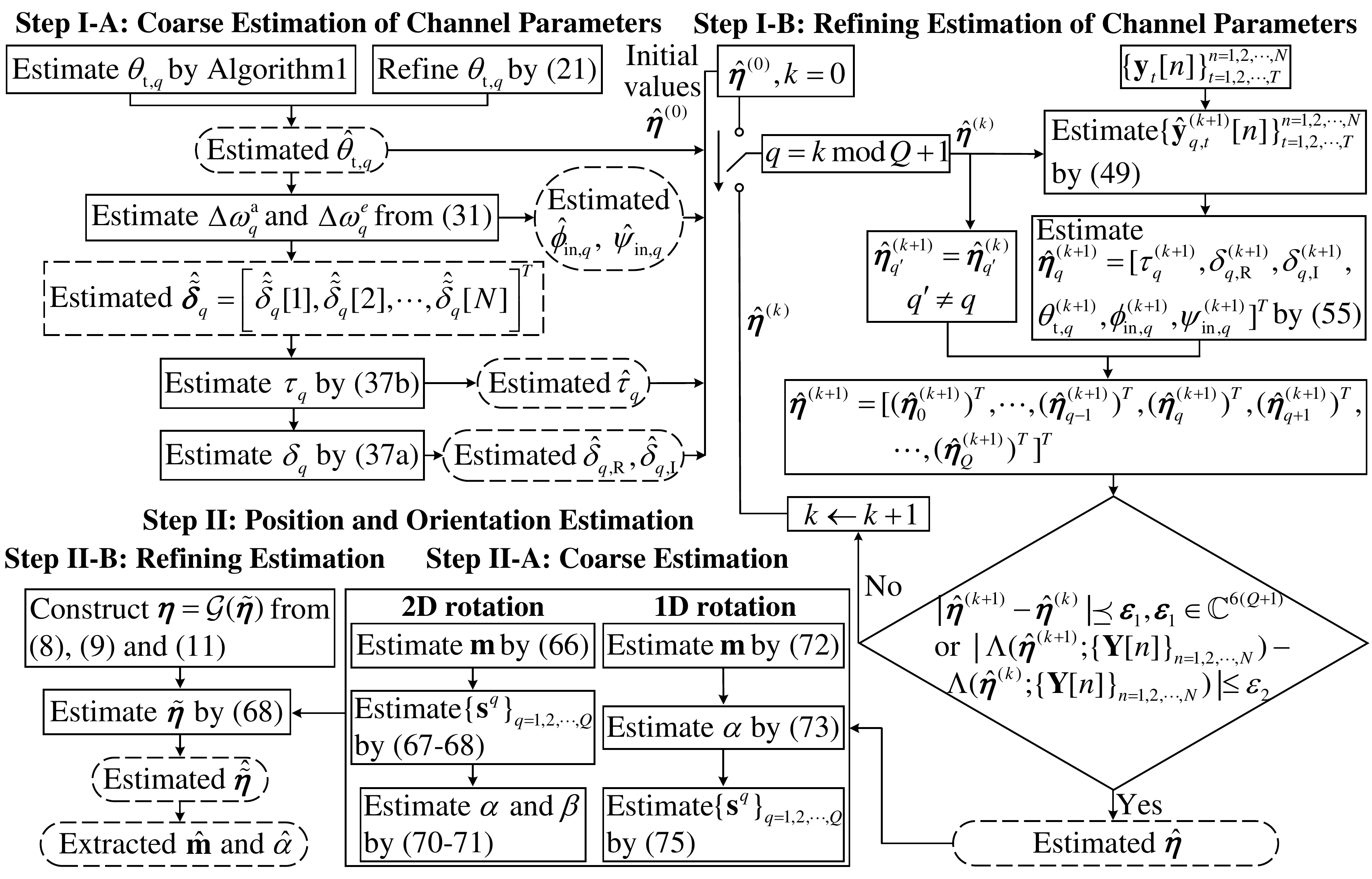}\caption{The flow chart of the proposed positioning algorithm.}
	\label{flowchart}
\end{figure}

The first parameter estimation problem is to estimate the channel
parameters in $\boldsymbol{{\bf \eta}}$ from the received signal,
which is solved by a coarse estimation process and a refining process
sequentially. In the coarse estimation process, to circumvent the
high-dimensional parameter estimation issue, we propose to estimate the
channel parameters separately by compression sensing algorithms, thus
the phase shifts of the RIS are specifically configured as shown
in Fig. \ref{phaseProtocol}. Assume that the total number of time slots utilized
for localization is $T$. In the first $T_{1}$ time slots, the phase
shifts of the RIS are time-invariant. The remaining $T_{2}=T-T_{1}$ time slots
are divided into $\Upsilon$ time blocks, each of which has
$V$ time slots. The
phase shifts of the RIS are time-invariant in each time block, and are time-variant over different time blocks. The coarsely-estimated
channel parameters are provided as the initial values of the refining
process, where the SAGE algorithm is utilized. The SAGE algorithm
belongs to the ML algorithms, and approaches the performance of the ML algorithms
with greatly reduced complexity. 

The second parameter estimation problem is to estimate the position-related
parameters in $\tilde{\bm{{\eta}}}$ from the estimated channel parameters
$\hat{\boldsymbol{{\bf \eta}}}$. Due to the unavoidable estimation
error in $\hat{\boldsymbol{{\bf \eta}}}$, the geometric relationship
described in Section \ref{The Geometric Relationship} can only be
utilized to calculate the initial values for $\tilde{\bm{{\eta}}}$,
and then an equivalent ML algorithm is exploited to refine the estimation
for $\tilde{\bm{{\eta}}}$.

The flow chart of the proposed positioning algorithm
is given in Fig. \ref{flowchart}.

\section{Coarse Estimation of Channel Parameters}

\label{Coarse Estimation of Channel Parameters}
The main purpose of coarse estimation of channel parameters in this section is to provide good initial values for the SAGE algorithm in the refining process. Based on the designed configuration of RIS phase shifts, the channel parameters of AODs, AOAs and TOAs can be estimated separately.

The channel parameters of ${\bf \boldsymbol{\eta}}$ are coarsely estimated
in three sub-steps sequentially. In the first sub-step, we estimate the AOD $\theta_{{\rm t},q},q=0,1,2,\cdots,Q$
at the MS in the first $T_{1}$ time slots, where the phase shifts
of the RIS are time-invariant. Then, by using the estimated AODs $\theta_{{\rm t},q},q=0,1,2,\cdots,Q$, we estimate
$\omega_{q}\triangleq\cos\phi_{{\rm in},q}-\cos\phi_{{\rm out},0}$,
$\gamma_{q}\triangleq\sin\psi_{{\rm in},q}\sin\phi_{{\rm in},q}-\sin\psi_{{\rm out},0}\sin\phi_{{\rm out},0}$
and $\tilde{\delta}_{q}[n]$
in the overall $T$ time slots in the second sub-step, where $q=0,1,2,\cdots,Q$. Since
$\phi_{{\rm out},0}$ and $\psi_{{\rm out},0}$ can be calculated
directly according to \eqref{psi_out0_revised} and \eqref{phi_out0_revised}, $\phi_{{\rm in},q}$ and $\psi_{{\rm in},q}$ can be estimated
from $\omega_{q}$ and $\gamma_{q}$ in the second sub-step. In the third
sub-step, by using the estimated values of $\tilde{\delta}_{q}[n],n=1,2,\cdots,N$
in the second sub-step, the TOAs $\tau_{q}$ and channel gains
$\delta_{q}$, $q=0,1,2,\cdots,Q$ are estimated by the DFT approach
and a time rotation operation.

\subsection{Estimation of the AODs at the MS}

\label{Estimation of the AODs at the MS}

\subsubsection{Estimate AODs by the DCS-SOMP Algorithm}

By substituting \eqref{SprsChnnlHMR} into \eqref{ReceivdYt}, the
pilot signal received during the first $T_{1}$ time slots can be
written as
\begin{align}
	\mathbf{y}_{t}[n] & ={\bf H}_{{\rm RB}}[n]\text{\ensuremath{\bm{{\Omega}}}}_{0}{\bf A}_{{\rm R}}\bm{{\varLambda}}_{\textrm{MR}}[n]{\bf A}_{{\rm M}}^{H}\mathbf{x}_{t}[n]+\mathbf{z}_{t}[n], \forall n=1,\cdots, N,
\end{align}
where $t=1,2,\cdots,T_1$. The conjugate transpose
of $\mathbf{y}_{t}[n]$ can be expressed as
\begin{align}
	\mathbf{y}_{t}^{H}[n] =\mathbf{x}_{t}^{H}[n]{\bf A}_{{\rm M}}\text{\ensuremath{\bm{{\Gamma}}}}_{\textrm{MR}}[n]+\mathbf{z}_{t}^{H}[n],
\end{align}
where $\text{\ensuremath{\bm{{\Gamma}}}}_{\textrm{MR}}[n]\triangleq\bm{{\varLambda}}_{\textrm{MR}}^{H}[n]{\bf A}_{{\rm R}}^{H}\text{\ensuremath{\bm{{\Omega}}}}_{0}^{H}{\bf H}_{{\rm RB}}^{H}[n]$. Since $\bm{{\varLambda}}_{\textrm{MR}}[n]$ is a sparse matrix with
$(Q+1)$ nonzero elements, $\text{\ensuremath{\bm{{\Gamma}}}}_{\textrm{MR}}[n]\in\mathbb{C}^{G_{\textrm{r}}\times N_{\textrm{b}}}$
is a row sparse matrix with $(Q+1)$ nonzero rows. By collecting the
received $\mathbf{y}_{t}^{H}[n]$ in the $T_{1}$ time slots, we have
\begin{align}
	\mathbf{Y}_{1}^{H}[n] & =\mathbf{X}_{1}^{H}[n]{\bf A}_{{\rm M}}\text{\ensuremath{\bm{{\Gamma}}}}_{\textrm{MR}}[n]+\mathbf{Z}_{1}^{H}[n]\nonumber \\
	& =\mathbf{\bm{{\Theta}}}_{{\rm {M}}}[n]\text{\ensuremath{\bm{{\Gamma}}}}_{\textrm{MR}}[n]+\mathbf{Z}_{1}^{H}[n],\label{AODofMS_ReceivData}
\end{align}
where $\mathbf{Y}_{1}[n]=[\mathbf{y}_{1}[n],\mathbf{y}_{2}[n],\cdots,\mathbf{y}_{T_{1}}[n]]\in\mathbb{C}^{N_{\textrm{b}}\times T_{1}}$,
$\mathbf{X}_{1}[n]=[\mathbf{x}_{1}[n],\mathbf{x}_{2}[n],\cdots,\mathbf{x}_{T_{1}}[n]]\in\mathbb{C}^{N_{\textrm{m}}\times T_{1}}$,
$\mathbf{Z}_{1}[n]=[\mathbf{z}_{1}[n],\mathbf{z}_{2}[n],\cdots,\mathbf{z}_{T_{1}}[n]]\in\mathbb{C}^{N_{\textrm{b}}\times T_{1}}$,
and $\mathbf{\bm{{\Theta}}}_{{\rm {M}}}[n]\triangleq\mathbf{X}_{1}^{H}[n]{\bf A}_{{\rm M}}$.
Note that the $k$-th row of $\text{\ensuremath{\bm{{\Gamma}}}}_{\textrm{MR}}[n]$
is nonzero only when the nonzero element of $\bm{{\varLambda}}_{\textrm{MR}}[n]$
exists on the $k$-th row of $\bm{{\varLambda}}_{\textrm{MR}}^{H}[n]$.
The $k$-th row of $\bm{{\varLambda}}_{\textrm{MR}}^{H}[n]$ is
associated with the $k$-th column of the dictionary matrix ${\bf A}_{{\rm M}}$,
which corresponds to one of the transmit array steering vectors $\{{\bf a}_{{\rm M}}^{H}(\theta_{{\rm t},q})\}_{q=0,1,\cdots,Q}$.
Thus, the task of the first sub-step is to determine the $(Q+1)$ columns
of ${\bf A}_{{\rm M}}$ corresponding to non-zero rows of $\text{\ensuremath{\bm{{\Gamma}}}}_{\textrm{MR}}[n]$
based on the received signal matrix $\mathbf{Y}_{1}^{H}[n]$ in \eqref{AODofMS_ReceivData},
which is a sparse recovery problem that can be solved by the DCS-SOMP
algorithm \cite{tropp2005simultaneous} given in Algorithm \ref{DCS-OMP}. For simplicity, we assume that the pilot
signal $\mathbf{X}_{1}[n]$ is the same over all subcarriers, which
can be denoted by $\mathbf{X}_{1}$. To achieve the best performance of the DCS-SOMP algorithm, the columns in the equivalent dictionary matrix $\mathbf{\bm{{\Theta}}}_{{\rm {M}}}$ should be mutually-orthogonal, thus the pilot signal $\mathbf{X}_{1}$ should be properly designed. The analytic designing method for $\mathbf{X}_{1}$ can be found in \cite{zhou2022channel}, and a simpler designing method is to randomly generate the elements in $\mathbf{X}_{1}$ \cite{wei2021channel}.

\begin{algorithm}
	\caption{DCS-SOMP algorithm for AOD estimation at the MS}
	\label{DCS-OMP} \begin{algorithmic}[1] \STATE \textbf{Parameter
			Setting:} Initialize ${\bar{\bm{{\Theta}}}}_{{\rm {M}}}$ and
		$\bar{{\bf A}}_{{\rm M}}$ as empty matrices. $\mathcal{K}_{0}$ is
		chosen to be an empty set. Set $\mathbf{R}_{\textrm{res}}[n]=\mathbf{Y}_{1}^{H}[n]$, $\forall n$
		and ${\bm{{\Theta}}}_{{\rm {M}}}=\mathbf{X}_{1}^{H}{\bf A}_{{\rm M}}$.
		
		\FOR {$q\leq Q$}
		
		\STATE $\bm{{\Psi}}[n]=\mathbf{\bm{{\Theta}}}_{{\rm {M}}}^{H}\mathbf{R}_{\textrm{res}}[n]$, $\forall n$;
		
		\STATE $k=\arg\underset{l}{\max}{\sum}_{n=1}^{N}\{\bm{{\Psi}}[n]\bm{{\Psi}}^{H}[n]\}_{l,l}$;
		
		\STATE Update the set of AOD indices $\mathcal{K}_{0}=\mathcal{K}_{0}\cup\{k\}$;
		
		\STATE ${\bar{\bm{{\Theta}}}}_{{\rm {M}}}=[{\bar{\bm{{\Theta}}}}_{{\rm {M}}},{\bm{{\Theta}}}_{{\rm {M}}}(:,k)]$;
		
		\STATE $\bar{{\bf A}}_{{\rm M}}=[\bar{{\bf A}}_{{\rm M}},{\bf A}_{{\rm M}}(:,k)]$;
		
		\STATE $\text{\ensuremath{\bm{{\Gamma}}}}_{\textrm{MR}}[n]=({\bar{\bm{{\Theta}}}}_{{\rm {M}}}^{H}{\bar{\bm{{\Theta}}}}_{{\rm {M}}})^{-1}{\bar{\bm{{\Theta}}}}_{{\rm {M}}}^{H}\mathbf{R}_{\textrm{res}}[n]$, $\forall n$;
		
		\STATE $\mathbf{R}_{\textrm{res}}[n]=\mathbf{Y}_{1}^{H}[n]-{\bar{\bm{{\Theta}}}}_{{\rm {M}}}\text{\ensuremath{\bm{{\Gamma}}}}_{\textrm{MR}}[n]$, $\forall n$;
		
		\ENDFOR
		
		\STATE \textbf{Output:} $\mathcal{K}_{0},\bar{{\bf A}}_{{\rm M}}(:,q)=\mathbf{f}((-1+\frac{2(k-1)}{G_{\mathrm{m}}})\frac{d_{\textrm{{\rm {MS}}}}}{\lambda},N_{\mathrm{m}})$
		and $\hat{\theta}_{{\rm t},q}=\arcsin[-1+\frac{2(k-1)}{G_{\mathrm{m}}}]$.
		
	\end{algorithmic}
\end{algorithm}

\begin{remark} \label{Remark_DCS_OMP} According to \cite{paper16},
	to find an $m$-sparse complex signal (vector) of dimension $n$,
	the number of measurements $T$ must satisfy $T>8m-2$. The dimension
	of the equivalent sensing matrix ${\bm{{\Theta}}}_{{\rm {M}}}$
	is $T_{1}\times G_{\mathrm{m}}$, and the corresponding sparsity level
	is $(Q+1)$, thus the pilot overhead required for the AOD estimation should
	satisfy $T_{1}\geq8(Q+1)-2$. \end{remark}

\subsubsection{Refining the AODs by the MLE}

\label{Refine AODs by MLE}

Due to the fact that the resolution of the DCS-SOMP algorithm is limited by the
number of the pre-discretized grids in the dictionary matrix, there
exists a mismatch between the discrete estimated angle and the real
continuous angle. Moreover, the estimation errors in AODs will greatly impact the accuracy of estimations in the next two sub-steps. To reduce
the error propagation in the next two sub-steps, we adopt the maximum
likelihood estimation (MLE) to further refine the AOD estimations, where the first $T_{1}$ time slots are utilized, and the AODs estimated by the DCS-SOMP algorithm are taken as the initial
values.

According to \eqref{TotalChnnl} and \eqref{ReceivdYt}, the received
signal over the first $T_{1}$ time slots can be rewritten as
\begin{align}
	\mathbf{y}_{t}[n] & =\ensuremath{\sum\limits _{q=0}^{Q}\delta_{q}e^{-j2\pi\tau_{q}\frac{(n-1)B}{N}}{\bf a}_{{\rm B}}(\vartheta_{{\rm r},0}){\bf a}_{{\rm R}}^{H}(\omega_{{\rm out},0}^{{\rm a}},\omega_{{\rm out},0}^{{\rm e}})\text{\ensuremath{\bm{{\Omega}}}}_{0}{\bf a}_{{\rm R}}(\omega_{{\rm in},q}^{{\rm a}},\omega_{{\rm in},q}^{{\rm e}}){\bf a}_{{\rm M}}^{H}(\vartheta_{{\rm t},q}){\bf x}_{t}[n]+{\bf z}_{t}[n]}\nonumber \\
	& =\ensuremath{\sum\limits _{q=0}^{Q}\check{\delta}_{q}[n]\tilde{\bf {H}}_{q}{\bf x}_{t}[n]+{\bf z}_{t}[n]}\nonumber \\
	& =\ensuremath{\tilde{\bf {X}}_{t}[n]\check{\boldsymbol{{\bf \delta}}}[n]+{\bf z}_{t}[n]}, t=1,2,\cdots,T_1,\label{AODfine_ReceivdYt}
\end{align}
where $\check{\delta}_{q}[n]\triangleq\delta_{q}e^{-j2\pi\tau_{q}\frac{(n-1)B}{N}}{\bf a}_{{\rm R}}^{H}(\omega_{{\rm out},0}^{{\rm a}},\omega_{{\rm out},0}^{{\rm e}})\text{\ensuremath{\bm{{\Omega}}}}_{0}{\bf a}_{{\rm R}}(\omega_{{\rm in},q}^{{\rm a}},\omega_{{\rm in},q}^{{\rm e}})$,
$ \tilde{\bf{H}}_{q}\triangleq{\bf a}_{{\rm B}}(\vartheta_{{\rm r},0}){\bf a}_{{\rm M}}^{H}(\vartheta_{{\rm t},q})$,
$\check{\boldsymbol{{\bf \delta}}}[n]\triangleq[\check{\delta}_{0}[n],\check{\delta}_{1}[n],\cdots,\check{\delta}_{Q}[n]]^{T}$
and $\tilde{\bf {X}}_{t}[n]\triangleq[\tilde{\bf {H}}_{0}{\bf x}_{t}[n],\tilde{\bf {H}}_{1}{\bf x}_{t}[n],\cdots,\tilde{\bf {H}}_{Q}{\bf x}_{t}[n]]$.
The parameters to be estimated from \eqref{AODfine_ReceivdYt} include $\{\check{\boldsymbol{{\bf \delta}}}[n],\theta_{{\rm t},q}\}$,
${q=0,1,\cdots,Q,n=1,\cdots,N}$. The joint probability density function
of the received signal in \eqref{AODfine_ReceivdYt} can be expressed as
\begin{align}
	\ensuremath{\ensuremath{f(\{{\bf y}_{t}[n]\}_{n=1,\cdots,N}^{t=1,\cdots,T_{1}})}} & =\ensuremath{\ensuremath{\prod\limits _{n=1}^{N}\prod\limits _{t=1}^{T_{1}}\frac{1}{\pi^{N_{{\rm b}}}{\rm det}(\sigma^{2}{\bf I}_{N_{{\rm b}}})}\exp\{-\frac{1}{\sigma^{2}}({\bf y}_{t}[n]-\tilde{\bf {X}}_{t}[n]\check{\boldsymbol{{\bf \delta}}}[n])^{H}({\bf y}_{t}[n]-\tilde{\bf {X}}_{t}[n]\check{\boldsymbol{{\bf \delta}}}[n])\}}}.\label{MLFunc_AODfine_ReceivdYt}
\end{align}
By taking the natural logarithm of \eqref{MLFunc_AODfine_ReceivdYt}
and omitting the constant term, the logarithmic likelihood function
becomes
\begin{align}
	\ensuremath{\ln f} & =\ensuremath{-\sum\limits _{n=1}^{N}\sum\limits _{t=1}^{T_{1}}\left\Vert {\bf y}_{t}[n]-\tilde{\bf {X}}_{t}[n]\check{\boldsymbol{{\bf \delta}}}[n]\right\Vert _{2}^{2}}.\label{logMLFunc_AODfine_ReceivdYt}
\end{align}
By fixing $\ensuremath{\theta_{{\rm t},q},q=0,1,\cdots,Q}$, we take
the derivative of $\ln f$ with respect to the conjugation of $\ensuremath{\check{\boldsymbol{{\bf \delta}}}[n],n=1,\cdots,N}$
as \cite{zhang2017matrix}
\begin{align}
	\ensuremath{\ensuremath{\frac{\partial\ln f}{\partial\check{\boldsymbol{{\bf \delta}}}^{*}[n]}}} & =\ensuremath{\ensuremath{-\frac{\partial\sum\limits _{t=1}^{T_{1}}[({\bf y}_{t}[n]-\tilde{\bf {X}}_{t}[n]\check{\boldsymbol{{\bf \delta}}}[n])^{H}({\bf y}_{t}[n]-\tilde{\bf {X}}_{t}[n]\check{\boldsymbol{{\bf \delta}}}[n])]}{\partial\check{\boldsymbol{{\bf \delta}}}^{*}[n]}}}\nonumber \\
	& \ensuremath{=-\frac{\partial\sum\limits _{t=1}^{T_{1}}({\bf y}_{t}^{H}[n]{\bf y}_{t}[n]-{\bf y}_{t}^{H}[n]\tilde{\bf {X}}_{t}[n]\check{\boldsymbol{{\bf \delta}}}[n]-\check{\boldsymbol{{\bf \delta}}}^{H}[n]\tilde{\bf {X}}_{t}^{H}[n]{\bf y}_{t}[n]+\check{\boldsymbol{{\bf \delta}}}^{H}[n]{\tilde{\bf X}}_{t}^{H}[n]\tilde{\bf {X}}_{t}[n]\check{\boldsymbol{{\bf \delta}}}[n])}{\partial\check{\boldsymbol{{\bf \delta}}}^{*}[n]}}\nonumber \\
	& \ensuremath{=\sum\limits _{t=1}^{T_{1}}(\tilde{\bf {X}}_{t}^{H}[n]{\bf y}_{t}[n]-\tilde{\bf {X}}_{t}^{H}[n]\tilde{\bf {X}}_{t}[n]\check{\boldsymbol{{\bf \delta}}}[n])}.\label{Deriva_AODmleFine}
\end{align}
Then, by setting the derivative in \eqref{Deriva_AODmleFine} to $\mathbf{0}$, we
have
\begin{align}
	\sum\limits _{t=1}^{T_{1}}({\tilde{\bf X}}_{t}^{H}[n]{\bf y}_{t}[n]-{\tilde{\bf X}}_{t}^{H}[n]{\tilde{\bf X}}_{t}[n]\check{\boldsymbol{{\bf \delta}}}[n]) & \ensuremath{=\mathbf{0}}.\label{Equatn_Deriva_AODmleFine}
\end{align}
By solving \eqref{Equatn_Deriva_AODmleFine}, we have
\begin{align}
	\ensuremath{\check{\boldsymbol{{\bf \delta}}}[n]} & =\ensuremath{\ensuremath{[\sum\limits _{t=1}^{T_{1}}({\tilde{\bf X}}_{t}^{H}[n]{\tilde{\bf X}}_{t}[n])]^{-1}\sum\limits _{t=1}^{T_{1}}({\tilde{\bf X}}_{t}^{H}[n]{\bf y}_{t}[n])}}.\label{SolveDeriva_AODmleFine}
\end{align}
By substituting \eqref{SolveDeriva_AODmleFine} into \eqref{logMLFunc_AODfine_ReceivdYt},
the logarithmic likelihood function becomes
\begin{align}
	\ensuremath{\ln f} & \ensuremath{=-\sum\limits _{n=1}^{N}\sum\limits _{t=1}^{T_{1}}\left\Vert {\bf y}_{t}[n]-{\tilde{\bf X}}_{t}[n][\sum\limits _{t=1}^{T_{1}}({\tilde{\bf X}}_{t}^{H}[n]{\tilde{\bf X}}_{t}[n])]^{-1}\sum\limits _{t=1}^{T_{1}}({\tilde{\bf X}}_{t}^{H}[n]{\bf y}_{t}[n])\right\Vert _{2}^{2}}. \label{logMLf_AODmleFine} 
\end{align}

Then, the maximum likelihood estimate of ${{\boldsymbol{\theta}} _{\rm{t}}} \triangleq {[{\theta _{{\rm{t}},0}},{\theta _{{\rm{t}},1}}, \cdots ,{\theta _{{\rm{t}},Q}}]^T}$
becomes
\begin{align}
	\ensuremath{\hat{\boldsymbol{\theta}} _{\rm{t}}} & \ensuremath{=\ensuremath{\arg\mathop{\max}\limits _{{\boldsymbol{\theta}} _{\rm{t}}}\{-\sum\limits _{n=1}^{N}\sum\limits _{t=1}^{T_{1}}\left\Vert {\bf y}_{t}[n]-{ \tilde{\bf X}}_{t}[n][\sum\limits _{t=1}^{T_{1}}({\tilde{\bf X}}_{t}^{H}[n]{\tilde{\bf X}}_{t}[n])]^{-1}\sum\limits _{t=1}^{T_{1}}({\tilde{\bf X}}_{t}^{H}[n]{\bf y}_{t}[n])\right\Vert _{2}^{2}\}}}\nonumber \\
	& \ensuremath{=\arg\mathop{\min}\limits _{{\boldsymbol{\theta}} _{\rm{t}}}\sum\limits _{n=1}^{N}\sum\limits _{t=1}^{T_{1}}\left\Vert {\bf y}_{t}[n]-{\tilde{\bf X}}_{t}[n][\sum\limits _{t=1}^{T_{1}}({ \tilde{\bf X}}_{t}^{H}[n]{\tilde{\bf X}}_{t}[n])]^{-1}\sum\limits _{t=1}^{T_{1}}({\tilde{\bf X}}_{t}^{H}[n]{\bf y}_{t}[n])\right\Vert _{2}^{2}}.\label{AODmle_fine}
\end{align}

The maximum likelihood function in \eqref{AODmle_fine} can be further
simplified to reduce the computational complexity, which can be found
in Appendix \ref{Appendix_A}. The MLE in \eqref{AODmle_fine} becomes
a $(Q+1)$-dimensional search problem, and we adopt the AOD estimations obtained
by the DCS-SOMP algorithm to serve as the initial values for the MLE.

\subsection{Estimation of the AOAs at the RIS}

\label{Estimation of the AOAs at the RIS}

According to the RIS phase profile designed in Fig. \ref{phaseProtocol},
besides the first $T_{1}$ time slots, the remaining $T_{2}$ time slots
are divided into $\Upsilon$ time blocks, and each time block has $V$ time slots. By substituting the sparse representation
of channels in \eqref{SprsChnnlHRB} and \eqref{SprsChnnlHMR} into
\eqref{ReceivdYt}, the received signal at the $t$-th time slot of
the $i$-th time block can be written in the angle domain as
\begin{align}
	\mathbf{y}_{t,i}[n] & ={\bf H}_{{\rm RB}}[n]\text{\ensuremath{\bm{{\Omega}}}}_{i}{\bf H}_{{\rm MR}}[n]\mathbf{x}_{t}[n]+\mathbf{z}_{t,i}[n]\nonumber \\
	& ={\bf A}_{{\rm B}}\bm{{\varLambda}}_{\textrm{RB}}[n]\mathbf{A}_{{\rm R}}^{H}\text{\ensuremath{\bm{{\Omega}}}}_{i}{\bf A}_{{\rm R}}\bm{{\varLambda}}_{\textrm{MR}}[n]{\bf A}_{{\rm M}}^{H}\mathbf{x}_{t}[n]+\mathbf{z}_{t,i}[n],\label{ReceivdYt_RISangles_ToAs}
\end{align}
where $i=0,1,\cdots,\Upsilon$, $t=T_{1}+(i-1)V+1,\cdots,T_{1}+iV$
if $i\neq0$, and $t=1,2,\cdots,T_{1}$ if $i=0$.

By using the AODs estimated in Section \ref{Estimation of the AODs at the MS},
we can obtain a $N_{\textrm{b}}\times(Q+1)$ matrix $\bar{{\bf A}}_{{\rm M}}\triangleq$\\
$[{\bf a}_{{\rm M}}(\hat{\vartheta}_{{\rm t},0}),{\bf a}_{{\rm M}}(\hat{\vartheta}_{{\rm t},1}),\cdots,{\bf a}_{{\rm M}}(\hat{\vartheta}_{{\rm t},Q})]$
whose columns correspond to the non-zero columns of $\bm{{\varLambda}}_{\textrm{MR}}[n]$.
We define a $N_{\textrm{r}}\times(Q+1)$ matrix $\bar{{\bf A}}_{{\rm R,in}}\triangleq[{\bf a}_{{\rm R}}(\omega_{{\rm in},0}^{{\rm a}},\omega_{{\rm in},0}^{{\rm e}}),{\bf a}_{{\rm R}}(\omega_{{\rm in},1}^{{\rm a}},\omega_{{\rm in},1}^{{\rm e}}),\cdots,{\bf a}_{{\rm R}}(\omega_{{\rm in},Q}^{{\rm a}},\omega_{{\rm in},Q}^{{\rm e}})]$,
which contains the columns in the dictionary matrix ${\bf A}_{{\rm R}}$
corresponding to the non-zero rows of $\bm{{\varLambda}}_{\textrm{MR}}[n]$.
We collect the non-zero elements of $\bm{{\varLambda}}_{\textrm{MR}}[n]$
in a diagonal matrix as ${\bar{\bm{\varLambda}}}_{{\rm {MR}}}[n]\triangleq  \textrm{diag}\{\tilde{\delta}_{{\rm MR,}0}[n],\tilde{\delta}_{{\rm MR,}1}[n],\cdots,\tilde{\delta}_{{\rm MR,}Q}[n]\}$, where $\tilde{\delta}_{{\rm MR,}q}[n]\triangleq\delta_{{\rm MR,}q}e^{-j2\pi\tau_{{\rm MR,}q}\frac{(n-1)B}{N}}$,
$q=0,1,\cdots,Q$. The matrix ${\bar{\bm{\varLambda}}}_{\textrm{MR}}[n]\in\mathbb{C}^{(Q+1)\times(Q+1)}$
can be also obtained by eliminating all zero rows and columns of $\bm{{\varLambda}}_{\textrm{MR}}[n]\in\mathbb{C}^{G_{\mathrm{r}}\times G_{\mathrm{m}}}$.
The matrix $\bar{{\bf A}}_{{\rm M}}$ is estimated in Section \ref{Estimation of the AODs at the MS},
while the matrices $\bar{{\bf A}}_{{\rm R,in}}$ and ${\bar{\bm{\varLambda}}}_{{\rm {MR}}}[n]$
are unknown. Then, the subchannel matrix ${\bf H}_{{\rm MR}}[n]$ can be compactly expressed as
\begin{align}
	{\bf H}_{{\rm MR}}[n] & =\bar{{\bf A}}_{{\rm R,in}}{\bar{\bm{\varLambda}}}_{{\rm {MR}}}[n]{\bar{\bf A}}_{{\rm M}}^{H}.\label{ChnnlHMR_MatCompact}
\end{align}

The vector ${\bar{\bf a}}_{{\rm B}}\triangleq{\bf a}_{{\rm B}}(\vartheta_{{\rm r},0})$
is a column in the dictionary matrix ${\bf A}_{{\rm B}}$ corresponding
to the only non-zero row of $\bm{{\varLambda}}_{\textrm{RB}}[n]$,
and the vector $\bar{{\bf a}}_{{\rm R,out}}\triangleq{\bf a}_{{\rm R}}(\omega_{{\rm out},0}^{{\rm a}},\omega_{{\rm out},0}^{{\rm e}})$
is a column in the dictionary matrix $\mathbf{A}_{{\rm R}}$ corresponding
to the only non-zero column of $\bm{{\varLambda}}_{\textrm{RB}}[n]$.
Since the AOA $\theta_{{\rm r},0}$ at the BS and the AOD $\phi_{{\rm out},0},\psi_{{\rm out},0}$
at the RIS can be calculated in advance, the steering vectors ${\bar{\bf a}}_{{\rm B}}$
and $\bar{{\bf a}}_{{\rm R,out}}$ are known. The only one non-zero
element in $\bm{{\varLambda}}_{\textrm{RB}}[n]$ is $\tilde{\delta}_{{\rm RB,}0}[n]\triangleq\delta_{{\rm RB,}0}e^{-j2\pi\tau_{{\rm RB,}0}\frac{(n-1)B}{N}}$,
which is unknown. Then, the subchannel matrix ${\bf H}_{{\rm RB}}[n]$
can be compactly expressed as
\begin{align}
	{\bf H}_{{\rm RB}}[n] & ={\bar{\bf a}}_{{\rm B}}\tilde{\delta}_{{\rm RB,}0}[n]\bar{{\bf a}}_{{\rm R,out}}^{H}.\label{ChnnlHRB_MatCompact}
\end{align}

By substituting \eqref{ChnnlHMR_MatCompact} and \eqref{ChnnlHRB_MatCompact}
into \eqref{ReceivdYt_RISangles_ToAs}, the received signal can be
rewritten as
\begin{align}
	\mathbf{y}_{t,i}[n] & ={\bar{\bf a}}_{{\rm B}}\tilde{\delta}_{{\rm RB,}0}[n]\bar{{\bf a}}_{{\rm R,out}}^{H}\textrm{diag}\{\mathbf{g}_{i}\}\bar{{\bf A}}_{{\rm R,in}}{\bar{\bm{\varLambda}}}_{{\rm {MR}}}[n]{\bar{\bf A}}_{{\rm M}}^{H}\mathbf{x}_{t}[n]+\mathbf{z}_{t,i}[n]\nonumber \\
	& ={\bar{\bf a}}_{{\rm B}}\tilde{\delta}_{{\rm RB,}0}[n]\mathbf{g}_{i}^{T}\bar{{\bf A}}_{{\rm R,in-out}}{\bar{\bm{\varLambda}}}_{{\rm {MR}}}[n]{\bar{\bf A}}_{{\rm M}}^{H}\mathbf{x}_{t}[n]+\mathbf{z}_{t,i}[n],\label{Receivdyt_RISangles_compact}
\end{align}
where $\bar{{\bf A}}_{{\rm R,in-out}}\triangleq\textrm{diag}\{\bar{{\bf a}}_{{\rm R,out}}^{H}\}\bar{{\bf A}}_{{\rm R,in}}$. The $q$-th column of $\bar{{\bf A}}_{{\rm R,in-out}}$ can be written as
\begin{align}
	\bar{{\bf A}}_{{\rm R,in-out}}(:,q)=& {\bf a}_{{\rm R}}(\omega_{{\rm in},q}^{{\rm a}},\omega_{{\rm in},q}^{{\rm e}})\circ{\bf a}_{{\rm R}}^{*}(\omega_{{\rm out},0}^{{\rm a}},\omega_{{\rm out},0}^{{\rm e}})\nonumber \\
	= & {\bf a}_{{\rm R}}(\triangle\omega_{q}^{{\rm a}},\triangle\omega_{q}^{{\rm e}}),\label{RISanglesDiffrc}
\end{align}
where $\triangle\omega_{q}^{{\rm a}}\triangleq\omega_{{\rm in},q}^{{\rm a}}-\omega_{{\rm out},0}^{{\rm a}}$
and $\triangle\omega_{q}^{{\rm e}}\triangleq\omega_{{\rm in},q}^{{\rm e}}-\omega_{{\rm out},0}^{{\rm e}}$.

By multiplying $\bar{\bf {a}}_{{\rm B}}^{H}/N_{\textrm{b}}$ on the
left hand side of $\mathbf{y}_{t,i}[n]$, we have
\begin{align}
	\check{y}_{t,i}[n] & ={\bar{\bf a}}_{{\rm B}}^{H}\mathbf{y}_{t,i}[n]/N_{\textrm{b}}\nonumber \\
	& =\mathbf{g}_{i}^{T}\bar{{\bf A}}_{{\rm R,in-out}}{\bar{\bm{\varLambda}}}_{{\rm {MRB}}}[n]{\bar{\bf A}}_{{\rm M}}^{H}\mathbf{x}_{t}[n]+\check{z}_{t,i}[n],\label{Recv_ytTaBconj}
\end{align}
where $\check{z}_{t,i}[n]={\bar{\bf a}}_{{\rm B}}^{H}\mathbf{z}_{t,i}[n]/N_{\textrm{b}}$, and ${\bar{\bm{\varLambda}}}_{{\rm {MRB}}}[n]\triangleq\tilde{\delta}_{{\rm RB,}0}[n]{\bar{\bm{\varLambda}}}_{{\rm {MR}}}[n]$
is a $(Q+1)\times(Q+1)$ diagonal matrix with the $q$-th element
given by $\tilde{\delta}_{q}[n]=\tilde{\delta}_{{\rm RB,}0}[n]\tilde{\delta}_{{\rm MR,}q}[n]$.

For $i\neq0$, by collecting the processed signal ${\bar{\bf a}}_{{\rm B}}^{H}\mathbf{y}_{t,i}[n]/N_{\textrm{b}}$,
$t=T_{1}+(i-1)V+1,\cdots,T_{1}+iV$ over the $V$
time slots of the $i$-th time block in a row vector, we have
\begin{align}
	({\check{\mathbf y}}_{i}[n])^{T} & =\mathbf{g}_{i}^{T}\bar{{\bf A}}_{{\rm R,in-out}}\bar{\bm{{\varLambda}}}_{{\rm {MRB}}}[n][\bar{{\bf A}}_{{\rm M}}^{H}\mathbf{x}_{T_{1}+(i-1)V+1}[n],\cdots,\bar{{\bf A}}_{{\rm M}}^{H}\mathbf{x}_{T_{1}+iV}[n]]+\check{\mathbf{z}}_{i}^{T}[n]\nonumber \\
	& \triangleq\mathbf{g}_{i}^{T}\bar{{\bf A}}_{{\rm R,in-out}}\bar{\bm{{\varLambda}}}_{{\rm {MRB}}}[n]\bar{{\bf B}}_{i}+\check{\mathbf{z}}_{i}^{T}[n],i\neq0
\end{align}
where  $\check{\mathbf{y}}_{i}[n]=[\check{y}_{T_{1}+(i-1)V+1,i}[n],\cdots,\check{y}_{T_{1}+iV,i}[n]]^{T}$, $\check{\mathbf{z}}_{i}[n]=[\check{z}_{T_{1}+(i-1)V+1,i}[n],\cdots,\check{z}_{T_{1}+iV,i}[n]]^{T}$ and $\bar{{\bf B}}_{i}\triangleq\bar{{\bf A}}_{{\rm M}}^{H}[\mathbf{x}_{T_{1}+(i-1)V+1}[n],\cdots,\mathbf{x}_{T_{1}+iV}[n]]$. 

For $i=0$, by collecting the processed signal ${\bar{\bf a}}_{{\rm B}}^{H}\mathbf{y}_{t,0}[n]/N_{\textrm{b}}$,
$t=1,2,\cdots,T_{1}$ over the $T_{1}$ time slots of the $0$-th
time block in a row vector, we have
\begin{align}
	({\check{\mathbf y}}_{0}[n])^{T} & =\mathbf{g}_{0}^{T}\bar{{\bf A}}_{{\rm R,in-out}}{\bar{\bm{\varLambda}}}_{{\rm {MRB}}}[n][\bar{{\bf A}}_{{\rm M}}^{H}\mathbf{x}_{1}[n],\bar{{\bf A}}_{{\rm M}}^{H}\mathbf{x}_{2}[n],\cdots,\bar{{\bf A}}_{{\rm M}}^{H}\mathbf{x}_{T_{1}}[n]]+\check{\mathbf{z}}_{0}^{T}[n]\nonumber \\
	& \triangleq\mathbf{g}_{0}^{T}\bar{{\bf A}}_{{\rm R,in-out}}{\bar{\bm{\varLambda}}}_{{\rm {MRB}}}[n]\bar{{\bf B}}_{0}+\check{\mathbf{z}}_{0}^{T}[n],
\end{align}
where ${\check{\mathbf y}}_{0}[n]=[\check{y}_{1,0}[n],\check{y}_{2,0}[n],\cdots,\check{y}_{T_{1},0}[n]]^{T}$,
$\check{\mathbf{z}}_{0}[n]=[\check{z}_{1,0}[n],\check{z}_{2,0}[n],\cdots,\check{z}_{T_{1},0}[n]]^{T}$,
and $\bar{{\bf B}}_{0}\triangleq\bar{{\bf A}}_{{\rm M}}^{H}[\mathbf{x}_{1}[n],\mathbf{x}_{2}[n],\cdots,\mathbf{x}_{T_{1}}[n]]$.
We further find the inverse matrix of $\bar{{\bf B}}_{i}\in\mathbb{C}^{Q\times V^{(i)}}$,
where $V^{(i)}=V$ if $i\neq0$ and $V^{(0)}=T_{1}$. The number of
time slots within each block $V^{(i)}$ should satisfy $V^{(i)}\geq Q$,
otherwise the right inverse matrix $\bar{{\bf B}}_{i}^{\dagger}=\bar{{\bf B}}_{i}^{H}(\bar{{\bf B}}_{i}\bar{{\bf B}}_{i}^{H})^{-1}$
does not exist.

The signal $({\check{\mathbf y}}_{i}[n])^{T}\in\mathbb{C}^{1\times V^{(i)}}$
can be further processed as
\begin{align}
	({\check{\mathbf y}}_{i}[n])^{T}\bar{{\bf B}}_{i}^{\dagger} & =\mathbf{g}_{i}^{T}\bar{{\bf A}}_{{\rm R,in-out}}{\bar{\bm{\varLambda}}}_{{\rm {MRB}}}[n]+\check{\mathbf{z}}_{i}^{T}[n]\bar{{\bf B}}_{i}^{\dagger}.
\end{align}
By collecting the processed signal $({\check{\mathbf y}}_{i}[n])^{T}\bar{{\bf B}}_{i}^{\dagger}$
of all the $(\Upsilon+1)$ time blocks, we have
\begin{align}
	{\check{\mathbf Y}}[n] & =[(\bar{{\bf B}}_{0}^{\dagger})^{T}{\check{\mathbf y}}_{0}[n],(\bar{{\bf B}}_{1}^{\dagger})^{T}{\check{\mathbf y}}_{1}[n],\cdots,(\bar{{\bf B}}_{\Upsilon}^{\dagger})^{T}{\check{\mathbf y}}_{\Upsilon}[n]]^{T} \nonumber \\
	&=\mathbf{G}^{T}\bar{{\bf A}}_{{\rm R,in-out}}{\bar{\bm{\varLambda}}}_{{\rm {MRB}}}[n]+{\check{\mathbf Z}}[n],
\end{align}
where $\mathbf{G}^{T}=[\mathbf{g}_{0},\mathbf{g}_{1},\cdots,\mathbf{g}_{\Upsilon}]^{T}$ and ${\check{\mathbf Z}}[n]=[(\bar{{\bf B}}_{0}^{\dagger})^T\check{\mathbf{z}}_{0},(\bar{{\bf B}}_{1}^{\dagger})^T\check{\mathbf{z}}_{1},\cdots,(\bar{{\bf B}}_{\Upsilon}^{\dagger})^T\check{\mathbf{z}}_{\Upsilon}]^T$.

Then, we further rewrite ${\check{\mathbf Y}}[n]=[\mathbf{r}_{0}[n],\mathbf{r}_{1}[n],\cdots,\mathbf{r}_{Q}[n]]$,
${\bar{\bm{\varLambda}}}_{{\rm {MRB}}}[n]=[\bm{{\lambda}}_{0}[n],\bm{{\lambda}}_{1}[n],\cdots,\bm{{\lambda}}_{Q}[n]]$,
and ${\check{\mathbf Z}}[n]=[\mathbf{n}_{0}[n],\mathbf{n}_{1}[n],\cdots,\mathbf{n}_{Q}[n]]$,
and have
\begin{subequations}
\begin{align}
	\mathbf{r}_{q}[n] & =\mathbf{G}^{T}\bar{{\bf A}}_{{\rm R,in-out}}\bm{{\lambda}}_{q}[n]+\mathbf{n}_{q}[n],\label{SprsProblm_RISangles}\\
	& =\mathbf{G}^{T}{\bf a}_{{\rm R}}(\triangle\omega_{q}^{{\rm a}},\triangle\omega_{q}^{{\rm e}})\tilde{\delta}_{q}[n]+\mathbf{n}_{q}[n].\label{RISqPathb}
\end{align}
\end{subequations}
Equation \eqref{SprsProblm_RISangles} can be cast as a sparse signal
recovery problem, which can be solved by the DCS-SOMP given in Algorithm \ref{DCS-OMP}.
The dimension of the equivalent dictionary matrix $\mathbf{G}^{T}\mathbf{A}_{{\rm R}}$
is $(\Upsilon+1)\times G_{\mathrm{r}}$, and the corresponding sparsity
level is 1, thus the pilot overhead required should satisfy $(\Upsilon+1)\geq8\times1-2$.
Since $Q\leq V^{(i)}$ and $T=\sum_{i=0}^{\Upsilon}V^{(i)}$, we have
$T\geq8(Q+1)-2+\Upsilon Q\geq13Q+6$. The phase shift matrix $\mathbf{G}$
of the RIS should be properly designed, similar to how the pilot signal $\mathbf{X}_{1}$ was designed in Section \ref{Estimation of the AODs at the MS}, to produce an orthogonal dictionary matrix.

Equation \eqref{SprsProblm_RISangles} gives the sparse signal recovery problem for the $q$-th path. To obtain the cascaded cosine $\triangle\omega_{q}^{{\rm e}}$, the cascaded sine $\triangle\omega_{q}^{{\rm a}}$ and the hybrid channel gain $\tilde{\delta}_{q}[n]$ for $q=0,1,2,\cdots,Q$,
the DCS-SOMP algorithm can be performed by $(Q+1)$ times to recover $(Q+1)$
independent sparse variables based on \eqref{SprsProblm_RISangles}. If the $k^{q}$-th column of the sensing matrix
$\mathbf{G}^{T}\mathbf{A}_{{\rm R}}$ is selected for the $q$-th
path, the column ${\bf A}_{{\rm R}}(:,k^{q})$ corresponds to the
RIS array response vectors ${\bf a}_{{\rm R}}(\triangle\omega_{q}^{{\rm a}},\triangle\omega_{q}^{{\rm e}})$
according to \eqref{RISanglesDiffrc}, and the projection coefficient
is the hybrid complex channel gain $\tilde{\delta}_{q}[n]$. Based
on \eqref{OverCompleteMatrxRIS}, the column ${\bf A}_{{\rm R}}(:,k^{q})$
can be written as
\begin{align}
	{\bf A}_{{\rm R}}(:,k^{q}) & =\mathbf{f}([-1+\frac{2(k_{\textrm{e}}^{q}-1)}{G_{\mathrm{e}}}]\frac{d_{\textrm{{\rm {RIS,e}}}}}{\lambda},N_{\textrm{e}})\otimes\mathbf{f}([-1+\frac{2(k_{\textrm{a}}^{q}-1)}{G_{\mathrm{a}}}]\frac{d_{\textrm{{\rm {RIS,a}}}}}{\lambda},N_{\textrm{a}}),\label{arrayVctrIndxed}
\end{align}
where $k^{q}=(k_{\textrm{e}}^{q}-1)G_{\mathrm{a}}+k_{\textrm{a}}^{q}$. Equivalently,
we have $k_{\textrm{e}}^{q}=\left\lceil k^{q}/G_{\mathrm{a}}\right\rceil $, $k_{\textrm{a}}^{q}=k^{q}-(k_{\textrm{e}}^{q}-1)G_{\mathrm{a}}$. Since \eqref{RISanglesDiffrc}
is equivalent to \eqref{arrayVctrIndxed}, we arrive at \begin{subequations}
	\label{SolvingAngles}
	\begin{align}
		\cos\phi_{{\rm in},q}-\cos\phi_{{\rm out},0} & =[-1+\frac{2(k_{\textrm{e}}^{q}-1)}{G_{\mathrm{e}}}],\label{CosDiff_RISin2out}\\
		\sin\psi_{{\rm in},q}\sin\phi_{{\rm in},q}-\sin\psi_{{\rm out},0}\sin\phi_{{\rm out},0} & =[-1+\frac{2(k_{\textrm{a}}^{q}-1)}{G_{\mathrm{a}}}].\label{SinSinDiff_RISin2out}
	\end{align}
\end{subequations}

Since $\phi_{{\rm out},0}$ and $\psi_{{\rm out},0}$ are known,
the $\phi_{{\rm in},q}$ and $\psi_{{\rm in},q}$ can be calculated
from \eqref{SolvingAngles}.

\subsection{Estimation of the TOAs and Channel Gains}

\label{Estimation of the TOAs and Channel Gains}

By using the DCS-SOMP algorithm, the nonzero element at the $q$-th row
and $q$-th column of ${\bar{\bm{\varLambda}}}_{{\rm {MRB}}}[n]$
(i.e., $({\bar{\bm{\varLambda}}}_{{\rm {MRB}}}[n])_{q,q}$) is the
estimated value for the complex channel gain $\tilde{\delta}_{q}[n]$
of the $q$-th cascaded path. Since $\tilde{\delta}_{q}[n]=\delta_{q}e^{-j2\pi\tau_{q}\frac{(n-1)B}{N}}$,
by stacking the $\tilde{\delta}_{q}[n]$ of all the $N$ subcarriers
in a vector, we have  $\tilde{\bm{{\delta}}}_{q}=[\tilde{\delta}_{q}[1],\tilde{\delta}_{q}[2],\cdots,\tilde{\delta}_{q}[N]]^{T}$
and $\tilde{\bm{{\delta}}}_{q} =\delta_{q}\mathbf{t}(\upsilon_{q}),q=0,1,\cdots,Q$, where $\mathbf{t}(\upsilon_{q})=[1,e^{-j2\pi\upsilon_{q}},\cdots,e^{-j2\pi(N-1)\upsilon_{q}}]^{T}$ and $\upsilon_{q}\triangleq\tau_{q}\frac{B}{N}$. We assume that
$\upsilon_{q}$ satisfies $0<\upsilon_{q}<1$.

By defining a normalized DFT matrix $\mathbf{U}_{N}$ with $[\mathbf{U}_{N}]_{m',m}=\frac{1}{\sqrt{N}}e^{-j\frac{2\pi}{N}(m'-1)(m-1)}$,
we have $\mathbf{z}_{q}\triangleq \mathbf{U}_{N}^{H}\tilde{\bm{{\delta}}}_{q}=\delta_{q}\mathbf{U}_{N}^{H}\mathbf{t}(\upsilon_{q})$, and the $m$-th element of $\mathbf{z}_{q}$ can be written as
\begin{align}
	[\mathbf{z}_{q}]_{m} & =\frac{\delta_{q}}{\sqrt{N}}\sum_{m'=1}^{N}e^{j\frac{2\pi}{N}(m'-1)(m-1)}e^{-j2\pi(m'-1)\upsilon_{q}}\nonumber \\
	& =\frac{\delta_{q}}{\sqrt{N}}e^{-j2\pi\{[\upsilon_{q}-\frac{(m-1)}{N}]\frac{N-1}{2}\}}\frac{\sin\{2\pi[\upsilon_{q}-\frac{(m-1)}{N}]\frac{N}{2}\}}{\sin\{2\pi[\upsilon_{q}-\frac{(m-1)}{N}]\frac{1}{2}\}}. \label{zq_m}
\end{align}

From \eqref{zq_m}, it is noted that if $N\rightarrow\infty$,
there always exists some integers $m_{q}=N\upsilon_{q}+1=\tau_{q}B+1$
such that $[\mathbf{z}_{q}]_{m}=\frac{\delta_{q}}{\sqrt{N}}$ while the
other elements of $\mathbf{z}_{q}$ are all zeros. Thus, we can find $m_{q}$ by 
	\begin{align}
		m_{q} & =\arg\underset{m}{\max}\left|\mathbf{e}_{m}^{T}\mathbf{U}_{N}^{H}\hat{\tilde{\bm{{\delta}}}}_{q}\right|,\label{m_q}
	\end{align}
where $\mathbf{e}_{m}$ is a vector comprising all zeros, except a
1 in the $m$-th entry, and $\hat{\tilde{\bm{{\delta}}}}_{q}$ is
the estimated value for $\tilde{\bm{{\delta}}}_{q}$. Then we obtain the coarsely-estimated TOA as $\hat{\tau}_{q}=(m_{q}-1)/B$.

The accuracy of
TOA estimations obtained above are limited by the resolution
of the DFT, thus we further refine the TOA estimation by utilizing the rotation property.
We define a rotation matrix as $\bm{{\Phi}}_{N}(\triangle\tau_{q})=\textrm{diag}\{1,e^{j2\pi\triangle\tau_{q}\frac{B}{N}},\cdots,e^{j2\pi(N-1)\triangle\tau_{q}\frac{B}{N}}\}$,
then we have
\begin{align}
	[\mathbf{U}_{N}^{H}\bm{{\Phi}}_{N}^{H}(\triangle\tau_{q})\hat{\tilde{\bm{{\delta}}}}_{q}]_{m} & =\frac{\delta_{q}}{\sqrt{N}}e^{-j2\pi[\tau_{q}\frac{B}{N}+\triangle\tau_{q}\frac{B}{N}-\frac{(m-1)}{N}]\frac{N-1}{2}\}}\frac{\sin\{2\pi[\tau_{q}\frac{B}{N}+\triangle\tau_{q}\frac{B}{N}-\frac{(m-1)}{N}]\frac{N}{2}\}}{\sin\{2\pi[\tau_{q}\frac{B}{N}+\triangle\tau_{q}\frac{B}{N}-\frac{(m-1)}{N}]\frac{1}{2}\}}.
\end{align}

It is noted that the power will concentrate on the $m_{q}$th row
without power leakage when the time rotation parameters satisfy $\triangle\tau_{q}=(m_{q}-1)/{B}-\tau_{q}$. The rotation parameter $\triangle\tau_{q}$ can be found by a one-dimensional search by
solving the following problem
\begin{align}
	\triangle\tau_{q} & =\arg\underset{\triangle\tau_{q}\in[-\frac{1}{2B},\frac{1}{2B}]}{\max}\left|\mathbf{U}_{m_{q},:}^{H}\bm{{\Phi}}_{N}^{H}(\triangle\tau_{q})\hat{\tilde{\bm{{\delta}}}}_{q}\right|. \label{rotationSrch}
\end{align}

When $\triangle\tau_{q}$ is obtained from \eqref{rotationSrch}, the TOA $\tau_{q}$ can be
estimated by  
\begin{align}
	\hat{\tau}_{q} & =\frac{(m_{q}-1)}{B}-\triangle\tau_{q}.\label{DeltTOA_TOA}
\end{align}

Then, according to the least squares (LS) estimation, the complex channel gain can be estimated by
\begin{align}
	\hat{\delta}_{q} & =\frac{\mathbf{t}^{H}(\hat{\upsilon}_{q})\hat{\tilde{\bm{{\delta}}}}_{q}}{N},\label{Opt_Chnnlgain}
\end{align}
where $\hat{\upsilon}_{q}=\hat{\tau}_{q}\frac{B}{N}$.

\section{Refining Estimation of Channel Parameters}

\label{Fine Estimation of Channel Parameters} Although the AOD estimations
are refined by the MLE after using the DCS-SOMP algorithm, only the
first $T_{1}$ time slots are utilized for estimation. The estimations
for AOD can be further refined by exploiting all the $T$ time slots.
Although the TOA estimations are refined based on the time rotation operation
after the DFT approach, these estimations rely on the hybrid complex
channel gains in $\tilde{\bm{{\delta}}}_{q}$ which are obtained from
the AOA estimation at the RIS by the DCS-SOMP algorithm. Thus, the
TOA estimations should be further refined due to the lack of high
accuracy of the coarsely-estimated hybrid channel gains. Moreover,
since the accuracy of azimuth AOA $\psi_{\mathrm{in},q}$ and elevation
AOA $\phi_{{\rm in},q}$ at the RIS is limited by the DCS-SOMP algorithm,
the AOAs at the RIS should also be further refined. Thus, we consider
to exploit the MLE to jointly refine all these channel parameters
by utilizing the received signal over all the $T$ time slots.

\subsection{MLE of All Channel Parameters} \label{MLE_all_param}

Similar to the manipulations in \eqref{Receivdyt_RISangles_compact},
the channel matrix in \eqref{TotalChnnl} can be rewritten as 
\begin{align}
	{\bf H}_{t}[n] & =\sum\limits _{q=0}^{Q}\delta_{q}e^{-j2\pi\tau_{q}\frac{(n-1)B}{N}}{\bf a}_{{\rm B}}(\vartheta_{{\rm r},0})\mathbf{g}_{t}^{T}\textrm{diag}\{{\bf a}_{{\rm R}}^{H}(\omega_{{\rm out},0}^{{\rm a}},\omega_{{\rm out},0}^{{\rm e}})\}{\bf a}_{{\rm R}}(\omega_{{\rm in},q}^{{\rm a}},\omega_{{\rm in},q}^{{\rm e}}){\bf a}_{{\rm M}}^{H}(\vartheta_{{\rm t},q})\nonumber \\
	& =\sum\limits _{q=0}^{Q}\delta_{q}e^{-j2\pi\tau_{q}\frac{(n-1)B}{N}}\mathbf{g}_{t}^{T}{\bf a}_{{\rm R}}(\triangle\omega_{q}^{{\rm a}},\triangle\omega_{q}^{{\rm e}}){\bf a}_{{\rm B}}(\vartheta_{{\rm r},0}){\bf a}_{{\rm M}}^{H}(\vartheta_{{\rm t},q}).\label{TotalChnnl_vrsn2}
\end{align}
By substituting \eqref{TotalChnnl_vrsn2} into \eqref{ReceivdYt},
the received signal can be written as 
\begin{align}
	{\bf y}_{t}[n] & =\sum\limits _{q=0}^{Q}\delta_{q}e^{-j2\pi\tau_{q}\frac{(n-1)B}{N}}\mathbf{g}_{t}^{T}{\bf a}_{{\rm R}}(\triangle\omega_{q}^{{\rm a}},\triangle\omega_{q}^{{\rm e}})\ensuremath{{\tilde{{\bf H}}}_{q}}{\bf x}_{t}[n]+{\bf z}_{t}[n],\label{stage2_ReceData}
\end{align}
where ${\tilde{{\bf H}}}_{q}\triangleq{\bf a}_{{\rm B}}(\vartheta_{{\rm r},0}){\bf a}_{{\rm M}}^{H}(\vartheta_{{\rm t},q})$.

By stacking the received signal ${\bf y}_{t}[n]$ at different time
slots (i.e., $t=1,\cdots,T$) in a columnwise order, we have 
\begin{equation}
	{\bf Y}[n]=\sum\limits _{q=0}^{Q}\delta_{q}e^{-j2\pi\tau_{q}\frac{(n-1)B}{N}}\ensuremath{{\tilde{{\bf H}}}_{q}}{\bf X}[n]\bm{{\Sigma}}_{q}+{\bf Z}[n],\label{stage2_ReceData_pilot}
\end{equation}
where $\ensuremath{{\bf Y}[n]\triangleq[{\bf y}_{1}[n],\cdots,{\bf y}_{T}[n]]}$,
$\ensuremath{{\bf X}[n]\triangleq[{\bf x}_{1}[n],\cdots,{\bf x}_{T}[n]]}$,
$\ensuremath{{\bf Z}[n]\triangleq[{\bf z}_{1}[n],\cdots,{\bf z}_{T}[n]]}$
and $\bm{{\Sigma}}_{q}\triangleq{\rm diag}\{\mathbf{g}_{1}^{T}{\bf a}_{{\rm R}}(\triangle\omega_{q}^{{\rm a}},\triangle\omega_{q}^{{\rm e}}),\cdots,\mathbf{g}_{T}^{T}{\bf a}_{{\rm R}}(\triangle\omega_{q}^{{\rm a}},\triangle\omega_{q}^{{\rm e}})\}$.
Then we vectorize the ${\bf Y}[n]$ in \eqref{stage2_ReceData_pilot},
and have 
\begin{align}
	{\bf y}[n] & =\sum\limits _{q=0}^{Q}\delta_{q}e^{-j2\pi\tau_{q}\frac{(n-1)B}{N}}(\bm{{\Sigma}}_{q}^{T}\otimes{\tilde{{\bf H}}}_{q}){\bf x}[n]+{\bf z}[n],\label{stage2_ReceData_pilot_Vec}
\end{align}
where ${\bf y}[n]\triangleq{\rm vec}({\bf Y}[n])$, ${\bf x}[n]\triangleq{\rm vec}({\bf X}[n])$
and $\ensuremath{{\bf z}[n]\triangleq{\rm vec}({\bf Z}[n])}$. The
\eqref{stage2_ReceData_pilot_Vec} is obtained from \eqref{stage2_ReceData_pilot} by using the identity
${{\rm vec}({\bf ABC})=({\bf C}^{T}\otimes{\bf A}){\rm vec}({\bf B})}$.
Then, based on the received signal in \eqref{stage2_ReceData_pilot_Vec},
the likelihood function of the random vector $\{{\bf y}[n]\}_{n=1,2,\cdots,N}$
conditioned on $\boldsymbol{{\bf \eta}}$ can be written as 
\begin{align}
	& f(\{{\bf y}[n]\}_{n=1,2,\cdots,N}|\boldsymbol{{\bf \eta}})\nonumber \\
	&=  \prod\limits _{n=1}^{N}\frac{1}{\pi^{TN_{{\rm b}}}{\rm det}(\sigma^{2}{\bf I}_{TN_{{\rm b}}})}\exp\left[-({\bf y}[n]-\boldsymbol{\mu}[n])^{H}{\bf \bm{{\Xi}}}^{-1}({\bf y}[n]-\boldsymbol{\mu}[n])\right].\label{joint_probability_density_allTimeSlots}
\end{align}
where $\boldsymbol{\mu}[n]\triangleq\sum_{q=0}^{Q}\delta_{q}e^{-j2\pi\tau_{q}\frac{(n-1)B}{N}}(\bm{{\Sigma}}_{q}^{T}\otimes{\tilde{{\bf H}}}_{q}){\bf x}[n]$ and ${\bf \bm{{\Xi}}}=\sigma^{2}{\bf I}_{TN_{{\rm b}}}$.

By taking the natural logarithm of \eqref{joint_probability_density_allTimeSlots}
and omitting the constant term, we have the log-likelihood functon
$\Lambda(\boldsymbol{{\bf \eta}};\{{\bf y}[n]\}_{n=1,2,\cdots,N})$
of $\boldsymbol{{\bf \eta}}$ given the observed signal ${\bf y}[n],n=1,2,\cdots,N$.
After some mathmatical simplifications, the $\Lambda(\boldsymbol{{\bf \eta}};\{{\bf y}[n]\}_{n=1,2,\cdots,N})$
can be finally expressed as 
\begin{align}
	& \Lambda(\boldsymbol{{\bf \eta}};\{{\bf Y}[n]\}_{n=1,2,\cdots,N})=2{\mathfrak{Re}}\left[\sum\limits _{q=0}^{Q}{\bf a}_{{\rm M}}^{H}(\vartheta_{{\rm t},q})\left(\sum\limits _{n=1}^{N}\delta_{q}e^{-j2\pi\tau_{q}\frac{(n-1)B}{N}}{\bf X}[n]\bm{{\Sigma}}_{q}{\bf Y}^{H}[n]\right){\bf a}_{{\rm B}}(\vartheta_{{\rm r},0})\right]\nonumber \\
	& \quad-N_{\mathrm{b}}\sum\limits _{q_{1}=0}^{Q}\sum\limits _{q_{2}=0}^{Q}\left[\delta_{q_{1}}^{*}\delta_{q_{2}}{\bf a}_{{\rm M}}^{H}(\vartheta_{{\rm t},q_{2}})\left(\sum\limits _{n=1}^{N}e^{j2\pi(\tau_{q_{1}}-\tau_{q_{2}})\frac{(n-1)B}{N}}{\bf X}[n]\bm{{\Sigma}}_{q_{2}}\bm{{\Sigma}}_{q_{1}}^{H}{\bf X}^{H}[n]\right){\bf a}_{{\rm M}}(\vartheta_{{\rm t},q_{1}})\right].\label{stage2_Allparameters_MLE}
\end{align}
The detailed derivation process is given in \eqref{stage2_Allparameters_MLE_Derivation}
of Appendix \ref{Appendices B.A}. Then the MLE for the channel parameters
in $\boldsymbol{{\bf \eta}}$ can be formulated by $\hat{\boldsymbol{{\bf \eta}}}=\arg\underset{\boldsymbol{{\bf \eta}}}{\max}{\:}\Lambda(\boldsymbol{{\bf \eta}};\{{\bf Y}[n]\}_{n=1,2,\cdots,N})$.
Obviously, the expression of $\Lambda(\boldsymbol{{\bf \eta}};\{{\bf Y}[n]\}_{n=1,2,\cdots,N})$
is complex, and the high dimensional nonlinear optimization for $\boldsymbol{{\bf \eta}}\in\mathbb{C}^{6(Q+1)}$
is computationally prohibitive. Since the MLE is intractable, we propose to exploit
the space alternating generalized expectation (SAGE) algorithm\cite{fessler1994space}
to update the parameters sequentially. The SAGE algorithm has been
widely utilized in the estimation problem for channel parameters \cite{fleury1999channel},
where the high-dimensional optimization process is circumvented by
several sequentially-performed low-dimensional maximization procedures.

\subsection{Solving the MLE by SAGE algorithm}

The derivation of the SAGE algorithm relies on two key notions of
the complete (unobservable) and incomplete (observable) data. The
received signal ${\bf y}_{t}[n]$ in \eqref{stage2_ReceData} is identified
as the incomplete data. The individual received signals ${\bf y}_{q,t}[n]$
from the $q$-th path, $q=0,1,\cdots,Q$, constitute a set of complete
data, which can be written as 
\begin{align}
	{\bf y}_{q,t}[n] & =\delta_{q}e^{-j2\pi\tau_{q}\frac{(n-1)B}{N}}\mathbf{g}_{t}^{T}{\bf a}_{{\rm R}}(\triangle\omega_{q}^{{\rm a}},\triangle\omega_{q}^{{\rm e}})\ensuremath{{\tilde{{\bf H}}}_{q}}{\bf x}_{t}[n]+{\bf z}_{q,t}[n],q=0,1,\cdots,Q,\label{y_qtComplteData}
\end{align}
where ${\bf z}_{q,t}[n]\in\mathbb{C}^{N_{\mathrm{b}}}$ represents
the noise of the $q$-th path satisfying ${\bf z}_{q,t}[n]\sim\mathcal{CN}(0,\sigma_{q}^{2}\mathbf{I}_{N_{{\rm b}}})$,
and $\sigma_{q}^{2}$ denotes the noise power. The mutually-independent ${\bf z}_{0,t}[n],{\bf z}_{1,t}[n],\cdots,{\bf z}_{Q,t}[n]$ form a decomposition of ${\bf z}_{t}[n]$
in \eqref{stage2_ReceData}, thus we have $\sum_{q=0}^{Q}\sigma_{q}^{2}=\sigma^{2}$.
The incomplete data ${\bf y}_{t}[n]$ are related to the complete
data $\{{\bf y}_{q,t}[n]\}_{q=0,1,\cdots,Q}$ by ${\bf y}_{t}[n]=\sum_{q=0}^{Q}{\bf y}_{q,t}[n]$.
Since ${\bf y}_{0,t}[n],{\bf y}_{1,t}[n],\cdots,{\bf y}_{Q,t}[n]$
are independent, the signals ${\bf y}_{q',t}[n]$, $q'\neq q$, are
irrelevant for the estimation of $\boldsymbol{{\bf \eta}}_{q}$. If
the complete data can be observed, the MLE of the parameters of the
$q$-th path in $\boldsymbol{{\bf \eta}}_{q}$ can be obtained from
the observed ${\bf y}_{q,t}[n]$ through $(\hat{\boldsymbol{{\bf \eta}}}_{q})_{\textrm{ML}}({\bf y}_{q,t}[n])=\arg\max L(\boldsymbol{{\bf \eta}}_{q};{\bf y}_{q,t}[n])$,
where $L(\boldsymbol{{\bf \eta}}_{q};{\bf y}_{q,t}[n])$ denotes the
likelihood function of $\boldsymbol{{\bf \eta}}_{q}$ given an observation
${\bf y}_{q,t}[n]$, and $(\hat{\boldsymbol{{\bf \eta}}}_{q})_{\textrm{ML}}({\bf y}_{q,t}[n])$
means the MLE of $\boldsymbol{{\bf \eta}}_{q}$ is a function of the
observation ${\bf y}_{q,t}[n]$. Since the ${\bf y}_{q,t}[n]$ cannot
be observed, it is usually estimated from the incomplete data ${\bf y}_{t}[n]$
and a previous estimation of $\boldsymbol{{\bf \eta}}_{q}$. Thus,
in the SAGE algorithms, we first formulate the MLE problem of a single
path, then estimate the complete data ${\bf y}_{q,t}[n]$, and finally
solve the MLE problem by utilizing the estimated ${\bf y}_{q,t}[n]$.
The three key steps are illustrated as follows.

\subsubsection{MLE of a Single Path}

We first derive the log-likelihood function of the channel parameters
$\boldsymbol{{\bf \eta}}_{q}$ assuming that ${\bf y}_{q,t}[n]$ is
observable. By stacking ${\bf y}_{q,t}[n]$ over $T$ time slots in
a columnwise order, we can obtain ${\bf Y}_{q}[n]\triangleq[{\bf y}_{q,1}[n],\cdots,{\bf y}_{q,T}[n]]$.
By using \eqref{y_qtComplteData} and similar to \eqref{stage2_ReceData_pilot},
${\bf Y}_{q}[n]$ can be written as 
\begin{equation}
	{\bf Y}_{q}[n]=\delta_{q}e^{-j2\pi\tau_{q}\frac{(n-1)B}{N}}{\tilde{{\bf H}}}_{q}{\bf X}[n]\bm{{\Sigma}}_{q}+{\bf Z}_{q}[n],\label{Subcarrier_T2pilot}
\end{equation}
where $\ensuremath{{\bf Z}_{q}[n]\triangleq[{\bf z}_{q,1}[n],\cdots,{\bf z}_{q,T}[n]]}$.
We vectorize the equality in \eqref{Subcarrier_T2pilot} to obtain
\begin{align}
	{\bf y}_{q}[n] & =\delta_{q}e^{-j2\pi\tau_{q}\frac{(n-1)B}{N}}(\bm{{\Sigma}}_{q}^{T}\otimes{\tilde{{\bf H}}}_{q}){\bf x}[n]+{\bf z}_{q}[n],\label{Subcarrier_T2pilot_vec}
\end{align}
where ${\bf y}_{q}[n]\triangleq{\rm vec}({\bf Y}_{q}[n])$ and ${\bf z}_{q}[n]\triangleq\mathrm{vec}({\bf Z}_{q}[n])$.

According to \eqref{Subcarrier_T2pilot_vec}, the joint probability
density function of ${\bf y}_{q}[n]$ over $N$ subcarriers is similar
to that in \eqref{joint_probability_density_allTimeSlots}. And the
corresponding log-likelihood function of $\boldsymbol{{\bf \eta}}_{q}$
for the observations $\{{\bf y}_{q}[n]\}_{n=1,\cdots,N}$ can be written
as
\begin{align}
	L(\boldsymbol{{\bf \eta}}_{q};\{{\bf y}_{q}[n]\}_{n=1,\cdots,N}) & \propto-\sum\limits _{n=1}^{N}\left\Vert {\bf y}_{q}[n]-\delta_{q}e^{-j2\pi\tau_{q}\frac{(n-1)B}{N}}(\bm{{\Sigma}}_{q}^{T}\otimes\tilde{{\bf {H}}}_{q}){\bf x}[n]\right\Vert _{2}^{2}\nonumber \\
	& =-\sum\limits _{n=1}^{N}\left\Vert {\bf Y}_{q}[n]-\delta_{q}e^{-j2\pi\tau_{q}\frac{(n-1)B}{N}}\tilde{{\bf {H}}}_{q}{\bf X}[n]\bm{{\Sigma}}_{q}\right\Vert _{F}^{2}.\label{stage2_lnMLE_omit}
\end{align}
where $\ensuremath{\propto}$ denotes equality up to irrelevant constants.
By further omitting the constant terms in \eqref{stage2_lnMLE_omit},
the $L(\boldsymbol{{\bf \eta}}_{q};\{{\bf y}_{q}[n]\}_{n=1,\cdots,N})$
can be simplified into 
\begin{align}
	L(\boldsymbol{{\bf \eta}}_{q};\{{\bf Y}_{q}[n]\}_{n=1,\cdots,N}) & =2{\mathfrak{Re}}\left[\delta_{q}{\bf a}_{{\rm M}}^{H}(\vartheta_{{\rm t},q})\left(\sum\limits _{n=1}^{N}e^{-j2\pi\tau_{q}\frac{(n-1)B}{N}}{\bf X}[n]\bm{{\Sigma}}_{q}{\bf Y}_{q}^{H}[n]\right){\bf a}_{{\rm B}}(\vartheta_{{\rm r},0})\right]\nonumber \\
	& \quad-N_{{\rm b}}\delta_{q}^{*}\delta_{q}{\bf a}_{{\rm M}}^{H}(\vartheta_{{\rm t},q})\left(\sum\limits _{n=1}^{N}{\bf X}[n]\bm{{\Sigma}}_{q}\bm{{\Sigma}}_{q}^{H}{\bf X}^{H}[n]\right){\bf a}_{{\rm M}}(\vartheta_{{\rm t},q}).\label{stage2_lnMLE}
\end{align}
The detailed derivation process for \eqref{stage2_lnMLE} is similar
as that in Appendix \ref{Appendices B.A}. Or equivalently, the \eqref{stage2_lnMLE}
is a special case of \eqref{stage2_Allparameters_MLE}, thus can be
obtained from \eqref{stage2_Allparameters_MLE} directly. 

Then the MLE of $\boldsymbol{{\bf \eta}}_{q}$ can be formulated as
\begin{align}
	(\hat{\boldsymbol{\eta}}_{q})_{\textrm{ML}}(\{{\bf Y}_{q}[n]\}_{n=1,\cdots,N}) & =\arg\underset{\boldsymbol{{\bf \eta}}_{q}}{\max}{\:}L(\boldsymbol{{\bf \eta}}_{q};\{{\bf Y}_{q}[n]\}_{n=1,\cdots,N}).\label{MLEforqParamtrs}
\end{align}

\subsubsection{Estimation of the Complete Data}

From \eqref{MLEforqParamtrs}, we find that to obtain the MLE of $\boldsymbol{{\bf \eta}}_{q}$,
the received signal ${\bf Y}_{q}[n]$ (equivalently the ${\bf y}_{q,t}[n]$)
should be given. Since ${\bf y}_{q,t}[n]$ is unobservable, it can
be estimated based on the observation ${\bf y}_{t}[n]$ of the incomplete
data and the previous estimation of $\boldsymbol{{\bf \eta}}$. Specifically,
in the $k$-th iteration, the received signal of the $q$-th path
can be estimated by its conditional expectation given the incomplete
data ${\bf y}_{t}[n]$ and the last estimation of ${\hat{\boldsymbol{\eta}}}^{(k-1)}$
as 
\begin{align}
	{\hat{{\bf y}}}_{q,t}^{(k)}[n] & =\mathbb{E}({\bf y}_{q,t}[n]\left|{\bf y}_{t}[n],{\hat{\boldsymbol{\eta}}}^{(k-1)}\right.).\label{ExpctnForlthCompltData_StageII}
\end{align}
In the SAGE algorithm, \eqref{ExpctnForlthCompltData_StageII} can
be explicitly calculated by\cite{paper17} 
\begin{align}
	{\hat{{\bf y}}}_{q,t}^{(k)}[n]= & {\bf y}_{t}[n]-\sum\limits _{q'=0}^{q-1}\delta_{q'}^{(k)}e^{-j2\pi\tau_{q'}^{(k)}\frac{(n-1)B}{N}}{\bf g}^{T}_{t}{\bf a}_{{\rm R}}((\triangle\omega_{q'}^{{\rm a}})^{(k)},(\triangle\omega_{q'}^{{\rm e}})^{(k)})\ensuremath{{\tilde{{\bf H}}}_{q'}^{(k)}}{\bf x}_{t}[n]\nonumber \\
	& -\sum\limits _{q''=q+1}^{Q}\delta_{q''}^{(k-1)}e^{-j2\pi\tau_{q''}^{(k-1)}\frac{(n-1)B}{N}}{\bf g}^{T}_{t}{\bf a}_{{\rm R}}((\triangle\omega_{q''}^{{\rm a}})^{(k-1)},(\triangle\omega_{q''}^{{\rm e}})^{(k-1)})\ensuremath{{\tilde{{\bf H}}}_{q''}^{(k-1)}}{\bf x}_{t}[n],\label{Received_signal_reconstruction_StageII}
\end{align}
where ${{\tilde{{\bf H}}}_{q'}^{(k)}={\bf a}_{{\rm B}}(\vartheta_{{\rm r},0}^{(k)}){\bf a}_{{\rm M}}^{H}(\vartheta_{{\rm t},q'}^{(k)})}$
and ${\tilde{{\bf H}}}_{q''}^{(k-1)}={\bf a}_{{\rm B}}(\vartheta_{{\rm r},0}^{(k-1)}){\bf a}_{{\rm M}}^{H}(\vartheta_{{\rm t},q''}^{(k-1)})$.

\subsubsection{Solving the MLE by Coordinate Ascent Method}

With the estimated ${\hat{{\bf y}}}_{q,t}^{(k)}[n]$ above, we have
$\hat{{\bf Y}}_{q}^{(k)}[n]=[{\hat{{\bf y}}}_{q,1}^{(k)}[n],{\hat{{\bf y}}}_{q,2}^{(k)}[n],\cdots,{\hat{{\bf y}}}_{q,T}^{(k)}[n]]$.
By substituting the estimated ${\hat{{\bf Y}}}_{q}^{(k)}[n]$ into
\eqref{MLEforqParamtrs}, the MLE of $\boldsymbol{{\bf \eta}}_{q}$
can be rewritten as 
\begin{align}
	(\hat{\boldsymbol{\eta}}_{q})_{\textrm{ML}}(\{{\hat{{\bf Y}}}_{q}^{(k)}[n]\}_{n=1,\cdots,N}) & =\arg\underset{\boldsymbol{{\bf \eta}}_{q}}{\max}{\:}L(\boldsymbol{{\bf \eta}}_{q};\{{\hat{{\bf Y}}}_{q}^{(k)}[n]\}_{n=1,\cdots,N}).\label{MLEforqParamtrs_vrsn2}
\end{align}
When solving the MLE problem in \eqref{MLEforqParamtrs_vrsn2}, we
can first find the closed-form solution for the channel gain $\delta_{q}$
in ${\boldsymbol{\eta}}_{q}=[\tau_{q},\delta_{q,\textrm{R}},\delta_{q,\textrm{I}},\theta_{{\rm t},q},\phi_{{\rm in},q},\psi_{{\rm in},q}]^{T}$.
Specifically, by fixing $\{\tau_{q},\theta_{{\rm t},q},\phi_{{\rm in},q},\psi_{{\rm in},q}\}$,
we take the derivative of $L(\boldsymbol{{\bf \eta}}_{q};\{{\hat{{\bf Y}}}_{q}^{(k)}[n]\}_{n=1,\cdots,N})$
with respect to the conjugate of $\delta_{q}$ as \begin{subequations}
	\begin{align}
		& \frac{\partial L(\boldsymbol{{\bf \eta}}_{q};\{{\hat{{\bf Y}}}_{q}^{(k)}[n]\}_{n=1,\cdots,N})}{\partial\delta_{q}^{*}}=-\frac{\partial\sum\limits _{n=1}^{N}\left\Vert {\bf \hat{Y}}_{q}^{(k)}[n]-\delta_{q}e^{-j2\pi\tau_{q}\frac{(n-1)B}{N}}{\tilde{{\bf H}}}_{q}{\bf X}[n]\bm{{\Sigma}}_{q}\right\Vert _{F}^{2}}{\partial\delta_{q}^{*}},\label{stage2deltap_Derivation_propor}\\
		& \overset{(a)}{=}\sum\limits _{n=1}^{N}\left\{ e^{j2\pi\tau_{q}\frac{(n-1)B}{N}}{\rm tr}[({\tilde{{\bf H}}}_{\mathit{q}}{\bf X}[n]\bm{{\Sigma}}_{q})^{H}{\hat{{\bf Y}}}_{q}^{(k)}[n]]\right\} -\delta_{q}\sum\limits _{n=1}^{N}{\rm tr}[({\tilde{{\bf H}}}_{\mathit{q}}{\bf X}[n]\bm{{\Sigma}}_{q})^{H}({\tilde{{\bf H}}}_{\mathit{q}}{\bf X}[n]\bm{{\Sigma}}_{q})],\label{stage2deltap_Derivation}
	\end{align}
\end{subequations} where the \eqref{stage2deltap_Derivation_propor}
is due to \eqref{stage2_lnMLE} $\propto$ \eqref{stage2_lnMLE_omit},
and the detailed derivation of step $(a)$ can be found in \eqref{Detailed_derivation_stage2}
of Appendix \ref{Appendices B.B}. By setting the derivative in \eqref{stage2deltap_Derivation}
to zero, we can build an equation \cite{zhang2017matrix}, and the
solution for $\delta_{q}$ can be readily obtained as 
\begin{align}
	\delta_{q} & =\frac{\sum\limits _{n=1}^{N}\left\{ e^{j2\pi\tau_{q}\frac{(n-1)B}{N}}{\rm tr}[({\tilde{{\bf H}}}_{q}{\bf X}[n]\bm{{\Sigma}}_{q})^{H}{\hat{{\bf Y}}}_{q}^{(k)}[n]]\right\} }{\sum\limits _{n=1}^{N}{\rm tr}[({\tilde{{\bf H}}}_{q}{\bf X}[n]\bm{{\Sigma}}_{q})^{H}({\tilde{{\bf H}}}_{q}{\bf X}[n]\bm{{\Sigma}}_{q})]}\nonumber \\
	& \overset{(b)}{=}\frac{{\bf a}_{{\rm B}}^{H}(\vartheta_{{\rm r,0}})\left\{ \sum\limits _{n=1}^{N}(e^{j2\pi\tau_{q}\frac{(n-1)B}{N}}{\hat{{\bf Y}}}_{q}^{(k)}[n]\bm{{\Sigma}}_{q}^{H}{\bf X}^{H}[n])\right\} {\bf a}_{{\rm M}}(\vartheta_{{\rm t,\mathit{q}}})}{N_{{\rm b}}{\bf a}_{{\rm M}}^{H}(\vartheta_{{\rm t,\mathit{q}}})\left\{ \sum\limits _{n=1}^{N}({\bf X}[n]\bm{{\Sigma}}_{q}\bm{{\Sigma}}_{q}^{H}{\bf X}^{H}[n])\right\} {\bf a}_{{\rm M}}(\vartheta_{{\rm t,\mathit{q}}})},\label{stage2deltap_MLE}
\end{align}
where the detailed simplification of step $(b)$ in \eqref{stage2deltap_MLE}
can be found in \eqref{Detailed_deltap_stage2} of Appendix \ref{Appendices B.C}.
By substituting \eqref{stage2deltap_MLE} into $L(\boldsymbol{{\bf \eta}}_{q};\{{\hat{{\bf Y}}}_{q}^{(k)}[n]\}_{n=1,\cdots,N})$,
we obtain the likelihood function of $\bar{\boldsymbol{\eta}}_{q}=[\tau_{q},\theta_{{\rm t},q},\phi_{{\rm in},q},\psi_{{\rm in},q}]^{T}$ as
follows: 
\begin{align}
	F(\bar{\boldsymbol{\eta}}_{q};\{{\hat{{\bf Y}}}_{q}^{(k)}[n]\}_{n=1,\cdots,N}) & \overset{(c)}{=}\frac{\left|{\bf a}_{{\rm B}}^{H}(\vartheta_{{\rm r},0})\left\{ \sum\limits _{n=1}^{N}(e^{j2\pi\tau_{q}\frac{(n-1)B}{N}}{\hat{{\bf Y}}}_{q}^{(k)}[n]\bm{{\Sigma}}_{q}^{H}{\bf X}^{H}[n])\right\} {\bf a}_{{\rm M}}(\vartheta_{{\rm t},q})\right|^{2}}{N_{{\rm b}}{\bf a}_{{\rm M}}^{H}(\vartheta_{{\rm t},q})\left\{ \sum\limits _{n=1}^{N}({\bf X}[n]\bm{{\Sigma}}_{q}\bm{{\Sigma}}_{q}^{H}{\bf X}^{H}[n])\right\} {\bf a}_{{\rm M}}(\vartheta_{{\rm t},q})},\label{stage2angle_MLE}
\end{align}
where the detailed simplification of step $(c)$ can be found in \eqref{stage2_LikeliHood_simp}
of Appendix \ref{Appendices B.D}. Thus the maximum likelihood estimation
of $\bar{\boldsymbol{\eta}}_{q}$ becomes 
\begin{equation}
	\hat{\bar{\boldsymbol{\eta}}}_{q}^{(k)}=\arg\mathop{\max}\limits _{\bar{\boldsymbol{\eta}}_{q}}F(\bar{\boldsymbol{\eta}}_{q};\{{\hat{{\bf Y}}}_{q}^{(k)}[n]\}_{n=1,\cdots,N}).\label{stage2angle_iteration}
\end{equation}

To further reduce the complexity in \eqref{stage2angle_iteration},
we aim at replacing the multiple-dimension optimization to compute
the MLE of the parameters in $\bar{\boldsymbol{\eta}}_{q}$ by multiple
separate one-dimensional optimizations, where each parameter in $\bar{\boldsymbol{\eta}}_{q}$
is estimated individually. By denoting the objective function of \eqref{stage2angle_iteration}
by $F(\bar{\boldsymbol{\eta}}_{q})$ for notation convenience, the
coordinate-wise updating procedure of the parameter estimates of the
$q$-th path can be performed sequentially as follows: \begin{subequations}
	\label{Update_parameter_StageII} 
	\begin{align}
		\hat{\tau}_{q}^{(k)} & =\arg\mathop{\max}\limits _{\tau_{q}}F(\tau_{q},\hat{\theta}_{{\rm t},q}^{(k-1)},\hat{\phi}_{{\rm in},q}^{(k-1)},\hat{\psi}_{{\rm in},q}^{(k-1)}),\label{tau_k}\\
		\hat{\theta}_{{\rm t},q}^{(k)} & =\arg\mathop{\max}\limits _{\theta_{\mathrm{t},q}}F(\hat{\tau}_{q}^{(k)},\theta_{\mathrm{t},q},\hat{\phi}_{{\rm in},q}^{(k-1)},\hat{\psi}_{{\rm in},q}^{(k-1)}),\label{theta_t_k}\\
		\hat{\phi}_{{\rm in},q}^{(k)} & =\arg\mathop{\max}\limits _{\phi_{{\rm in},q}}F(\hat{\tau}_{q}^{(k)},\hat{\theta}_{{\rm t},q}^{(k)},\phi_{{\rm in},q},\hat{\psi}_{{\rm in},q}^{(k-1)}),\label{phi_k}\\
		\hat{\psi}_{{\rm in},q}^{(k)} & =\arg\mathop{\max}\limits _{\psi_{{\rm in},q}}F(\hat{\tau}_{q}^{(k)},\hat{\theta}_{{\rm t},q}^{(k)},\hat{\phi}_{{\rm in},q}^{(k)},\psi_{{\rm in},q}),\label{psi_k}\\
		\hat{\delta}_{q}^{(k)} & =\frac{{\bf a}_{{\rm B}}^{H}(\vartheta_{\mathrm{r},0})\left\{ \sum\limits _{n=1}^{N}(e^{j2\pi\hat{\tau}_{q}^{(k)}\frac{(n-1)B}{N}}\hat{{\bf {Y}}}_{q}^{(k)}[n]({\bf \Sigma}_{q}^{(k)})^{H}{\bf X}^{H}[n])\right\} {\bf a}_{{\rm M}}(\hat{\vartheta}_{{\rm t},q}^{(k)})}{N_{{\rm b}}{\bf a}_{{\rm M}}^{H}(\hat{\vartheta}_{{\rm t},q}^{(k)})\left\{ \sum\limits _{n=1}^{N}({\bf X}[n]({\bf \Sigma}_{q}^{(k)})({\bf \Sigma}_{q}^{(k)})^{H}{\bf X}^{H}[n])\right\} {\bf a}_{{\rm M}}(\hat{\vartheta}_{{\rm t},q}^{(k)})},\label{delta_k}
	\end{align}
\end{subequations} where the $\hat{\delta}_{q}^{(k)}$ in \eqref{delta_k}
is obtained by sustituting the estimated values in \eqref{tau_k},
\eqref{theta_t_k}, \eqref{phi_k} and \eqref{psi_k} into \eqref{stage2deltap_MLE}.

The SAGE algorithm first allows the splitting of the joint optimization
for superimposed $(Q+1)$ paths into separate optimization for a single
path, then for each single path, not all parameters but only a single
parameter related to the path is estimated while keeping the estimates
of the other parameters fixed. The iteration process in the SAGE algorithm is terminated when the distance
between the estimated $\boldsymbol{{\bf \eta}}$ returned by the SAGE
algorithm at two consecutive iteration steps is below a predefined
threshold or when the value sequence of ML $\Lambda(\boldsymbol{{\bf \eta}};\{{\bf Y}[n]\}_{n=1,2,\cdots,N})$
has stabilized. The overall SAGE algorithm to estimate the channel
parameters in $\boldsymbol{{\bf \eta}}$ is shown in Algorithm \ref{SAGE_algorithm_step},
where the channel parameters in $\boldsymbol{\eta}$ estimated in
Section \ref{Coarse Estimation of Channel Parameters} are taken as
the initial values. 
\begin{algorithm}
	\caption{The SAGE algorithm to estimate parameters $\boldsymbol{{\bf \eta}}$}
	\label{SAGE_algorithm_step}
	
	\begin{algorithmic}[1] 
		\REQUIRE (1) The received signal $\{{\bf y}_{t}[n]\}_{t=1,\cdots,T}^{n=1,\cdots,N}$;
		(2) Initial values of all channel parameters in ${\hat{\boldsymbol{\eta}}}^{(0)}=[({\hat{\boldsymbol{\eta}}}_{0}^{(0)})^{T},({\hat{\boldsymbol{\eta}}}_{1}^{(0)})^{T},\cdots,({\hat{\boldsymbol{\eta}}}_{q}^{(0)})^{T},\cdots,({\hat{\boldsymbol{\eta}}}_{Q}^{(0)})^{T}]^{T}$
		, where ${\hat{\boldsymbol{\eta}}}_{q}^{(0)}=[\tau_{q}^{(0)},\delta_{q,\textrm{R}}^{(0)},\delta_{q,\textrm{I}}^{(0)},\theta_{{\rm t},q}^{(0)},\phi_{{\rm in},q}^{(0)},\psi_{{\rm in},q}^{(0)}]^{T}$;
		(3) The error thresholds $\boldsymbol{{\bf \varepsilon}}_{1}\in\mathbb{C}^{6(Q+1)\times1}$ and $\varepsilon_{2}$; (4) $k=0$;
		
		\WHILE {$1$}
		
		\STATE $q=k{\ }{\rm mod}(Q+1)$;
		
		\STATE Estimate $ \hat{\bf{y}}_{q,t}^{(k+1)}[n],t=1,\cdots,T,n=1,\cdots,N$,
		from $\{{\bf y}_{t}[n]\}_{t=1,\cdots,T}^{n=1,\cdots,N}$ with $\hat{\boldsymbol{\eta}}^{(k)}$ according to \eqref{Received_signal_reconstruction_StageII};
		
		\STATE Estimate the channel parameters in ${\hat{\boldsymbol{\eta}}}_{q}^{(k+1)}=[\tau_{q}^{(k+1)},\delta_{q,\textrm{R}}^{(k+1)},\delta_{q,\textrm{I}}^{(k+1)},\theta_{{\rm t},q}^{(k+1)},\phi_{{\rm in},q}^{(k+1)},\psi_{{\rm in},q}^{(k+1)}]^{T}$ based on the estimated $\hat{\bf {y}}_{q,t}^{(k)}[n]$ sequentially according to \eqref{Update_parameter_StageII};
		
		\STATE Update $\hat{\boldsymbol{\eta}}^{(k+1)}\!\!=\![(\hat{\boldsymbol{\eta}}_{0}^{(k)}\!)^{T}\!,\!\cdots\!,\!(\hat{\boldsymbol{\eta}}_{q-1}^{(k)}\!)^{T}\!,\!(\hat{\boldsymbol{\eta}}_{q}^{(k+1)}\!)^{T}\!,\!(\hat{\boldsymbol{\eta}}_{q+1}^{(k)}\!)^{T},\cdots,(\hat{\boldsymbol{\eta}}_{Q}^{(k)})^{T}]^{T}$;
		
		\IF{$\left|{\hat{\boldsymbol{\eta}}}^{(k+1)}\!-\!{\hat{\boldsymbol{\eta}}}^{(k)}\right|\!\preceq\!\boldsymbol{{\bf \varepsilon}}_{1}$
			or $\left|\Lambda({\hat{\boldsymbol{\eta}}}^{(k+1)};\{{\bf Y}[n]\}_{n=1,2,\cdots,N})\!-\!\Lambda({\hat{\boldsymbol{\eta}}}^{(k)};\{{\bf Y}[n]\}_{n=1,2,\cdots,N})\right|\!\leqslant\!\varepsilon_{2}$}
		
		\STATE break;
		
		\ELSE
		
		\STATE${k}\leftarrow{k}+1$;
		
		\ENDIF
		
		\ENDWHILE
		
		\ENSURE Estimated channel parameters $\hat{\boldsymbol{\eta}}=[\hat{\boldsymbol{\eta}}_{0}^{T},\hat{\boldsymbol{\eta}}_{1}^{T},\cdots,\hat{\boldsymbol{\eta}}_{q}^{T},\cdots,\hat{\boldsymbol{\eta}}_{Q}^{T}]^{T}$.
		
	\end{algorithmic}
\end{algorithm}

\section{Estimation of Position-Related Parameters}
\label{Position and Rotation Angle Estimation}
Based on the refined estimates of $\hat{\boldsymbol{\eta}}$,
the final parameters in $\tilde{\bm{{\eta}}}$ are first coarsely estimated in closed form to provide the initial values for the refined estimating process. 

\subsection{Coarse Estimation for Scenarios with 2D Rotation of MS}
\label{CoarseEst_Postn_2Drott}
According to the geometric relationships described in Fig. \ref{figpositioning}, we can formulate $\hat{{\bf m}}$ in closed-form expressions by using the estimated channel parameters in $\hat{\bm{{\eta}}}$ as
	\begin{align}
		{\hat{{\bf m}}} & ={\bf r}-(\hat{\tau}_{0}c-\left\Vert {\bf r}-{\bf b}\right\Vert _{2})\hat{{\bf {e}}}_{0},\label{MScordVlosVlos_2Drott}
	\end{align}
	where $\hat{{\bf {e}}}_{q}=[\sin\hat{\phi}_{{\rm in},q}\cos\hat{\psi}_{{\rm in},q},\sin\hat{\phi}_{{\rm in},q}\sin\hat{\psi}_{{\rm in},q},\cos\hat{\phi}_{{\rm in},q}]^{T}$, $q=0,1,\cdots,Q$.
	
	To find the closed-form expression for ${\bf s}^{q}$, we construct the following system of equations.
	\begin{subequations}
		\label{EqutnSetFord_sr_q} 
		\begin{align}
			{\bf s}^{q} & ={\bf r}-{d}_{sr}^{q}\hat{{\bf e}}_{q},{d}_{sr}^{q}=\left\Vert {\bf r}-{\bf s}^{q}\right\Vert _{2},\label{SqSubEq1}\\
			(\hat{\tau}_{q}c-\left\Vert {\bf r}-{\bf b}\right\Vert _{2}) & ={d}_{sr}^{q}+{d}_{ms}^{q},{d}_{ms}^{q}=\left\Vert {\bf s}^{q}\!-\!{\hat{{\bf m}}}\right\Vert _{2}.\label{SqSubEq2}
		\end{align}
	\end{subequations}
	
	In \eqref{EqutnSetFord_sr_q}, the $\hat{{\bf {e}}}_{q}$	and $\hat{d}_{mqr}\triangleq(\hat{\tau}_{q}c-\left\Vert {\bf r}-{\bf b}\right\Vert _{2})$ are known based on the estimated channel parameters,	and the ${\hat{{\bf m}}}$ is known by using \eqref{MScordVlosVlos_2Drott}. Thus, the only unknown variable in \eqref{EqutnSetFord_sr_q} is essentially ${d}_{sr}^{q}$, which is the distance from the $q$th	scatterer to the RIS. By solving \eqref{EqutnSetFord_sr_q}, we can obtain
	\begin{align}
		\hat{d}_{sr}^{q} & =(\hat{d}_{mqr}^{2}-\hat{d}_{rm}^{2})/\{2[\hat{d}_{mqr}-({\bf r}-\hat{{\bf m}})^{T}\hat{{\bf {e}}}_{q}]\}, \label{hatd_sr}
	\end{align}
	where $\hat{d}_{rm}\triangleq\left\Vert {\bf r}\!-\!{\hat{{\bf m}}}\right\Vert _{2}$. By substituting $\hat{d}_{sr}^{q}$ into \eqref{SqSubEq1}, the $\hat{\bf s}^{q}$ can be obtained. 
	
	To find the closed-form expression for $\alpha$,$\beta$, we can
	first obtain $\mathbf{e}_{\textrm{rot}}$ by solving the following
	system of equations. \begin{subequations} \label{EqtnsForRotArry-FrminRevised}
		\begin{align}
			({\bf r}-\hat{{\bf m}})^{T}\mathbf{e}_{\textrm{rot}} & =\sin\hat{\theta}_{t,0}\left\Vert {\bf r}-\hat{{\bf m}}\right\Vert _{2},\label{Smplfy1}\\
			(\hat{{\bf s}}^{q}-\hat{{\bf m}})^{T}\mathbf{e}_{\textrm{rot}} & =\sin\hat{\theta}_{t,q}\left\Vert \hat{{\bf s}}^{q}-\hat{{\bf m}}\right\Vert _{2},\label{Smplfy2}\\
			\mathbf{e}_{\textrm{rot}}^{T}\mathbf{e}_{\textrm{rot}} & =1.\label{Smplfy3}
		\end{align}
	\end{subequations} After obtaining $\hat{\mathbf{e}}_{\textrm{rot}}=[\hat{e}_{\textrm{rot}}^{x},\hat{e}_{\textrm{rot}}^{y},\hat{e}_{\textrm{rot}}^{z}]^{T}$,
	we have 
	\begin{align}
		& \qquad\qquad\qquad\qquad\hat{\beta}=\arccos\hat{e}_{\textrm{rot}}^{z},\\
		& \!\!\!\!\hat{\alpha}\!=\!\left\{ \begin{array}{c}
			\!\!\!\!\!-\!\arctan(\hat{e}_{\textrm{rot}}^{y}/\hat{e}_{\textrm{rot}}^{x}),\textrm{if}-\!\pi/2\!<\!\!\arctan(\hat{e}_{\textrm{rot}}^{y}/\hat{e}_{\textrm{rot}}^{x})\!\!<\!0,\\
			\!\!\!\!\!\pi\!-\!\arctan(\hat{e}_{\textrm{rot}}^{y}/\hat{e}_{\textrm{rot}}^{x}),\textrm{if}\,\,0\!\leq\!\arctan(\hat{e}_{\textrm{rot}}^{y}/\hat{e}_{\textrm{rot}}^{x})\!\!<\!\!\pi/2.
		\end{array}\right.\label{alpha_Recvr}
	\end{align}
	where the definition of $\mathbf{e}_{\textrm{rot}}$ and the assumed
	$-\pi<-\hat{\alpha}<0$ in Section \ref{System Model} are utilized.
	It is noted that if there are no scatterers, the system of equations
	\eqref{EqtnsForRotArry-FrminRevised} are degenerated, and the closed-form solutions
	cannot be obtained. 
	
	The detailed solving process for \eqref{EqutnSetFord_sr_q}
	and \eqref{EqtnsForRotArry-FrminRevised} can be found in Appendix \ref{Appendices_B_add.A} and \ref{Appendices_B_add.B}, respectively. From the solving process of \eqref{EqtnsForRotArry-FrminRevised},
	it is found that three cases of solutions may occur: ``no solution'',
	``unique solution'', and ``two solutions''. Then, the closed-form estimation of the final parameters $\ensuremath{\hat{\tilde{\bm{{\eta}}}}}$
	becomes unstable due to the possibility of multiple solution cases
	for $\mathbf{e}_{\textrm{rot}}$. In contrast, when the MS has an one-dimensional (1D) rotation, the closed-form solutions are notably more stable. 

The closed-form estimates of the final parameters only exploit
some of the channel parameters, and do not extract the full information
in all the channel parameters. For example, the ${\bf m}$ is estimated
through \eqref{MScordVlosVlos_2Drott} by only using the channel parameters
in the MS-RIS path. Actually, the channel parameters in the MS-scatterer$q$-RIS
path also also carry information regarding the MS coordinates. Thus,
further refining the estimates of the final parameters in $\tilde{\bm{{\eta}}}$
is needed by using the closed-form solutions as the initial values.

\subsection{Coarse Estimation for Scenarios with 1D Rotation of MS}\label{Position-Coarse-Estimation}
When the MS is assumed to rotate on a plane parallel to the ${x-o-y}$ plane, we have $\beta=\pi/2 {\textrm{ rad}}$, and the 1D unknown rotation angle is $\alpha$. Then the $\mathbf{e}_{\textrm{rot}}$ become $\mathbf{e}_{\textrm{rot}}\triangleq[\cos\alpha,-\sin\alpha,0]^{T}$. According to the geometric relationship
described in Fig. \ref{figpositioning}, we can find $\hat{{\bf m}}$
and $\hat{\alpha}$ in closed-form expressions by using the estimated
channel parameters in $\hat{\bm{{\eta}}}$ as
\begin{align}
	{\hat{{\bf m}}} & ={\bf r}+(\hat{\tau}_{0}c-\left\Vert {\bf r}-{\bf b}\right\Vert _{2})[-\sin\hat{\phi}_{{\rm in},0}\cos\hat{\psi}_{{\rm in},0},-\sin\hat{\phi}_{{\rm in},0}\sin\hat{\psi}_{{\rm in},0},-\cos\hat{\phi}_{{\rm in},0}]^{T},\label{MScordVlosVlos}\\
	{\hat{\alpha}} & =(2\pi-\hat{\psi}_{{\rm in},0})-\arccos[(\left\Vert {\bf m}-{\bf r}\right\Vert _{2}\sin\hat{\theta}_{{\rm t},0})/(\left\Vert {\bf m}-{\bf r}\right\Vert _{2}\sin\hat{\phi}_{{\rm in},0})]\nonumber \\
	& =(2\pi-\hat{\psi}_{{\rm in},0})-\arccos(\sin\hat{\theta}_{{\rm t},0}/\sin\hat{\phi}_{{\rm in},0}).\label{RotaVlosVlos}
\end{align}
To find the closed-form expression for ${\bf s}^{q}$, we can build
the following set of equations: \begin{subequations} \label{Scattrq}
	\begin{align}
		{\hat{{\bf s}}}^{q} & ={\bf r}+{\hat{d}}_{sr}[-\sin\hat{\phi}_{{\rm in},q}\cos\hat{\psi}_{{\rm in},q},-\sin\hat{\phi}_{{\rm in},q}\sin\hat{\psi}_{{\rm in},q},-\cos\hat{\phi}_{{\rm in},q}]^{T},\\
		(\hat{\tau}_{q}c-\left\Vert {\bf r}-{\bf b}\right\Vert _{2}) & ={\hat{d}}_{sr}+{\hat{d}}_{ms},\\
		\left\Vert {\hat{{\bf s}}}^{q}-{\hat{{\bf m}}}\right\Vert _{2}\sin{\hat{\theta}}_{{\rm t},q} & =({\hat{s}}_{x}^{q}-{\hat{m}}_{x})\cos{\hat{\alpha}}-({\hat{s}}_{y}^{q}-{\hat{m}}_{y})\sin{\hat{\alpha}}.
	\end{align}
\end{subequations} The set of equations in \eqref{Scattrq} can be readily solved. After some mathematical derivations, the closed-form
solution for the scatterer coordinates in ${\bf s}^{q}$ can be represented
as \begin{subequations} \label{ScattrqCordinates}
	\begin{align}
		\ensuremath{\hat{s}_{x}^{q}} & =\ensuremath{\frac{(AD+r_{x})\sin\hat{\theta}_{{\rm t},q}+\hat{m}_{x}A\cos\hat{\alpha}+(r_{y}A-r_{x}B-\hat{m}_{y}A)\sin\hat{\alpha}}{\sin\hat{\theta}_{{\rm t},q}+A\cos\hat{\alpha}-B\sin\hat{\alpha}}},\\
		\ensuremath{\hat{s}_{y}^{q}} & =\ensuremath{r_{y}+(\hat{s}_{x}^{q}-r_{x})\frac{B}{A}},\\
		\ensuremath{\hat{s}_{z}^{q}} & =\ensuremath{r_{z}+(\hat{s}_{x}^{q}-r_{x})\frac{C}{A}},
	\end{align}
\end{subequations} where $A=-\sin{\hat{\phi}_{{\rm {in}},q}}\cos{\hat{\psi}_{{\rm {in}},q}}$,
$B=-\sin{\hat{\phi}_{{\rm {in}},q}}\sin{\hat{\psi}_{{\rm {in}},q}}$,
$C=-\cos{\hat{\phi}_{{\rm {in}},q}}$ and $D={\hat{\tau}_{q}}c-{\left\Vert {{\bf {r}}-{\bf {b}}}\right\Vert _{2}}$.

\subsection{Refining Estimation}
Although the closed-form solutions for $\tilde{\bm{{\eta}}}$ can be obtained
	as shown in Section \ref{CoarseEst_Postn_2Drott} and Section \ref{Position-Coarse-Estimation},
they are inaccurate due to the estimation error of $\boldsymbol{{\bf \eta}}$. Moreover, the
closed-form estimates of the final parameters only exploit some
of the channel parameters, and do not extract the full information
in all the channel parameters. Thus, further refining the estimates of the final
parameters in $\tilde{\bm{{\eta}}}$ is needed by
using the closed-form solutions as the initial values.

The equations in Section \ref{The Geometric Relationship} describe the relationship between $\bm{{\eta}}$ and $\tilde{\bm{{\eta}}}$. Since the coordinates
of the RIS and the BS are assumed to be known, the angles of the RIS-BS
link are taken as known values, which means that the equations in
\eqref{AnlgesRIS-BS-VLOS} are not estimated but calculated in advance.
The expressions in \eqref{AnglesMS-RIS-VLOS}, \eqref{AnglesMS-RIS-NLOS}
and \eqref{Distances} describe a mapping $\bm{{\eta}}=\mathcal{G}(\tilde{\bm{{\eta}}})$.
Thus, the estimated $\hat{\tilde{\bm{{\eta}}}}$ can be obtained by
\begin{align}
	\hat{\tilde{\bm{{\eta}}}} & =\arg\underset{\tilde{\bm{{\eta}}}}{\min}{\ }[\hat{\bm{{\eta}}}-\mathcal{G}(\tilde{\bm{{\eta}}})]^{T}\mathbf{J}_{\hat{\bm{{\eta}}}}[\hat{\bm{{\eta}}}-\mathcal{G}(\tilde{\bm{{\eta}}})],\label{Estor1}
\end{align}
where $\mathbf{J}_{\hat{\bm{{\eta}}}}$ is the fisher information matrix
(FIM) for the channel parameters $\boldsymbol{{\bf \eta}}$ calculated
by substituting the estimated value $\hat{\bm{{\eta}}}$. The FIM
for $\boldsymbol{{\bf \eta}}$ is given in \eqref{FIM_stage2_rewritten}
of Section \ref{subsec:FIM-for-ChnnlParmtrs}. The estimator in \eqref{Estor1}
is asymptotically (w.r.t. $T\times N$) equivalent to the ML estimation
of the transformed parameter $\tilde{\bm{{\eta}}}$\cite{stoica1989reparametrization}\cite{swindlehurst1998maximum}.
The problem in \eqref{Estor1} is a nonlinear least squares problem,
which can be solved by the Levenberg-Marquardt Method \cite{levenberg1944method,marquardt1963algorithm},
which can be initialized in Section \ref{CoarseEst_Postn_2Drott} or \ref{Position-Coarse-Estimation}.

\section{Position and Orientation Estimation Fundamental Bounds}

\label{Position and Orientation Estimation Fundermental Bounds}

To provide a benchmark for the accuracy of the proposed estimation algorithms, we derive the error bounds for the formulated estimation problems.

\subsection{CRB for Channel Parameters\label{subsec:FIM-for-ChnnlParmtrs}}

We first derive the FIM ${\bf J}_{\boldsymbol{\eta}}$ of the channel
parameters in $\boldsymbol{{\bf \eta}}$. The joint probability density function of the vectorized signal ${\bf y}[n]$ is given in \eqref{joint_probability_density_allTimeSlots}. By taking the natural logarithm of \eqref{joint_probability_density_allTimeSlots} and omitting
the constant terms, the log-likelihood functon
$\Lambda(\boldsymbol{{\bf \eta}};\{{\bf y}[n]\}_{n=1,\cdots,N})$
of $\boldsymbol{{\bf \eta}}$ given the observed signal $\{{\bf y}[n]\}_{n=1,\cdots,N}$ can be written as
\begin{align}
	\Lambda(\boldsymbol{{\bf \eta}};\{{\bf y}[n]\}_{n=1,\cdots,N}) & \propto-\sum\limits _{n=1}^{N}\left\{ ({\bf y}[n]-\boldsymbol{\mu}[n])^{H}{\bf \bm{{\Xi}}}^{-1}({\bf y}[n]-\boldsymbol{\mu}[n])\right\},\label{stage2_ML}
\end{align}
where $\boldsymbol{\mu}[n]=\tilde{\bf {H}}[n]{\bf x}[n]$ and $\tilde{\bf {H}}[n]\triangleq\sum\limits _{q=0}^{Q}\delta_{q}e^{-j2\pi\tau_{q}\frac{(n-1)B}{N}}(\bm{{\Sigma}}_{q}^{T}\otimes \tilde{\bf {H}}_{q})$.

Defining $\hat{\boldsymbol{\eta}}$ as the unbiased estimator of $\boldsymbol{\eta}$,
the mean squared error (MSE) of $\hat{\boldsymbol{\eta}}$ is bounded
as  \begin{align}
	\mathbb{E}\left[(\hat{\boldsymbol{\eta}}-\boldsymbol{\eta})(\hat{\boldsymbol{\eta}}-\boldsymbol{\eta})^{T}\right] & \succeq{\bf J}_{\boldsymbol{\eta}}^{-1},
\end{align}
where ${\bf J}_{\boldsymbol{\eta}}\in\mathbb{C}^{6(Q+1)\times6(Q+1)}$
is the FIM of $\boldsymbol{\eta}$. The element at the $u$-th row and the
$v$-th column of ${\bf J}_{\boldsymbol{\eta}}$ can be represented
as \cite{kay1993fundamentals}
\begin{align}
	[{\bf J}_{\boldsymbol{\eta}}]_{u,v} & =-\mathbb{E}\left\{ \frac{\partial^{2}\Lambda(\boldsymbol{{\bf \eta}};\{{\bf y}[n]\}_{n=1,\cdots,N})}{\partial[\boldsymbol{\eta}]_{u}\partial[\boldsymbol{\eta}]_{v}}\right\} \nonumber \\
	& =\mathbb{E}\left\{ \frac{\partial \Lambda(\boldsymbol{{\bf \eta}};\{{\bf y}[n]\}_{n=1,\cdots,N})}{\partial[\boldsymbol{\eta}]_{u}}\frac{\partial \Lambda(\boldsymbol{{\bf \eta}};\{{\bf y}[n]\}_{n=1,\cdots,N})}{\partial[\boldsymbol{\eta}]_{v}}\right\} .\label{FIM_stage2}
\end{align}
It is noted that the $(u,u)$-th element of ${\bf J}_{\boldsymbol{\eta}}^{-1}$
(i.e., $[{\bf J}_{\boldsymbol{\eta}}^{-1}]_{u,u}$) is the CRB for
the $u$-th parameter of ${\bf \boldsymbol{\eta}}$.

By substituting \eqref{stage2_ML} into \eqref{FIM_stage2}, \eqref{FIM_stage2}
can be rewritten as
\begin{align}
	[{\bf J}_{\boldsymbol{\eta}}]_{u,v} & \overset{(d)}{=}\sum\limits _{n=1}^{N}\left\{ 2\mathfrak{Re}[\frac{\partial\boldsymbol{\mu}^{H}[n]}{\partial[\boldsymbol{\eta}]_{u}}\bm{{\Xi}}^{-1}\frac{\partial\boldsymbol{\mu}[n]}{\partial[\boldsymbol{\eta}]_{v}}]+{\rm Tr}[\bm{{\Xi}}^{-1}\frac{\partial{\bf \bm{{\Xi}}}}{\partial[\boldsymbol{\eta}]_{u}}\bm{{\Xi}}^{-1}\frac{\partial{\bf \bm{{\Xi}}}}{\partial[\boldsymbol{\eta}]_{v}}]\right\} \nonumber \\
	& \overset{(e)}{=}\frac{2}{\sigma^{2}}\sum\limits _{n=1}^{N}\mathfrak{Re}\left\{ {\bf x}^{H}[n]\frac{\partial \tilde{\bf {H}}^{H}[n]}{\partial[\boldsymbol{\eta}]_{u}}\frac{\partial \tilde{\bf {H}}[n]}{\partial[\boldsymbol{\eta}]_{v}}{\bf x}[n]\right\} ,\label{FIM_stage2_rewritten}
\end{align}
where the step $(d)$ can refer to Appendix 15C of \cite{kay1993fundamentals}.
The step $(e)$ is obtained because ${\bf \bm{{\Xi}}}$ is a constant
matrix. From \eqref{FIM_stage2_rewritten}, we find that the key to
obtain the FIM ${\bf J}_{\boldsymbol{\eta}}$ is to calculate $\frac{\partial \tilde{\bf {H}}[n]}{\partial[\boldsymbol{\eta}]_{u}},u=1,2,\cdots,6(Q+1)$,
which are derived in Appendix \ref{Appendices C.A}.

\subsection{Position and Orientation Error Bounds}
Through a transformation
of variables from $\boldsymbol{{\bf \eta}}$ to $\tilde{\bm{{\eta}}}$, the FIM ${\bf J}_{\tilde{\bm{{\eta}}}}\in\mathbb{C}^{[5(Q+1)+2]\times[5(Q+1)+2]}$ of the
final position parameters in $\tilde{\bm{{\eta}}}$
can be determined from ${\bf J}_{\boldsymbol{\eta}}\in\mathbb{C}^{[6(Q+1)]\times[6(Q+1)]}$
as
\begin{align}
	{\bf J}_{\tilde{\bm{{\eta}}}} & =\mathbf{T}{\bf J}_{\boldsymbol{\eta}}\mathbf{T}^{T},\label{FIM_PEB_OEB}
\end{align}
where $\mathbf{T}\triangleq\frac{\partial\boldsymbol{\eta}^{T}}{\partial\tilde{\bm{{\eta}}}}\in\mathbb{C}^{[5(Q+1)+2]\times[6(Q+1)]}$
is the transformation matrix. To obtain $\mathbf{T}$,
the relationships between $\boldsymbol{{\bf \eta}}$ and $\tilde{\bm{{\eta}}}$
described in Section \ref{The Geometric Relationship} are utilized.
The specific derivations for $\mathbf{T}$ can be
	found in Appendix \ref{Appendices C.B}.

By defining $\mathbf{c}_{\tilde{\bm{{\eta}}}}\triangleq\textrm{diag}[{\bf J}_{\tilde{\bm{{\eta}}}}^{-1}]$,
	the elements in $\mathbf{c}_{\tilde{\bm{{\eta}}}}$ represent the
	theoretical lower bounds of the MSEs of the corresponding
	elements in estimated $\hat{\tilde{\bm{{\eta}}}}$. We can  equivalently
	rewrite $\mathbf{c}_{\tilde{\bm{{\eta}}}}=[\mathbf{c}_{\mathbf{h}_{0}}^{T},\mathbf{c}_{\mathbf{h}_{1}}^{T},\cdots,\mathbf{c}_{\mathbf{h}_{Q}}^{T},\mathbf{c}_{\tilde{{\bf {m}}}}^{T},\mathbf{c}_{\mathbf{s}^{1}}^{T},\cdots,\mathbf{c}_{\mathbf{s}^{Q}}^{T}]^{T}$,
	where $\mathbf{c}_{\mathbf{h}_{q}}\triangleq[c_{\delta_{q,\textrm{R}}},c_{\delta_{q,\textrm{I}}}]^{T}$,
	$\mathbf{c}_{\tilde{{\bf {m}}}}\triangleq[c_{m_{x}},c_{m_{y}},c_{m_{z}},c_{\alpha},c_{\beta}]^{T}$,
	and $\mathbf{c}_{\mathbf{s}^{q}}\triangleq[c_{s_{x}^{q}},c_{s_{y}^{q}},c_{s_{z}^{q}}]^{T}$.
	Then, the PEB of the MS is obtained by $\textrm{PEB}=\sqrt{c_{m_{x}}+c_{m_{y}}+c_{m_{z}}}$,
	and the PEB for the scatterer is  $\textrm{PEB\_scttr}=\sqrt{(1/Q)\sum_{q=1}^{Q}(c_{s_{x}^{q}}+c_{s_{y}^{q}}+c_{s_{z}^{q}})}$.
	The OEB is obtained as $\mathrm{OEB}=\sqrt{c_{\alpha}+c_{\beta}}$.

In scenarios with MS's 1D rotation, the $\beta$ is removed from $\tilde{\bm{{\eta}}}$, and the specific derivations for the revised $\mathbf{T} \in\mathbb{C}^{[5(Q+1)+1]\times[6(Q+1)]}$ can be
	obtained from Appendix \ref{Appendices C.B}. With the revised  $\mathbf{T}$, the ${\bf J}_{\tilde{\bm{{\eta}}}}$ is calculated by \eqref{FIM_PEB_OEB}. And the revised OEB becomes $\mathrm{OEB}=\sqrt{c_{\alpha}}$.

\section{Complexity Analysis}
	\label{Complexity}
The computational complexity of the proposed algorithm arises from
three main parts: coarse channel parameter estimation, refined channel
parameter estimation using the SAGE algorithm, and positioning parameter
estimation. 
\subsection{Complexity of Coarse Channel Parameter Estimation}
Firstly, we analyze the computational complexity of the coarse
estimation of channel parameres presented in Section \ref{Coarse Estimation of Channel Parameters}, which comprises the following stages.

\subsubsection{Estimation of the AODs at the MS}

Firstly, the AODs at the MS are coarsely estimated by the DCS-OMP
algorithm in Algorithm \ref{DCS-OMP}. The computational complexity of the DCS-SOMP
algorithm for coarsely estimating AOD is $\mathcal{O}((Q+1){{T}_{1}}{{G}_{\text{m}}}{{N}_{\text{b}}})\approx\mathcal{O}(Q{{T}_{1}}{{G}_{\text{m}}}{{N}_{\text{b}}})$ \cite{he_pilot_2016},\cite{lin_channel_sparsity_2022}, where $Q+1$ is the number of iterations (i.e., the sparseness
of the channel), ${{T}_{1}}$ is the number of time slots, ${{G}_{\text{m}}}$
is the the number of discretized angles in the dictionary, and ${{N}_{\text{b}}}$
is the number of BS antennas.

Then, the AODs are refined by the MLE. The AODs are refined to break
the resolution of the DCS-SOMP algorithm by solving the MLE in Problem
\eqref{AODmle_fine}. The corresponding logarithmic likelihood
function is given by \eqref{logMLf_AODmleFine}. Asymptotically speaking, we observe
that the complexity in performing the one dimensional (1D) optimization
on AODs required by the MLE in \eqref{AODmle_fine} is on the order of $\mathcal{O}(S_{{{\theta}_{\text{t}}}}^{\text{c}}N{{T}_{1}}{{N}_{\text{b}}}{{Q}^{2}})+\mathcal{O}(S_{{{\theta}_{\text{t}}}}^{\text{c}}N{{Q}^{3}})+\mathcal{O}(S_{{{\theta}_{\text{t}}}}^{\text{c}}N{{T}_{1}}{{N}_{\text{b}}}Q{{N}_{\text{m}}})$,
where $S_{{{\theta}_{\text{t}}}}^{\text{c}}$ denotes the number of
evaluation points along the searching dimension (the number of grids
if the grid-search method is utilized). Under the assumption that
$T_{1}N_{{\rm {b}}}N_{{\rm {m}}}$ is significantly larger than $Q^{2}$
and that $Q$ is small, the overall complexity reduces to $\mathcal{O}(S_{{{\theta}_{\text{t}}}}^{\text{c}}N{{T}_{1}}{{N}_{\text{b}}}Q{{N}_{\text{m}}})$.

\subsubsection{Estimation of the AOAs at the RIS}

To coarsely estimate the AOAs at the RIS, the DCS-SOMP is performed
on \eqref{SprsProblm_RISangles}. Then, the computational complexity
is $\mathcal{O}((Q+1)(\Upsilon+1){{G}_{\text{r}}})\approx\mathcal{O}(Q\Upsilon{{G}_{\text{r}}})$\cite{peng_channel_2022}.

\subsubsection{Estimation of the TOAs and Channel Gains}

The TOA ${{\tau}_{q}}$ is estimated by \eqref{DeltTOA_TOA}. The complexity of obtaining ${{m}_{q}}$ by ${{m}_{q}}=\arg\underset{m}{\mathop{\max}}\,\left|\mathbf{i}_{m}^{T}\mathbf{U}_{N}^{H}{{\widehat{\widetilde{\delta}}}_{q}}\right|$
is $\mathcal{O}((Q+1){{N}^{2}})$. The complexity of searching the
rotation parameter $\vartriangle{{\tau}_{q}}$ by \eqref{rotationSrch} is $\mathcal{O}((Q+1)S_{\tau}^{\text{c}}N)$,
where $S_{\tau}^{\text{c}}$ denotes the number of evaluation points
along the searching dimension (the number of grids if the grid-search
method is utilized). Then, it is observed from \eqref{DeltTOA_TOA} that the
complexity of estimating the TOA is on the order of $\mathcal{O}(Q{{N}^{2}})+\mathcal{O}(QS_{\tau}^{\text{c}}N)$
\cite{zhou2022channel}. Since $S_{\tau}^{\text{c}}\geq N$, the second term dominates,
and the overall complexity simplifies to $\mathcal{O}(QS_{\tau}^{\text{c}}N)$.

The channel gain ${{\delta}_{q}}$ is estimated in \eqref{Opt_Chnnlgain} according
to the least squares (LS) estimation. It is observed from \eqref{Opt_Chnnlgain}
that the complexity of estimating the channel gain is on the order
of $\mathcal{O}(QN)$.

In summary, the overall complexity of the coarse estimation for channel
parameters is given by
\begin{equation}
	\mathcal{O}_{\text{chnl}}^{\text{c}}=\mathcal{O}(Q{{T}_{1}}{{G}_{\text{m}}}{{N}_{\text{b}}})+\mathcal{O}(S_{{{\theta}_{\text{t}}}}^{\text{c}}N{{T}_{1}}{{N}_{\text{b}}}Q{{N}_{\text{m}}})+\mathcal{O}(Q\Upsilon{{G}_{\text{r}}})+\mathcal{O}(QS_{\tau}^{\text{c}}N).
\end{equation}
\subsection{Complexity of Refined Channel Parameter Estimation}
Then, we analyze the computational complexity of the proposed
SAGE algorithm to refine the estimation of the channel parameters,
also in comparison to the plain joint maximum likelihood estimator
(MLE) of all channel parameters. 
\subsubsection{The Complexity of the Plain MLE}
The MLE of all channel parameters is derived in Section \ref{MLE_all_param}. The log-likelihood
function $\widehat{\boldsymbol{\eta}}=\arg\underset{\mathbf{\boldsymbol{\eta}}}{\mathop{\max}}\,\;\Lambda(\boldsymbol{\eta};{{\{\mathbf{Y}[n]\}}_{n=1,2,\cdots,N}})$
is finally expressed in \eqref{stage2_Allparameters_MLE}. Asymptotically
speaking, we observe that the complexity of calculating the log-likelihood
function $\Lambda(\boldsymbol{\eta};{{\{\mathbf{Y}[n]\}}_{n=1,2,\cdots,N}})$
is on the order of $\mathcal{O}(QNTN_{E}^{'})+\mathcal{O}({{Q}^{2}}NT{{N}_{\text{r}}})+\mathcal{O}({{Q}^{2}}NTN_{\text{m}}^{2})$,
where the $N_{E}^{'}\triangleq{{N}_{\text{r}}}+{{N}_{\text{b}}}+{{N}_{\text{m}}}{{N}_{\text{b}}}$
is a term related to the number of elements at MS, RIS, and BS. The
MLE of all channel parameters requires the high-dimensional nonlinear
optimization for $\boldsymbol{\eta}\in{{\mathbb{C}}^{6(Q+1)}}$. Let
$S$ denotes the number of evaluation points per dimension of $\boldsymbol{\eta}\in{{\mathbb{C}}^{6(Q+1)}}$,
assumed to be the same for all the $6(Q+1)$ dimensions for the sake
of exposition. Then, the complexity in performing the $6(Q+1)$ dimensions
optimization for $\boldsymbol{\eta}\in{{\mathbb{C}}^{6(Q+1)}}$ required
by the plain joint MLE of all channel parameters is on the order of
\cite{fascista_ris-aided_2022}
	\begin{align}\mathcal{O}_{\text{PMLE}} & =\mathcal{O}({{S}^{6(Q+1)}}QNTN_{E}^{'})+\mathcal{O}({{S}^{6(Q+1)}}{{Q}^{2}}NT{{N}_{\text{r}}})+\mathcal{O}({{S}^{6(Q+1)}}{{Q}^{2}}NTN_{\text{m}}^{2}) \nonumber \\ 
		& =\mathcal{O}[{{S}^{6(Q+1)}}QNT(N_{E}^{'}+Q{{N}_{\text{r}}}+QN_{\text{m}}^{2})],
	\end{align}
which is computationally prohibitive.
\subsubsection{The Complexity of the Proposed SAGE Algorithm}
On the other hand, the computational complexity of the proposed SAGE
algorithm is mainly contributed by the step 3 and step 4 of the proposed
SAGE algorithm in Algorithm \ref{SAGE_algorithm_step}. In step3, the complexity to estimate
$\widehat{\mathbf{y}}_{q,t}^{(k+1)}[n],t=1,\cdots,T,n=1,\cdots,N$
according to \eqref{Received_signal_reconstruction_StageII} is on the order of
$\mathcal{O}(QNTN_{\text{E}}^{'})$. In step4, the coordinate-wise
updating procedure of the parameter estimates of the $q$-th path
is performed sequentially according to \eqref{tau_k}-\eqref{delta_k}. The \eqref{tau_k}-\eqref{psi_k}
is performing the 1D optimization by using \eqref{stage2angle_MLE}. The complexity
to calculate the likelihood function $F(\bar{\boldsymbol{\eta}}_{q};{{\{\widehat{\mathbf{Y}}_{q}^{(k+1)}[n]\}}_{n=1,\cdots,N}})$
of $\bar{\boldsymbol{\eta}}_{q}$ in \eqref{stage2angle_MLE} is $\mathcal{O}(NTN_{\text{E}}^{''})$,
where $N_{\text{E}}^{''}\triangleq{{N}_{\text{r}}}+{{N}_{\text{b}}}+{{N}_{\text{m}}}{{N}_{\text{b}}}+N_{\text{m}}^{2}$.
Similarly, we let $S$ denotes the number of evaluation points for
\eqref{tau_k}-\eqref{psi_k}, assumed to be the same for all the coordinate-wise
optimizations. The complexity of performing 1D optimization from \eqref{tau_k} to \eqref{psi_k} is on the order of $\mathcal{O}(SNTN_{\text{E}}^{''})$.
It is noted that if the iterative optimization algorithms (such as
the golden section search, Nelder-Mead, and quasi-newton methods),
instead of the exhaustive grid search method, are utilized to achieve
the 1D optimization, the above $S$ can be taken as the iteration
number. The update of the complex channel gain $\hat{\delta}_{q}^{(k)}$
according to \eqref{delta_k} does not need the 1D optimization, thus only
requires the complexity of $\mathcal{O}(NTN_{\text{E}}^{''})$. This
contribution is negligible compared to the four 1D optimizations in
\eqref{tau_k} to \eqref{psi_k}. Thus, the overall complexity of step 4 in
Algorithm \ref{SAGE_algorithm_step} is $\mathcal{O}(SNTN_{\text{E}}^{''})$. Then, for each
iteration of the SAGE algorithm, the overall complexity is given by
$\mathcal{O}(QNTN_{\text{E}}^{'})+\mathcal{O}(SNTN_{\text{E}}^{''})$,
where the first term and the second term correspond to the complexity
of step 3 and step 4, respectively. Since $SN_{\text{E}}^{''}\ge QN_{\text{E}}^{'}$,
the second term $\mathcal{O}(SNTN_{\text{E}}^{''})$ dominates the
first, and the overall complexity simplifies to $\mathcal{O}(SNTN_{\text{E}}^{''})$
for each iteration of the SAGE algorithm. If the iteration number
of SAGE algorithm is $K(Q+1)$, the overall complexity of the SAGE
algorithm is $\mathcal{O}{_{\text{SAGE}}}=\mathcal{O}(KQSNTN_{\text{E}}^{''})$
\cite{fascista_ris-aided_2022}.
\subsubsection{Complexity Comparison between the Plain MLE and SAGE Algorithm}
By comparing the joint MLE of all channel parameters with the proposed
SAGE algorithm in terms of the computational complexity, it is observed
that the complexity of the SAGE algorithm has been greatly reduced.
On one hand, the SAGE algorithm splits the joint optimization for
superimposed $Q+1$ paths into separate optimizations for a single
path, which can reduce the complexity by the order of $Q+1$. On the
other hand, in the SAGE algorithm, only a single parameter related
to the path is estimated for every single path while keeping the estimates
of the other parameters fixed, rather than estimating all parameters.
This manipulation can reduce the complexity by a factor of six (six
unknown parameters per single path). Moreover, the SAGE algorithm
can be further accelerated with lower complexity. For example, the
1D optimization in the SAGE algorithm can be performed more efficiently
by using some one-dimensional unconstrained optimization algorithms
with fast convergence (e.g., superlinear and quadratic convergence)
and low complexity. The log-likelihood function can be computed more
efficiently with some mathematical reformulation.
\subsection{Complexity of Positioning Parameter Estimation}
Finally, we analyze the computational complexity of estimating
the position-related parameters. 

When positioning the MS based on the estimated channel parameters,
there are two stages of coarse estimation and refined estimation.
In the coarse estimation stage, the coordinates of the MS and scatterers
as well as the rotation angle of the MS are obtained via closed-form
expressions, resulting in negligible complexity. In the refined estimation
stage, the Levenberg--Marquardt (L-M) method is employed to solve
Problem \eqref{Estor1}, which is a weighted nonlinear
least squares problem. For each iteration of the L-M method, the complexity
is $\mathcal{O}({{m}^{2}}n+m{{n}^{2}}+{{n}^{3}})$ \cite{levenberg1944method,marquardt1963algorithm,dennis1996numerical}, where
$m=6(Q+1)$ is the length of the observation vector (channel parameter
vector), and $n=5Q+6$ is the length of the position-related parameter
vector. Consequently, the total complexity complexity of the positioning
process is $\mathcal{O}({{N}_{\text{I}}}{{m}^{2}}n+{{N}_{\text{I}}}m{{n}^{2}}+{{N}_{\text{I}}}{{n}^{3}})$,
with $N_{I}$ denoting the number of L-M iterations. Since both $m$
and $n$ scale linearly with $Q$, each term $m^{2}n$, $mn^{2}$,
and $n^{3}$ is of order $\mathcal{O}(N_{I}Q^{3})$. Therefore, the
overall complexity simplifies to ${{\mathcal{O}}_{\text{Postn}}}=\mathcal{O}({{N}_{I}}{{Q}^{3}})$.

In summary, the total complexity of the overall positioning algorithm
is given by the sum of complexity of coarse channel parameter estimation,
refined channel parameter estimation, and positioning parameter estimation,
which is $\mathcal{O}_{\text{chnl}}^{\text{c}}+\mathcal{O}{_{\text{SAGE}}}+{{\mathcal{O}}_{\text{Postn}}}$.

\section{Simulation Results}
\label{Simulation}

\subsection{Simulation Setup}

\begin{figure}
	\centering 
	\includegraphics[width=4in]{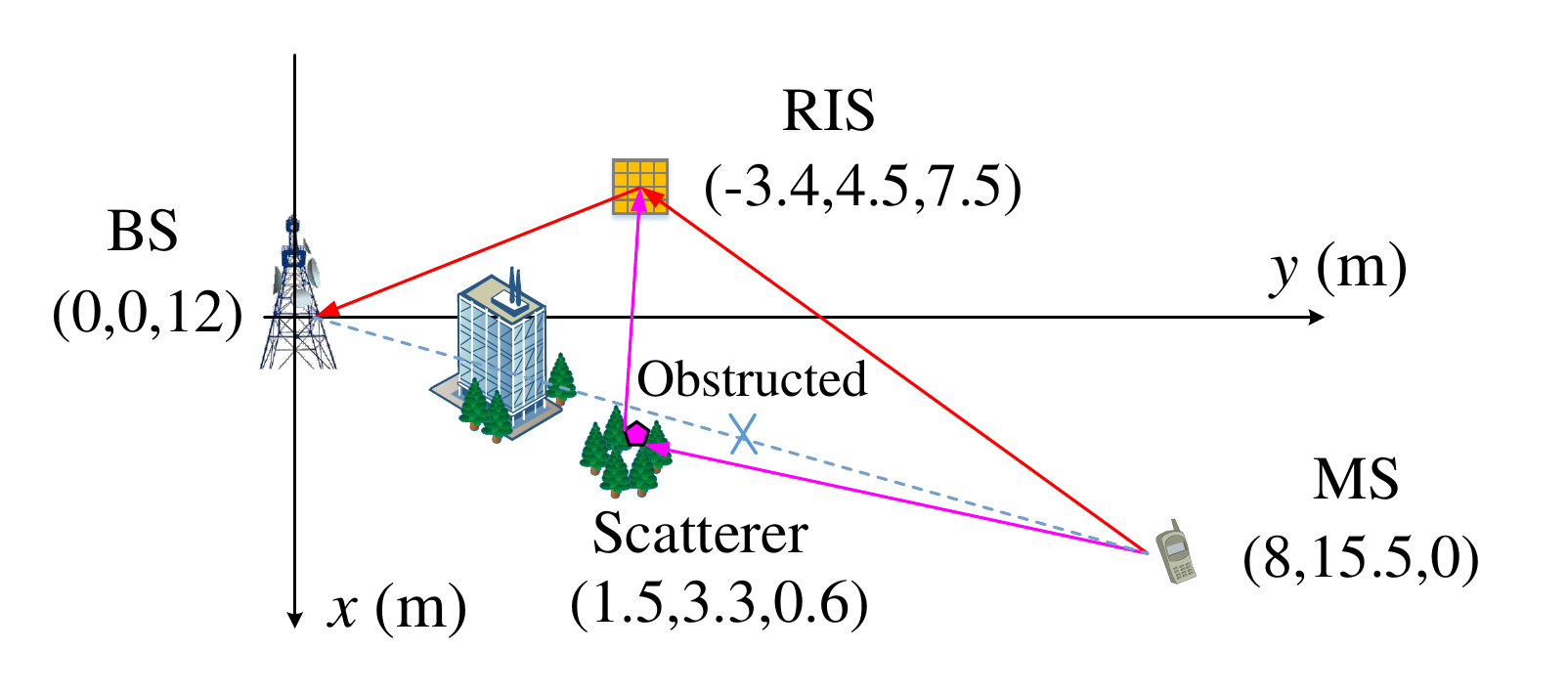}\caption{Relative position of different nodes.}
	\vspace{-0.3cm}
	\label{figRelaPosition}
\end{figure}

We set $f_{c}=4.9${\:}GHz, $B=20${\:}MHz, $c=3\times10^{8}${\:}m/s
and $N=20$. The noise power density is $-174$ dBm/Hz. The numbers
of transmit and receive antennas are $N_{\mathrm{m}}=16$
and $N_{\mathrm{b}}=40$, respectively, and the antenna spacing
is $d_{\textrm{{\rm {BS}}}}=d_{\textrm{{\rm {MS}}}}=\lambda/2$. The
number of reflecting elements of the RIS is $N_{\mathrm{r}}=N_{{\rm {a}}}N_{{\rm {e}}}=100$,
where $N_{{\rm {a}}}=N_{{\rm {e}}}=10$. The element
spacing at the RIS is $d_{\textrm{{\rm {RIS,a}}}}=d_{\textrm{{\rm {RIS,e}}}}=\lambda/2$.
The total number of time slots is $T=37$,
in which $T_{1}=16$, $T_{2}=21$, $\Upsilon=7$, and $V=3$. The maximum iteration number for the SAGE algorithm is $K=20$. The numbers of discretized grids are $G_{\mathrm{m}}=128$,
$G_{\mathrm{a}}=12$ and $G_{\mathrm{e}}=12$. The rotation angles at MS are $\alpha=132^{\circ}$ and $\beta=128^{\circ}$.
We assume that there is one scatterer \footnote{Here, we exclusively consider the scatterer with the strongest signal strength. Other weaker scatterers contribute little to localization while introducing redundant parameters, which increases computational load and degrades estimation performance. Addressing MS localization in dense and strong-scattering environments requires additional system resources and advanced techniques, an area we intend to explore in future research.} in the link from the MS to the RIS. The relative
positions of different nodes are shown in Fig. \ref{figRelaPosition}.

The complex channel gain $\delta_{q}$ is distributed as $\mathcal{CN}(0,10^{\frac{-{\rm {PL}}_{q}}{10}})$. In accordance with \cite{chen_multi-ris-enabled_2024}, the path loss in dB of the MS-RIS-BS link
	can be modeled as ${\rm {PL}}_{0}=-20\log10[{{\lambda}^{2}}/(16{{\pi}^{2}}{{d}_{\text{MR}}}{{d}_{\text{RB}}})]$,
	where $d_{\mathrm{MR}}$ and $d_{\mathrm{RB}}$ are the three-dimensional
	distances (in meters) of the MS-RIS link and RIS-BS link, respectively.
	The path loss in dB of the MS-(Scatterer-$q$)-RIS-BS link can be
	modeled as ${\rm {PL}}_{q}=-20\log10[(\sqrt{4\pi{{c}_{q}}}{{\lambda}^{2}})/(64{{\pi}^{3}}d_{ms}^{q}d_{sr}^{q}{{d}_{\text{RB}}})]$,
	where the $d_{ms}^{q}$ and $d_{sr}^{q}$ are the three-dimensional
	distances (in meters) of the MS-(Scatterer-$q$) link and (Scatterer-$q$)-RIS
	link, respectively. The ${c}_{q}$ is the radar cross section (RCS)
	coefficient of the $q$th scatterer, which depends on the scatterer
	type. The root mean square error (RMSE) of the estimation
algorithm is calculated over 1000 Monte Carlo simulation runs. The RMSEs are evaluated for transmit powers of [5, 10, 15, 20, 25, 30, 35, 40] dBm, which correspond to received SNRs of [-4.025, 0.975, 5.975, 10.975, 15.975, 20.975, 25.975, 30.975] dB, respectively.

\subsection{Comparison between the Coarse and Refined Estimation of the Channel
	Parameters}
\begin{figure*}
	\centering
	\subfigure[Real part of $\delta$]{ \includegraphics[width=0.3\columnwidth]{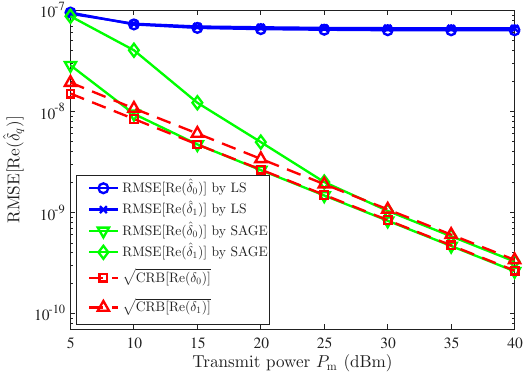}\label{RMSE_CRB_delta_Re}
	} \subfigure[Imaginary part of $\delta$]{ \includegraphics[width=0.3\columnwidth]{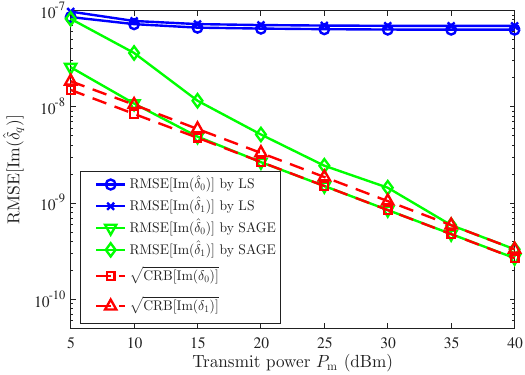}\label{RMSE_CRB_delta_Im}
	} \subfigure[TOA $\tau$]{
		\includegraphics[width=0.3\columnwidth]{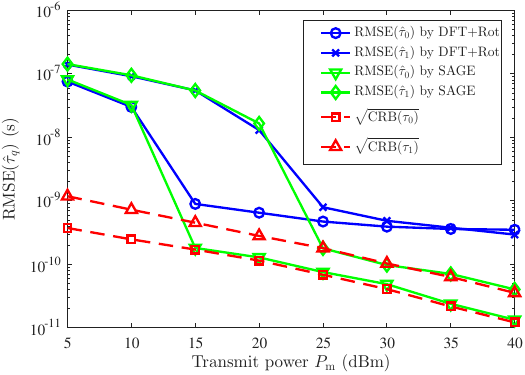}\label{RMSE_CRB_tau}}
	\subfigure[AOD $\theta_{\rm t}$]{
		\includegraphics[width=0.3\columnwidth]{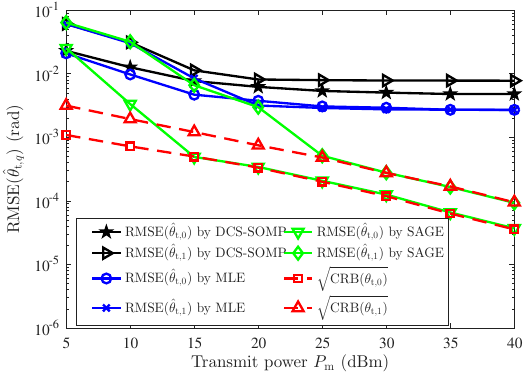}\label{RMSE_CRB_thetat}}
	\subfigure[Elevation AOA $\phi_{\rm in}$ at the RIS]{\label{RMSE_CRB_phi} \includegraphics[width=0.3\columnwidth]{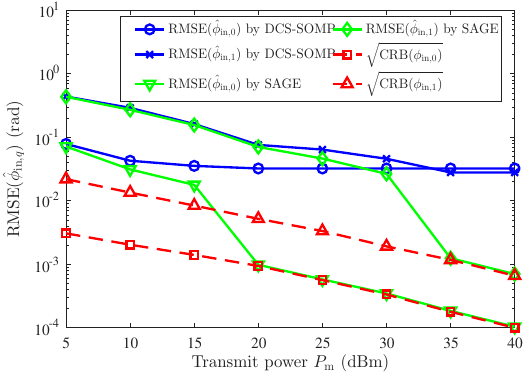}}
	\subfigure[Azimuth AOA $\psi_{\rm in}$ at the RIS]{\label{RMSE_CRB_psi} \includegraphics[width=0.3\columnwidth]{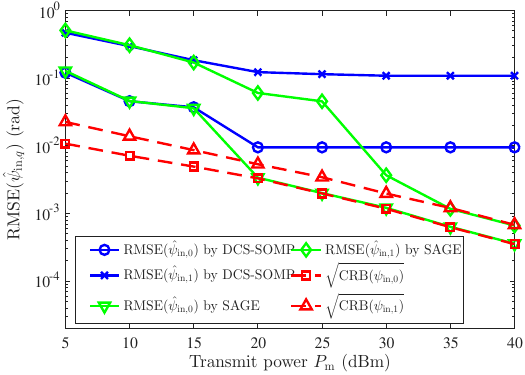}
	}
	\caption{The RMSE of channel parameters estimated by different algorithms versus
			the transmit power.}
	\vspace{-0.3cm}
	\label{RMSE_vs_CRB}
\end{figure*}
The channel parameters in $\boldsymbol{\eta}$ include the complex
channel gain, the AOD at the MS, the azimuth AOA at the RIS, the elevation
AOA at the RIS, and the TOA. There parameters are first coarsely estimated,
then refined by the SAGE algorithm. To evaluate the estimation performance
for channel parameters, the RMSEs of different estimation algorithms
versus transmit power are shown in Fig. \ref{RMSE_vs_CRB}, where
CRLB of different channel parameters are given as benchmarks. Fig.
\ref{RMSE_CRB_delta_Re} and Fig. \ref{RMSE_CRB_delta_Im} show the
estimation performance for the real part and imaginary part of the complex
channel gain, respectively. Fig. \ref{RMSE_CRB_tau} evaluates the
estimation performance for the TOA, and Fig. \ref{RMSE_CRB_thetat}
evaluates the estimation performance for the AOD. The estimation performance
of the azimuth AOA and elevation AOD at the RIS is respectively shown in
Fig. \ref{RMSE_CRB_phi} and \ref{RMSE_CRB_psi}.

From these figures, we can have the following observations and conclusions.
Firstly, it is observed that in all sub-figures of Fig. \ref{RMSE_vs_CRB},
the RMSE of the channel parameters for both the coarse estimation
and refining estimation decreases with the transmit power, which means
that the estimation accuracy can be improved
with transmit power. Secondly, in Fig. \ref{RMSE_CRB_thetat},
by comparing the black line marked by $\star$ (or $\rhd$) with the blue line marked by $\circ$ (or $\times$), we
find that the RMSEs of the refined AODs by MLE are lower than those
of the coarsely-estimated AODs by the DCS-SOMP algorithm, and the performance
gap becomes significant at high transmit power. This verifies
the effectiveness of refining AODs by MLE, which mitigates the error
propagation into AOA/TOA estimation, and generate better initial
values for the SAGE algorithm. Thirdly, by comparing the blue line marked by $\circ$ (or $\times$) with the green line marked by $\triangledown$ (or $\lozenge$) in all the subfigures of Fig. \ref{RMSE_vs_CRB},
it can be seen that the RMSEs of channel parameters by coarse estimation
are greatly reduced after the refining estimation by the SAGE algorithm,
and the performance gap increases with the transmit power. This verifies
the effectiveness of the refining estimation by the SAGE algorithm. Fourthly, by comparing the green line marked by $\triangledown$ (or $\lozenge$) with the red line marked by $\square$ (or $\triangle$) of all subfigures in Fig. \ref{RMSE_vs_CRB}, it is
observed that after the refining estimation concatenated by the coarse
estimation, the RMSEs of all channel parameters by the SAGE algorithm
gradually approach their corresponding CRLBs as the transmit power
increases, which validates the good estimation performance of our
proposed estimation algorithms.
\begin{figure*}[h]
	\centering
	\begin{minipage}{0.45\columnwidth}
		\centering
		\includegraphics[width=\linewidth]{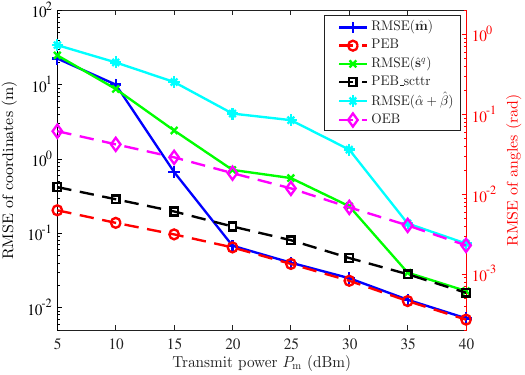}
		\caption{The RMSE of coordinates and angles versus transmit power with a 2D rotational MS.}
		\label{RMSE_vs_PEB_OEB_2Drott}
	\end{minipage}
	\hfill
	\begin{minipage}{0.45\columnwidth}
		\centering
		\includegraphics[width=\linewidth]{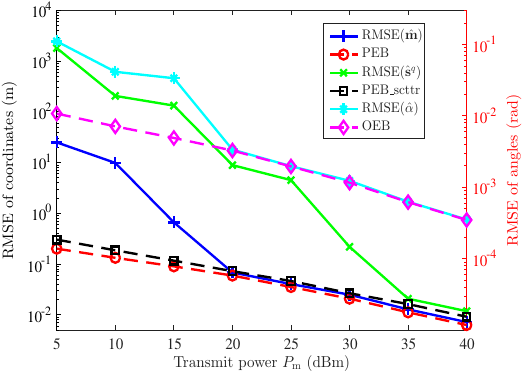}
		\caption{The RMSE of coordinates and angles versus transmit power with a 1D rotational MS.}
		\label{RMSE_vs_PEB_OEB_1Drott}
	\end{minipage}
	\vspace{-0.3cm}
\end{figure*}
\subsection{Positioning Performance Evaluation}

The coordinates and rotation angles of the MS can be further estimated from the estimated channel parameters. Fig. \ref{RMSE_vs_PEB_OEB_2Drott} presents the RMSEs of the MS coordinates and rotation angles versus the transmit power for scenarios with 2D rotation of the MS. It is observed that as the transmit power increases, the RMSE($\hat{\mathbf{m}}$), RMSE($\hat{\mathbf{s}}^q$), and RMSE($\hat{\alpha}+\hat{\beta}$) gradually decrease and eventually approach the PEB and OEB, respectively, which verifies the superiority of the proposed positioning algorithm. The trend of the RMSE curves for the positioning parameters is consistent with that for the channel parameters. Additionally, the RMSE of the scatterer is higher than that of the MS, indicating that scatterer mapping suffers from larger error compared to MS localization. The results in Fig. \ref{RMSE_vs_PEB_OEB_2Drott} also reveal that the proposed algorithm not only can achieve the positioning of the MS, but also can achieve the mapping of the surrounding environment. 
	
	For scenarios with 1D rotation of the MS, Fig. \ref{RMSE_vs_PEB_OEB_1Drott} shows the RMSEs of the MS coordinates and rotation angles versus the transmit power. It is observed that the estimation performance of the proposed algorithm for the MS, scatterer, and rotation angle is similar to that in Fig. \ref{RMSE_vs_PEB_OEB_2Drott}, leading to analogous conclusions. Moreover, compared with the 2D rotation case in Fig. \ref{RMSE_vs_PEB_OEB_2Drott}, the RMSE of the rotation angle estimation in Fig. \ref{RMSE_vs_PEB_OEB_1Drott} converges to the OEB more rapidly, demonstrating better estimation performance.

\section{Summary and Conclusion}
\label{Conclusion}
We proposed a two-step position and orientation estimation algorithm for an MS in an RIS-aided MIMO-OFDM system over dispersive channels. It harnesses NLoS links for 3D positioning by mapping scatterers as valid information rather than noise. To tackle the high-dimensional estimation challenge, this work employs parameter decoupling and a combination of coarse and refined estimation. Accurate estimation is achieved using compressed sensing, SAGE, geometric equation solving, LS, and other methods. Simulations validate the algorithm's effectiveness by comparing with the CRLB/PEB/OEB.
\vspace{-0.3cm}

\begin{appendices}
	
	\section{Simplification of the ML Function for Refining AODs}
	
	\label{Appendix_A}
	
	Since we assume that the pilot signal is the same over all subcarriers,
	$\ensuremath{{\tilde{{\bf X}}}_{t}[n]}$ can be written as $\ensuremath{{\tilde{{\bf X}}}_{t}=[{\tilde{{\bf H}}}_{0}{\bf x}_{t},{\tilde{{\bf H}}}_{1}{\bf x}_{t},\cdots,{\tilde{{\bf H}}}_{Q}{\bf x}_{t}]}$.
	Then \eqref{AODmle_fine} could be rewritten as
	\begin{align}
		\ensuremath{\hat{\boldsymbol{\theta}}_{{\rm t}}} & =\arg\mathop{\min}\limits _{{\boldsymbol{\theta}}_{{\rm t}}}\sum\limits _{n=1}^{N}\sum\limits _{t=1}^{T_{1}}\left\Vert {\bf y}_{t}[n]-{\tilde{{\bf X}}}_{t}[\sum\limits _{t=1}^{T_{1}}({\tilde{{\bf X}}}_{t}^{H}{\tilde{{\bf X}}}_{t})]^{-1}\sum\limits _{t=1}^{T_{1}}({\tilde{{\bf X}}}_{t}^{H}{\bf y}_{t}[n])\right\Vert _{2}^{2}.\label{AODmle_fine_simply}
	\end{align}
	The term $\sum\limits _{t=1}^{T_{1}}({\tilde{{\bf X}}}_{t}^{H}{\bf y}_{t}[n])$
	in \eqref{AODmle_fine_simply} can be simplified as follows:
	\begin{align}
		\sum\limits _{t=1}^{T_{1}}({\tilde{{\bf X}}}_{t}^{H}{\bf y}_{t}[n]) & =\sum\limits _{t=1}^{T_{1}}\left[\begin{array}{c}
			{\bf x}_{t}^{H}{\tilde{{\bf H}}}_{0}^{H}{\bf y}_{t}[n]\\
			{\bf x}_{t}^{H}{\tilde{{\bf H}}}_{1}^{H}{\bf y}_{t}[n]\\
			\vdots\\
			{\bf x}_{t}^{H}{\tilde{{\bf H}}}_{Q}^{H}{\bf y}_{t}[n]
		\end{array}\right]\nonumber \\
		& =\sum\limits _{t=1}^{T_{1}}[{\bf y}_{t}^{H}[n]{\tilde{{\bf H}}}_{0}{\bf x}_{t},{\bf y}_{t}^{H}[n]{\tilde{{\bf H}}}_{1}{\bf x}_{t},\cdots,{\bf y}_{t}^{H}[n]{\tilde{{\bf H}}}_{Q}{\bf x}_{t}]^{H}\nonumber \\
		& =\sum\limits _{t=1}^{T_{1}}[{\bf y}_{t}^{H}[n]{\bf a}_{{\rm B,0}}{\bf a}_{{\rm M,0}}^{H}{\bf x}_{t},{\bf y}_{t}^{H}[n]{\bf a}_{{\rm B,0}}{\bf a}_{{\rm M,1}}^{H}{\bf x}_{t},\cdots,{\bf y}_{t}^{H}[n]{\bf a}_{{\rm B,0}}{\bf a}_{{\rm M,}Q}^{H}{\bf x}_{t}]^{H}\nonumber \\
		& =[{\bf a}_{{\rm M,0}}^{H}(\sum\limits _{t=1}^{T_{1}}{\bf x}_{t}{\bf y}_{t}^{H}[n]){\bf a}_{{\rm B,0}},{\bf a}_{{\rm M,1}}^{H}(\sum\limits _{t=1}^{T_{1}}{\bf x}_{t}{\bf y}_{t}^{H}[n]){\bf a}_{{\rm B,0}},\cdots,{\bf a}_{{\rm M,}Q}^{H}(\sum\limits _{t=1}^{T_{1}}{\bf x}_{t}{\bf y}_{t}^{H}[n]){\bf a}_{{\rm B,0}}]^{H}\nonumber \\
		& =\left[\begin{array}{c}
			{\bf a}_{{\rm B,0}}^{H}(\sum\limits _{t=1}^{T_{1}}{\bf y}_{t}[n]{\bf x}_{t}^{H}){\bf a}_{{\rm M,0}}\\
			{\bf a}_{{\rm B,0}}^{H}(\sum\limits _{t=1}^{T_{1}}{\bf y}_{t}[n]{\bf x}_{t}^{H}){\bf a}_{{\rm M,1}}\\
			\vdots\\
			{\bf a}_{{\rm B,0}}^{H}(\sum\limits _{t=1}^{T_{1}}{\bf y}_{t}[n]{\bf x}_{t}^{H}){\bf a}_{{\rm M,}Q}
		\end{array}\right]\nonumber \\
		& =({\bf a}_{{\rm B,0}}^{H}(\sum\limits _{t=1}^{T_{1}}{\bf y}_{t}[n]{\bf x}_{t}^{H}){\bf A}_{{\rm M,}Q})^{T}\nonumber \\
		& ={\bf A}_{{\rm M,}Q}^{T}(\sum\limits _{t=1}^{T_{1}}{\bf x}_{t}^{*}{\bf y}_{t}^{T}[n]){\bf a}_{{\rm B,0}}^{*},\label{AODmle_fine_simply_part1}
	\end{align}
	where $\ensuremath{{\tilde{{\bf H}}}_{q}{\!}\triangleq{\!}{\bf a}_{{\rm B}}(\vartheta_{{\rm r},0}){\bf a}_{{\rm M}}^{H}(\vartheta_{{\rm t},q})}$,
	$\ensuremath{{\bf a}_{{\rm B,0}}{\!}\triangleq{\!}{\bf a}_{{\rm B}}(\vartheta_{{\rm r},0})}$,
	$\ensuremath{{\bf a}_{{\rm M},q}{\!}\triangleq{\!}{\bf a}_{{\rm M}}(\vartheta_{{\rm t},q})}$
	and $\ensuremath{{\bf A}_{{\rm M,}Q}{\!}\triangleq{\!}[\ensuremath{{\bf a}_{{\rm M,0}},{\bf a}_{{\rm M,1}},\cdots,{\bf a}_{{\rm M,}Q}}]}$.
	
	The term ${\tilde{{\bf X}}}_{t}[\sum\limits _{t=1}^{T_{1}}({\tilde{{\bf X}}}_{t}^{H}{\tilde{{\bf X}}}_{t})]^{-1}$
	in \eqref{AODmle_fine_simply} can be simplified as
	\begin{align}
		\ensuremath{{\tilde{{\bf X}}}_{t}^{H}{\tilde{{\bf X}}}_{t}} & \ensuremath{=\left[\begin{array}{c}
				{\bf x}_{t}^{H}{\bf a}_{{\rm M,0}}\\
				{\bf x}_{t}^{H}{\bf a}_{{\rm M,1}}\\
				\vdots\\
				{\bf x}_{t}^{H}{\bf a}_{{\rm M,}Q}
			\end{array}\right]{\bf a}_{{\rm B,0}}^{H}{\bf a}_{{\rm B,0}}[{\bf a}_{{\rm M,0}}^{H}{\bf x}_{t},{\bf a}_{{\rm M,1}}^{H}{\bf x}_{t},\cdots,{\bf a}_{{\rm M,}Q}^{H}{\bf x}_{t}]}\nonumber \\
		& \ensuremath{{\rm =}N_{\mathrm{b}}\left[\begin{array}{cccccc}
				{\bf x}_{t}^{H}{\bf a}_{{\rm M,0}}{\bf a}_{{\rm M,0}}^{H}{\bf x}_{t} & {\bf x}_{t}^{H}{\bf a}_{{\rm M,0}}{\bf a}_{{\rm M,1}}^{H}{\bf x}_{t} & \cdots & {\bf x}_{t}^{H}{\bf a}_{{\rm M,0}}{\bf a}_{{\rm M,}j}^{H}{\bf x}_{t} & \cdots & {\bf x}_{t}^{H}{\bf a}_{{\rm M,0}}{\bf a}_{{\rm M,}Q}^{H}{\bf x}_{t}\\
				{\bf x}_{t}^{H}{\bf a}_{{\rm M,1}}{\bf a}_{{\rm M,0}}^{H}{\bf x}_{t} & {\bf x}_{t}^{H}{\bf a}_{{\rm M,1}}{\bf a}_{{\rm M,1}}^{H}{\bf x}_{t} & \cdots & {\bf x}_{t}^{H}{\bf a}_{{\rm M,1}}{\bf a}_{{\rm M,}j}^{H}{\bf x}_{t} & \cdots & {\bf x}_{t}^{H}{\bf a}_{{\rm M,1}}{\bf a}_{{\rm M,}Q}^{H}{\bf x}_{t}\\
				\vdots & \vdots & \ddots & \vdots & \ddots & \vdots\\
				{\bf x}_{t}^{H}{\bf a}_{{\rm M,}i}{\bf a}_{{\rm M,0}}^{H}{\bf x}_{t} & {\bf x}_{t}^{H}{\bf a}_{{\rm M,}i}{\bf a}_{{\rm M,1}}^{H}{\bf x}_{t} & \cdots & {\bf x}_{t}^{H}{\bf a}_{{\rm M,}i}{\bf a}_{{\rm M,}j}^{H}{\bf x}_{t} & \cdots & {\bf x}_{t}^{H}{\bf a}_{{\rm M,}i}{\bf a}_{{\rm M,}Q}^{H}{\bf x}_{t}\\
				\vdots & \vdots & \ddots & \vdots & \ddots & \vdots\\
				{\bf x}_{t}^{H}{\bf a}_{{\rm M,}Q}{\bf a}_{{\rm M,0}}^{H}{\bf x}_{t} & {\bf x}_{t}^{H}{\bf a}_{{\rm M,}Q}{\bf a}_{{\rm M,1}}^{H}{\bf x}_{t} & \cdots & {\bf x}_{t}^{H}{\bf a}_{{\rm M,}Q}{\bf a}_{{\rm M,}j}^{H}{\bf x}_{t} & \cdots & {\bf x}_{t}^{H}{\bf a}_{{\rm M,}Q}{\bf a}_{{\rm M,}Q}^{H}{\bf x}_{t}
			\end{array}\right]}\nonumber \\
		& \ensuremath{\mathop{{\rm =}}\limits ^{(g)}N_{\mathrm{b}}\left[\begin{array}{cccccc}
				{\bf a}_{{\rm M,0}}^{H}{\bf x}_{t}{\bf x}_{t}^{H}{\bf a}_{{\rm M,0}} & {\bf a}_{{\rm M,1}}^{H}{\bf x}_{t}{\bf x}_{t}^{H}{\bf a}_{{\rm M,0}} & \cdots & {\bf a}_{{\rm M,}j}^{H}{\bf x}_{t}{\bf x}_{t}^{H}{\bf a}_{{\rm M,0}} & \cdots & {\bf a}_{{\rm M,}Q}^{H}{\bf x}_{t}{\bf x}_{t}^{H}{\bf a}_{{\rm M,0}}\\
				{\bf a}_{{\rm M,0}}^{H}{\bf x}_{t}{\bf x}_{t}^{H}{\bf a}_{{\rm M,1}} & {\bf a}_{{\rm M,1}}^{H}{\bf x}_{t}{\bf x}_{t}^{H}{\bf a}_{{\rm M,1}} & \cdots & {\bf a}_{{\rm M,}j}^{H}{\bf x}_{t}{\bf x}_{t}^{H}{\bf a}_{{\rm M,1}} & \cdots & {\bf a}_{{\rm M,}Q}^{H}{\bf x}_{t}{\bf x}_{t}^{H}{\bf a}_{{\rm M,1}}\\
				\vdots & \vdots & \ddots & \vdots & \ddots & \vdots\\
				{\bf a}_{{\rm M,0}}^{H}{\bf x}_{t}{\bf x}_{t}^{H}{\bf a}_{{\rm M,}i} & {\bf a}_{{\rm M,1}}^{H}{\bf x}_{t}{\bf x}_{t}^{H}{\bf a}_{{\rm M,}i} & \cdots & {\bf a}_{{\rm M,}j}^{H}{\bf x}_{t}{\bf x}_{t}^{H}{\bf a}_{{\rm M,}i} & \cdots & {\bf a}_{{\rm M,}Q}^{H}{\bf x}_{t}{\bf x}_{t}^{H}{\bf a}_{{\rm M,}i}\\
				\vdots & \vdots & \ddots & \vdots & \ddots & \vdots\\
				{\bf a}_{{\rm M,0}}^{H}{\bf x}_{t}{\bf x}_{t}^{H}{\bf a}_{{\rm M,}Q} & {\bf a}_{{\rm M,1}}^{H}{\bf x}_{t}{\bf x}_{t}^{H}{\bf a}_{{\rm M,}Q} & \cdots & {\bf a}_{{\rm M,}j}^{H}{\bf x}_{t}{\bf x}_{t}^{H}{\bf a}_{{\rm M,}Q} & \cdots & {\bf a}_{{\rm M,}Q}^{H}{\bf x}_{t}{\bf x}_{t}^{H}{\bf a}_{{\rm M,}Q}
			\end{array}\right]}\nonumber \\
		& =\ensuremath{N_{\mathrm{b}}({\bf A}_{{\rm M,}Q}^{H}{\bf x}_{t}{\bf x}_{t}^{H}{\bf A}_{{\rm M,}Q}})^{T}\nonumber \\
		& =\ensuremath{N_{\mathrm{b}}{\bf A}_{{\rm M,}Q}^{T}{\bf x}_{t}^{*}{\bf x}_{t}^{T}{\bf A}_{{\rm M,}Q}^{*}},
	\end{align}
	where the step $(g)$ is obtained by ${\bf {x}}_{t}^{H}{\bf {a}}_{{\rm {M,}}i}^ {}{\bf {a}}_{{\rm {M,}}j}^{H}{{\bf {x}}_{t}}={\rm {Tr}}({\bf {a}}_{{\rm {M,}}j}^{H}{{\bf {x}}_{t}}{\bf {x}}_{t}^{H}{\bf {a}}_{{\rm {M,}}i}^ {})={\bf {a}}_{{\rm {M,}}j}^{H}{{\bf {x}}_{t}}{\bf {x}}_{t}^{H}{\bf {a}}_{{\rm {M,}}i}^ {}$.
	Then ${\tilde{{\bf X}}}_{t}[\sum\limits _{t=1}^{T_{1}}({\tilde{{\bf X}}}_{t}^{H}{\tilde{{\bf X}}}_{t})]^{-1}$
	could be rewritten as
	\begin{align}
		\ensuremath{{\tilde{{\bf X}}}_{t}[\sum\limits _{t=1}^{T_{1}}({\tilde{{\bf X}}}_{t}^{H}{\tilde{{\bf X}}}_{t})]^{-1}} & \ensuremath{=\frac{1}{N_{\mathrm{b}}}{\bf a}_{{\rm B,0}}[{\bf a}_{{\rm M,0}}^{H}{\bf x}_{t},{\bf a}_{{\rm M,1}}^{H}{\bf x}_{t},\cdots,{\bf a}_{{\rm M,}Q}^{H}{\bf x}_{t}][{\bf A}_{{\rm M,}Q}^{T}\sum\limits _{t=1}^{T_{1}}({\bf x}_{t}^{*}{\bf x}_{t}^{T}){\bf A}_{{\rm M,}Q}^{*}]^{-1}}\nonumber \\
		& =\frac{1}{N_{\mathrm{b}}}{\bf a}_{{\rm B,0}}{\bf x}_{t}^{T}{\bf A}_{{\rm M,}Q}^{*}[{\bf A}_{{\rm M,}Q}^{T}\sum\limits _{t=1}^{T_{1}}({\bf x}_{t}^{*}{\bf x}_{t}^{T}){\bf A}_{{\rm M,}Q}^{*}]^{-1},\label{AODmle_fine_simply_part2}
	\end{align}
	where $\ensuremath{[{\bf a}_{{\rm M,1}}^{H}{\bf x}_{t},{\bf a}_{{\rm M,2}}^{H}{\bf x}_{t},\cdots,{\bf a}_{{\rm M,}Q}^{H}{\bf x}_{t}]=({\bf A}_{{\rm M,}Q}^{H}{\bf x}_{t})^{T}={\bf x}_{t}^{T}{\bf A}_{{\rm {M,}}Q}^{*}}$.
	
	By using \eqref{AODmle_fine_simply_part1} and \eqref{AODmle_fine_simply_part2},
	the term $\ensuremath{{\tilde{{\bf X}}}_{t}[\sum\limits _{t=1}^{T_{1}}({\tilde{{\bf X}}}_{t}^{H}{\tilde{{\bf X}}}_{t})]^{-1}\sum\limits _{t=1}^{T_{1}}({\tilde{{\bf X}}}_{t}^{H}{\bf y}_{t}[n])}$
	in \eqref{AODmle_fine_simply} could be simplified as
	\begin{align}
		& \ensuremath{{\tilde{{\bf X}}}_{t}[\sum\limits _{t=1}^{T_{1}}({\tilde{{\bf X}}}_{t}^{H}{\tilde{{\bf X}}}_{t})]^{-1}\sum\limits _{t=1}^{T_{1}}({\tilde{{\bf X}}}_{t}^{H}{\bf y}_{t}[n])}\nonumber \\
		= & \ensuremath{\frac{1}{N_{\mathrm{b}}}{\bf a}_{{\rm B,0}}{\bf x}_{t}^{T}{\bf A}_{{\rm M,}Q}^{*}[{\bf A}_{{\rm M,}Q}^{T}\sum\limits _{t=1}^{T_{1}}({\bf x}_{t}^{*}{\bf x}_{t}^{T}){\bf A}_{{\rm M,}Q}^{*}]^{-1}[{\bf A}_{{\rm M,}Q}^{T}\sum\limits _{t=1}^{T_{1}}({\bf x}_{t}^{*}{\bf y}_{t}^{T}[n]){\bf a}_{{\rm B,0}}^{*}]}\nonumber \\
		= & \ensuremath{\frac{1}{N_{\mathrm{b}}}{\bf a}_{{\rm B,0}}{\bf a}_{{\rm B,0}}^{H}(\sum\limits _{t=1}^{T_{1}}{\bf y}_{t}[n]{\bf x}_{t}^{H}){\bf A}_{{\rm M,}Q}[{\bf A}_{{\rm M,}Q}^{H}\sum\limits _{t=1}^{T_{1}}({\bf x}_{t}{\bf x}_{t}^{H}){\bf A}_{{\rm M,}Q}]^{-1}{\bf A}_{{\rm M,}Q}^{H}{\bf x}_{t}}\nonumber \\
		= & \ensuremath{\frac{1}{N_{\mathrm{b}}}{\bf a}_{{\rm B,0}}{\bf a}_{{\rm B,0}}^{H}{\bf B}[n]{\bf A}_{{\rm M,}Q}({\bf A}_{{\rm M,}Q}^{H}\ensuremath{{\bf C}}{\bf A}_{{\rm M,}Q})^{-1}{\bf A}_{{\rm M,}Q}^{H}{\bf x}_{t}}\nonumber \\
		= & \ensuremath{\frac{1}{N_{\mathrm{b}}}{\bf a}_{{\rm B,0}}{\bf a}_{{\rm B,0}}^{H}{\bf B}[n]\mathbf{D}(\boldsymbol{\theta}_{\mathrm{t}}){\bf x}_{t}},\label{AODmle_fine_simply_part3}
	\end{align}
	where \begin{subequations}
		\begin{align}
			{\bf B}[n] & \triangleq\sum\limits _{t=1}^{T_{1}}{\bf y}_{t}[n]{\bf x}_{t}^{H}=[{\bf y}_{1}[n],{\bf y}_{2}[n],\cdots,{\bf y}_{T_{1}}[n]]\left[\begin{array}{c}
				{\bf x}_{1}^{H}\\
				{\bf x}_{2}^{H}\\
				\vdots\\
				{\bf x}_{T_{1}}^{H}
			\end{array}\right]={\bf Y}_{1}[n]{\bf X}_{1}^{H},\\
			{\bf C} & \triangleq\sum\limits _{t=1}^{T_{1}}{\bf x}_{t}{\bf x}_{t}^{H}={\bf X}_{1}{\bf X}_{1}^{H},\\
			{\bf D}(\boldsymbol{\theta}_{{\rm t}}) & \triangleq\ensuremath{{\bf A}_{{\rm M,}Q}({\bf A}_{{\rm M,}Q}^{H}{\bf CA}_{{\rm M,}Q})^{-1}{\bf A}_{{\rm M,}Q}^{H}}.
		\end{align}
	\end{subequations}
	
	By substituting \eqref{AODmle_fine_simply_part3} into \eqref{AODmle_fine_simply},
	and defining $\mathbf{E}\triangleq\frac{1}{N_{\mathrm{b}}}{\bf a}_{{\rm B,0}}{\bf a}_{{\rm B,0}}^{H}$,
	we have
	\begin{align}
		& \ensuremath{\sum\limits _{n=1}^{N}\sum\limits _{t=1}^{T_{1}}\left\Vert {\bf y}_{t}[n]-{\tilde{{\bf X}}}_{t}[\sum\limits _{t=1}^{T_{1}}({\tilde{{\bf X}}}_{t}^{H}{\tilde{{\bf X}}}_{t})]^{-1}\sum\limits _{t=1}^{T_{1}}({\tilde{{\bf X}}}_{t}^{H}{\bf y}_{t}[n])\right\Vert _{2}^{2}}\nonumber \\
		= & \ensuremath{\sum\limits _{n=1}^{N}\sum\limits _{t=1}^{T_{1}}\left\Vert {\bf y}_{t}[n]-\frac{1}{N_{\mathrm{b}}}{\bf a}_{{\rm B,0}}{\bf a}_{{\rm B,0}}^{H}{\bf B}[n]{\bf D}(\boldsymbol{\theta}_{{\rm t}}){\bf x}_{t}\right\Vert _{2}^{2}}\nonumber \\
		= & \ensuremath{\sum\limits _{n=1}^{N}\sum\limits _{t=1}^{T_{1}}\left\Vert {\bf y}_{t}[n]-{\bf EB}[n]{\bf D}(\boldsymbol{\theta}_{{\rm t}}){\bf x}_{t}\right\Vert _{2}^{2}}\nonumber \\
		= & \ensuremath{\sum\limits _{n=1}^{N}\sum\limits _{t=1}^{T_{1}}({\bf y}_{t}[n]-{\bf EB}[n]{\bf D}(\boldsymbol{\theta}_{{\rm t}}){\bf x}_{t})^{H}({\bf y}_{t}[n]-{\bf EB}[n]{\bf D}(\boldsymbol{\theta}_{{\rm t}}){\bf x}_{t})}\nonumber \\
		= & \ensuremath{\sum\limits _{n=1}^{N}\sum\limits _{t=1}^{T_{1}}\{{\bf y}_{t}^{H}[n]{\bf y}_{t}[n]-2\mathfrak{Re}({\bf y}_{t}^{H}[n]{\bf EB}[n]{\bf D}(\boldsymbol{\theta}_{{\rm t}}){\bf x}_{t})+{\bf x}_{t}^{H}{\bf D}^{H}(\boldsymbol{\theta}_{{\rm t}})({\bf EB}[n])^{H}{\bf EB}[n]{\bf D}(\boldsymbol{\theta}_{{\rm t}}){\bf x}_{t}\}}\nonumber \\
		\propto & \ensuremath{\sum\limits _{n=1}^{N}\sum\limits _{t=1}^{T_{1}}\{-2\mathfrak{Re}({\bf y}_{t}^{H}[n]{\bf EB}[n]{\bf D}(\boldsymbol{\theta}_{{\rm t}}){\bf x}_{t})+{\bf x}_{t}^{H}{\bf D}^{H}(\boldsymbol{\theta}_{{\rm t}})({\bf EB}[n])^{H}{\bf EB}[n]{\bf D}(\boldsymbol{\theta}_{{\rm t}}){\bf x}_{t}\}}\nonumber \\
		= & \ensuremath{-2\mathfrak{Re}\{\sum\limits _{n=1}^{N}\sum\limits _{t=1}^{T_{1}}{\rm Tr}({\bf D}(\boldsymbol{\theta}_{{\rm t}}){\bf x}_{t}{\bf y}_{t}^{H}[n]{\bf EB}[n])\}}+\sum\limits _{n=1}^{N}\sum\limits _{t=1}^{T_{1}}\ensuremath{{\rm Tr}[({\bf EB}[n])^{H}{\bf EB}[n]{\bf D}(\boldsymbol{\theta}_{{\rm t}}){\bf x}_{t}{\bf x}_{t}^{H}{\bf D}^{H}(\boldsymbol{\theta}_{{\rm t}})]}\nonumber \\
		= & \ensuremath{{\!}-{\!}2\mathfrak{Re}\{{\rm Tr}({\bf D}(\boldsymbol{\theta}_{{\rm t}})\sum\limits _{n=1}^{N}\sum\limits _{t=1}^{T_{1}}{\bf x}_{t}{\bf y}_{t}^{H}[n]{\bf EB}[n])\}}{\!}+{\!}{\rm Tr}\{[\sum\limits _{n=1}^{N}({\bf EB}[n])^{H}{\bf EB}[n]]{\bf D}(\boldsymbol{\theta}_{{\rm t}})[\sum\limits _{t=1}^{T_{1}}{\bf x}_{t}{\bf x}_{t}^{H}]{\bf D}^{H}(\boldsymbol{\theta}_{{\rm t}})\}\nonumber \\
		\mathop{{\rm =}}\limits ^{(h)} & \ensuremath{-2\mathfrak{Re}\{{\rm Tr}({\bf D}(\boldsymbol{\theta}_{{\rm t}})\sum\limits _{n=1}^{N}({\bf B}[n])^{H}{\bf EB}[n])\}}+{\rm Tr}\{(\sum\limits _{n=1}^{N}({\bf B}[n])^{H}{\bf EB}[n]){\bf D}(\boldsymbol{\theta}_{{\rm t}})\mathbf{C}{\bf D}^{H}(\boldsymbol{\theta}_{{\rm t}})\},\label{AODmle_fine_simply_final}
	\end{align}
	where $\mathbf{E}^{H}\mathbf{E}=\mathbf{E}$ is utilized in step $(h)$.
	
	\section{Simplifications and Derivations in the SAGE Algorithm}
	
	\subsection{Simplification of the ML Function for All Channel Parameters in \eqref{stage2_Allparameters_MLE}}
	
	\label{Appendices B.A}
	
	By taking the natural logarithm of \eqref{joint_probability_density_allTimeSlots},
	and omitting the constant term, we get
	\begin{align}
		& {\rm ln}f({\bf y}[1],{\bf y}[2],\cdots,{\bf y}[N]|\boldsymbol{{\bf \eta}})\nonumber \\
		\propto & -\sum\limits _{n=1}^{N}({\bf y}[n]-\sum\limits _{q=0}^{Q}\delta_{q}e^{-j2\pi\tau_{q}\frac{(n-1)B}{N}}(\bm{{\Sigma}}_{q}^{T}\otimes{\tilde{{\bf H}}}_{q}){\bf x}[n])^{H}({\bf y}[n]-\sum\limits _{q=0}^{Q}\delta_{q}e^{-j2\pi\tau_{q}\frac{(n-1)B}{N}}(\bm{{\Sigma}}_{q}^{T}\otimes{\tilde{{\bf H}}}_{q}){\bf x}[n])\nonumber \\
		= & -\sum\limits _{n=1}^{N}\left\Vert {\bf y}[n]-\sum\limits _{q=0}^{Q}\delta_{q}e^{-j2\pi\tau_{q}\frac{(n-1)B}{N}}(\bm{{\Sigma}}_{q}^{T}\otimes{\tilde{{\bf H}}}_{q}){\bf x}[n]\right\Vert _{2}^{2}\nonumber \\
		= & -\sum\limits _{n=1}^{N}\left\Vert {\bf Y}[n]-\sum\limits _{q=0}^{Q}\delta_{q}e^{-j2\pi\tau_{q}\frac{(n-1)B}{N}}\ensuremath{{\tilde{{\bf H}}}_{q}}{\bf X}[n]\bm{{\Sigma}}_{q}\right\Vert _{F}^{2}\nonumber \\
		= & -\sum\limits _{n=1}^{N}{\rm tr}\left[({\bf Y}[n]-\sum\limits _{q=0}^{Q}\delta_{q}e^{-j2\pi\tau_{q}\frac{(n-1)B}{N}}\ensuremath{{\tilde{{\bf H}}}_{q}}{\bf X}[n]\bm{{\Sigma}}_{q})^{H}({\bf Y}[n]-\sum\limits _{q=0}^{Q}\delta_{q}e^{-j2\pi\tau_{q}\frac{(n-1)B}{N}}\ensuremath{{\tilde{{\bf H}}}_{q}}{\bf X}[n]\bm{{\Sigma}}_{q})\right]\nonumber \\
		= & -\sum\limits _{n=1}^{N}{\rm tr}({\bf Y}^{H}[n]{\bf Y}[n])+\sum\limits _{n=1}^{N}{\rm tr}\left[\sum\limits _{q=0}^{Q}\delta_{q}e^{-j2\pi\tau_{q}\frac{(n-1)B}{N}}{\bf Y}^{H}[n]{\tilde{{\bf H}}}_{q}{\bf X}[n]\bm{{\Sigma}}_{q}\right]\nonumber \\
		& +\sum\limits _{n=1}^{N}{\rm tr}\left[\sum\limits _{q=0}^{Q}\delta_{q}^{*}e^{j2\pi\tau_{q}\frac{(n-1)B}{N}}\bm{{\Sigma}}_{q}^{H}{\bf X}^{H}[n]{\tilde{{\bf H}}}_{q}^{H}{\bf Y}[n]\right]\nonumber \\
		& -\sum\limits _{n=1}^{N}{\rm tr}\left[(\sum\limits _{q=0}^{Q}\delta_{q}^{*}e^{j2\pi\tau_{q}\frac{(n-1)B}{N}}\bm{{\Sigma}}_{q}^{H}{\bf X}^{H}[n]{\tilde{{\bf H}}}_{q}^{H})(\sum\limits _{q=0}^{Q}\delta_{q}e^{-j2\pi\tau_{q}\frac{(n-1)B}{N}}{\tilde{{\bf H}}}_{q}{\bf X}[n]\bm{{\Sigma}}_{q}))\right].\label{stage2_Allparameters_MLE_Derivation_omit}
	\end{align}
By further omitting the constant terms in \eqref{stage2_Allparameters_MLE_Derivation_omit}, the \eqref{stage2_Allparameters_MLE_Derivation_omit} can be simplified into
    \begin{align}
		& \Lambda(\boldsymbol{{\bf \eta}};\{{\bf Y}[n]\}_{n=1,2,\cdots,N})\nonumber \\
		= & \sum\limits _{n=1}^{N}{\rm tr}\left[\sum\limits _{q=0}^{Q}\delta_{q}e^{-j2\pi\tau_{q}\frac{(n-1)B}{N}}{\bf Y}^{H}[n]{\tilde{{\bf H}}}_{q}{\bf X}[n]\bm{{\Sigma}}_{q}\right]+\sum\limits _{n=1}^{N}{\rm tr}\left[\sum\limits _{q=0}^{Q}\delta_{q}^{*}e^{j2\pi\tau_{q}\frac{(n-1)B}{N}}\bm{{\Sigma}}_{q}^{H}{\bf X}^{H}[n]{\tilde{{\bf H}}}_{q}^{H}{\bf Y}[n]\right]\nonumber \\
		& -\sum\limits _{n=1}^{N}{\rm tr}\left[(\sum\limits _{q=0}^{Q}\delta_{q}^{*}e^{j2\pi\tau_{q}\frac{(n-1)B}{N}}\bm{{\Sigma}}_{q}^{H}{\bf X}^{H}[n]{\tilde{{\bf H}}}_{q}^{H})(\sum\limits _{q=0}^{Q}\delta_{q}e^{-j2\pi\tau_{q}\frac{(n-1)B}{N}}{\tilde{{\bf H}}}_{q}{\bf X}[n]\bm{{\Sigma}}_{q}))\right]\nonumber \\
		\mathop{{\rm =}}\limits ^{(i)} & 2{\mathfrak{Re}}\left\{ \sum\limits _{n=1}^{N}{\rm tr}\left[\sum\limits _{q=0}^{Q}(\delta_{q}e^{-j2\pi\tau_{q}\frac{(n-1)B}{N}}{\bf Y}^{H}[n]{\tilde{{\bf H}}}_{q}{\bf X}[n]\bm{{\Sigma}}_{q})\right]\right\} \nonumber \\
		& -\sum\limits _{n=1}^{N}{\rm tr}(\sum\limits _{q_{1}=0}^{Q}\sum\limits _{q_{2}=0}^{Q}\delta_{q_{1}}^{*}\delta_{q_{2}}e^{j2\pi(\tau_{q_{1}}-\tau_{q_{2}})\frac{(n-1)B}{N}}\bm{{\Sigma}}_{q_{1}}^{H}{\bf X}^{H}[n]{\tilde{{\bf H}}}_{q_{1}}^{H}{\tilde{{\bf H}}}_{q_{2}}{\bf X}[n]\bm{{\Sigma}}_{q_{2}})\nonumber \\
		= & 2{\mathfrak{Re}}\left\{ \sum\limits _{n=1}^{N}\sum\limits _{q=0}^{Q}\delta_{q}e^{-j2\pi\tau_{q}\frac{(n-1)B}{N}}{\rm tr(}{\bf Y}^{H}[n]{\tilde{{\bf H}}}_{q}{\bf X}[n]\bm{{\Sigma}}_{q})\right\} \nonumber \\
		& -\sum\limits _{n=1}^{N}\sum\limits _{q_{1}=0}^{Q}\sum\limits _{q_{2}=0}^{Q}\delta_{q_{1}}^{*}\delta_{q_{2}}e^{j2\pi(\tau_{q_{1}}-\tau_{q_{2}})\frac{(n-1)B}{N}}{\rm tr}(\bm{{\Sigma}}_{q_{1}}^{H}{\bf X}^{H}[n]{\tilde{{\bf H}}}_{q_{1}}^{H}{\tilde{{\bf H}}}_{q_{2}}{\bf X}[n]\bm{{\Sigma}}_{q_{2}})\nonumber \\
		= & 2{\mathfrak{Re}}\left\{ \sum\limits _{n=1}^{N}\sum\limits _{q=0}^{Q}(\delta_{q}e^{-j2\pi\tau_{q}\frac{(n-1)B}{N}}{\bf a}_{{\rm M}}^{H}(\vartheta_{{\rm t},q}){\bf X}[n]\bm{{\Sigma}}_{q}{\bf Y}^{H}[n]{\bf a}_{{\rm B}}(\vartheta_{{\rm r},0}))\right\} \nonumber \\
		& -\sum\limits _{n=1}^{N}\sum\limits _{q_{1}=0}^{Q}\sum\limits _{q_{2}=0}^{Q}(\delta_{q_{1}}^{*}\delta_{q_{2}}e^{j2\pi(\tau_{q_{1}}-\tau_{q_{2}})\frac{(n-1)B}{N}}{\bf a}_{{\rm B}}^{H}(\vartheta_{{\rm r},0}){\bf a}_{{\rm B}}(\vartheta_{{\rm r},0}){\bf a}_{{\rm M}}^{H}(\vartheta_{{\rm t},q_{2}}){\bf X}[n]\bm{{\Sigma}}_{q_{2}}\bm{{\Sigma}}_{q_{1}}^{H}{\bf X}^{H}[n]{\bf a}_{{\rm M}}(\vartheta_{{\rm t},q_{1}}))\nonumber \\
		= & 2{\mathfrak{Re}}\left\{ \sum\limits _{q=0}^{Q}{\bf a}_{{\rm M}}^{H}(\vartheta_{{\rm t},q})\left(\sum\limits _{n=1}^{N}\delta_{q}e^{-j2\pi\tau_{q}\frac{(n-1)B}{N}}{\bf X}[n]\bm{{\Sigma}}_{q}{\bf Y}^{H}[n]\right){\bf a}_{{\rm B}}(\vartheta_{{\rm r},0})\right\} \nonumber \\
		& -N_{\mathrm{b}}\sum\limits _{q_{1}=0}^{Q}\sum\limits _{q_{2}=0}^{Q}\left\{ \delta_{q_{1}}^{*}\delta_{q_{2}}{\bf a}_{{\rm M}}^{H}(\vartheta_{{\rm t},q_{2}})\left(\sum\limits _{n=1}^{N}e^{j2\pi(\tau_{q_{1}}-\tau_{q_{2}})\frac{(n-1)B}{N}}{\bf X}[n]\bm{{\Sigma}}_{q_{2}}\bm{{\Sigma}}_{q_{1}}^{H}{\bf X}^{H}[n]\right){\bf a}_{{\rm M}}(\vartheta_{{\rm t},q_{1}})\right\} ,\label{stage2_Allparameters_MLE_Derivation}
	\end{align}
	where the step $(i)$ refers to the identity ${\rm tr}({\bf A}^{H})=[{\rm tr}({\bf A})]^{*}$.
	
	\subsection{Derivative of ML function of a Single Path with Respect to Channel
		Gain in \eqref{stage2deltap_Derivation}}
	
	\label{Appendices B.B}
	
	The specific derivation process to obtain \eqref{stage2deltap_Derivation}
	is given as follows.
	\begin{align}
		& \frac{\partial L(\boldsymbol{{\bf \eta}}_{q};\{{\hat{{\bf Y}}}_{q}^{(k)}[n]\}_{n=1,\cdots,N})}{\partial\delta_{q}^{*}}\nonumber \\
		\mathop{{\rm =}}\limits & -\frac{\partial\sum\limits _{n=1}^{N}\left\Vert {\hat{{\bf Y}}}_{q}^{(k)}[n]-\delta_{q}e^{-j2\pi\tau_{q}\frac{(n-1)B}{N}}{\tilde{{\bf H}}}_{q}{\bf X}[n]\bm{{\Sigma}}_{q}\right\Vert _{F}^{2}}{\partial\delta_{q}^{*}}\nonumber \\
		= & -\frac{\partial\sum\limits _{n=1}^{N}{\rm tr}\left\{ ({\hat{{\bf Y}}}_{q}^{(k)}[n]-\delta_{q}e^{-j2\pi\tau_{q}\frac{(n-1)B}{N}}{\tilde{{\bf H}}}_{q}{\bf X}[n]\bm{{\Sigma}}_{q})^{H}({\hat{{\bf Y}}}_{q}^{(k)}[n]-\delta_{q}e^{-j2\pi\tau_{q}\frac{(n-1)B}{N}}{\tilde{{\bf H}}}_{q}{\bf X}[n]\bm{{\Sigma}}_{q})\right\} }{\partial\delta_{q}^{*}}\nonumber \\
		= & -\frac{\partial\sum\limits _{n=1}^{N}{\rm tr}\left\{ ({\hat{{\bf Y}}}_{q}^{(k)}[n])^{H}{\hat{{\bf Y}}}_{q}^{(k)}[n]\right\} }{\partial\delta_{q}^{*}}+\frac{\partial\sum\limits _{n=1}^{N}{\rm tr}\left\{ \delta_{q}e^{-j2\pi\tau_{q}\frac{(n-1)B}{N}}({\hat{{\bf Y}}}_{q}^{(k)}[n])^{H}{\tilde{{\bf H}}}_{q}{\bf X}[n]\bm{{\Sigma}}_{q}\right\} }{\partial\delta_{q}^{*}}\nonumber \\
		& +\frac{\partial\sum\limits _{n=1}^{N}{\rm tr}\left\{ \delta_{q}^{*}e^{j2\pi\tau_{q}\frac{(n-1)B}{N}}({\tilde{{\bf H}}}_{q}{\bf X}[n]\bm{{\Sigma}}_{q})^{H}{\hat{{\bf Y}}}_{q}^{(k)}[n]\right\} }{\partial\delta_{q}^{*}}\nonumber \\
		& -\frac{\partial\sum\limits _{n=1}^{N}{\rm tr}\left\{ \delta_{q}^{*}\delta_{q}e^{j2\pi\tau_{q}\frac{(n-1)B}{N}}e^{-j2\pi\tau_{q}\frac{(n-1)B}{N}}({\tilde{{\bf H}}}_{q}{\bf X}[n]\bm{{\Sigma}}_{q})^{H}{\tilde{{\bf H}}}_{q}{\bf X}[n]\bm{{\Sigma}}_{q}\right\} }{\partial\delta_{q}^{*}}\nonumber \\
		= & -\frac{\partial\sum\limits _{n=1}^{N}{\rm tr}\left\{ ({\hat{{\bf Y}}}_{q}^{(k)}[n])^{H}{\hat{{\bf Y}}}_{q}^{(k)}[n]\right\} }{\partial\delta_{q}^{*}}+\frac{\partial\delta_{q}\sum\limits _{n=1}^{N}{\rm tr}\left\{ e^{-j2\pi\tau_{q}\frac{(n-1)B}{N}}({\hat{{\bf Y}}}_{q}^{(k)}[n])^{H}{\tilde{{\bf H}}}_{q}{\bf X}[n]\bm{{\Sigma}}_{q}\right\} }{\partial\delta_{q}^{*}}\nonumber \\
		& +\frac{\partial\delta_{q}^{*}\sum\limits _{n=1}^{N}{\rm tr}\left\{ e^{j2\pi\tau_{q}\frac{(n-1)B}{N}}({\tilde{{\bf H}}}_{q}{\bf X}[n]\bm{{\Sigma}}_{q})^{H}{\hat{{\bf Y}}}_{q}^{(k)}[n]\right\} }{\partial\delta_{q}^{*}}-\frac{\partial\delta_{q}^{*}\delta_{q}\sum\limits _{n=1}^{N}{\rm tr}\left\{ ({\tilde{{\bf H}}}_{q}{\bf X}[n]\bm{{\Sigma}}_{q})^{H}{\tilde{{\bf H}}}_{q}{\bf X}[n]\bm{{\Sigma}}_{q}\right\} }{\partial\delta_{q}^{*}}\nonumber \\
		\mathop{{\rm =}}\limits ^{(j)} & \sum\limits _{n=1}^{N}\left\{ e^{j2\pi\tau_{q}\frac{(n-1)B}{N}}{\rm tr}[({\tilde{{\bf H}}}_{q}{\bf X}[n]\bm{{\Sigma}}_{q})^{H}{\hat{{\bf Y}}}_{q}^{(k)}[n]]\right\} -\delta_{q}\sum\limits _{n=1}^{N}{\rm tr}[({\tilde{{\bf H}}}_{q}{\bf X}[n]\bm{{\Sigma}}_{q})^{H}({\tilde{{\bf H}}}_{q}{\bf X}[n]\bm{{\Sigma}}_{q})],\label{Detailed_derivation_stage2}
	\end{align}
	where the step $(j)$ is obtained by using $\ensuremath{\frac{\partial\delta}{\partial\delta^{*}}=0}$,
	$\ensuremath{\frac{\partial\delta^{*}}{\partial\delta^{*}}=1}$ and
	$\ensuremath{\frac{\partial\delta^{*}\delta}{\partial\delta^{*}}=\frac{\partial\delta^{*}}{\partial\delta^{*}}\delta+\delta^{*}\frac{\partial\delta}{\partial\delta^{*}}=\delta}$,
	which can be referred to the formula (3.4.7) on page 173 in \cite{zhang2017matrix}.
	
	\subsection{Simplification for the Expression of Channel Gain in \eqref{stage2deltap_MLE} }
	
	\label{Appendices B.C}
	
	The specific simplification process of ${\delta_{q}}$ in \eqref{stage2deltap_MLE}
	is as follows.
	\begin{align}
		\delta_{q} & =\frac{\sum\limits _{n=1}^{N}\left\{ e^{j2\pi\tau_{q}\frac{(n-1)B}{N}}{\rm tr}(\bm{{\Sigma}}_{q}^{H}{\bf X}^{H}[n]{\tilde{{\bf H}}}_{q}^{H}{\hat{{\bf Y}}}_{q}^{(k)}[n])\right\} }{\sum\limits _{n=1}^{N}{\rm tr}(\bm{{\Sigma}}_{q}^{H}{\bf X}^{H}[n]{\tilde{{\bf H}}}_{q}^{H}{\tilde{{\bf H}}}_{q}{\bf X}[n]\bm{{\Sigma}}_{q}^{H})}\nonumber \\
		& =\frac{\sum\limits _{n=1}^{N}\left\{ e^{j2\pi\tau_{q}\frac{(n-1)B}{N}}{\rm tr}({\bf a}_{{\rm M}}(\vartheta_{{\rm t},q}){\bf a}_{{\rm B}}^{H}(\vartheta_{{\rm r},0}){\hat{{\bf Y}}}_{q}^{(k)}[n]\bm{{\Sigma}}_{q}^{H}{\bf X}^{H}[n])\right\} }{\sum\limits _{n=1}^{N}{\rm tr}({\bf a}_{{\rm M}}(\vartheta_{{\rm t},q}){\bf a}_{{\rm B}}^{H}(\vartheta_{{\rm r},0}){\bf a}_{{\rm B}}(\vartheta_{{\rm r},0}){\bf a}_{{\rm M}}^{H}(\vartheta_{{\rm t},q}){\bf X}[n]\bm{{\Sigma}}_{q}\bm{{\Sigma}}_{q}^{H}{\bf X}^{H}[n])}\nonumber \\
		& =\frac{\sum\limits _{n=1}^{N}\left\{ e^{j2\pi\tau_{q}\frac{(n-1)B}{N}}{\bf a}_{{\rm B}}^{H}(\vartheta_{{\rm r},0}){\hat{{\bf Y}}}_{q}^{(k)}[n]\bm{{\Sigma}}_{q}^{H}{\bf X}^{H}[n]{\bf a}_{{\rm M}}(\vartheta_{{\rm t},q})\right\} }{\sum\limits _{n=1}^{N}[N_{{\rm b}}{\bf a}_{{\rm M}}^{H}(\vartheta_{{\rm t},q}){\bf X}[n]\bm{{\Sigma}}_{q}\bm{{\Sigma}}_{q}^{H}{\bf X}^{H}[n]{\bf a}_{{\rm M}}(\vartheta_{{\rm t},q})]}\nonumber \\
		& =\frac{{\bf a}_{{\rm B}}^{H}(\vartheta_{{\rm r},0})\left\{ \sum\limits _{n=1}^{N}(e^{j2\pi\tau_{q}\frac{(n-1)B}{N}}{\hat{{\bf Y}}}_{q}^{(k)}[n]\bm{{\Sigma}}_{q}^{H}{\bf X}^{H}[n])\right\} {\bf a}_{{\rm M}}(\vartheta_{{\rm t},q})}{N_{{\rm b}}{\bf a}_{{\rm M}}^{H}(\vartheta_{{\rm t},q})\left\{ \sum\limits _{n=1}^{N}({\bf X}[n]\bm{{\Sigma}}_{q}\bm{{\Sigma}}_{q}^{H}{\bf X}^{H}[n])\right\} {\bf a}_{{\rm M}}(\vartheta_{{\rm t},q})}.\label{Detailed_deltap_stage2}
	\end{align}

	\subsection{Simplification of the ML Function in \eqref{stage2angle_MLE}}
	
	\label{Appendices B.D}
	
	By substituting ${\hat{{\bf Y}}}_{q}^{(k)}[n]$ into \eqref{stage2_lnMLE},
	we can obtain
	\begin{align}
		L(\boldsymbol{{\bf \eta}}_{q};\{{\hat{{\bf Y}}}_{q}^{(k)}[n]\}_{n=1,\cdots,N})= & 2{\mathfrak{Re}}\left[\delta_{q}{\bf a}_{{\rm M}}^{H}(\vartheta_{{\rm t},q})\left(\sum\limits _{n=1}^{N}e^{-j2\pi\tau_{q}\frac{(n-1)B}{N}}{\bf X}[n]\bm{{\Sigma}}_{q}(\hat{\bf {Y}}_{q}^{(k)}[n])^{H}\right){\bf a}_{{\rm B}}(\vartheta_{{\rm r},0})\right]\nonumber \\
		& -N_{{\rm b}}\delta_{q}^{*}\delta_{q}{\bf a}_{{\rm M}}^{H}(\vartheta_{{\rm t},q})\left(\sum\limits _{n=1}^{N}{\bf X}[n]\bm{{\Sigma}}_{q}\bm{{\Sigma}}_{q}^{H}{\bf X}^{H}[n]\right){\bf a}_{{\rm M}}(\vartheta_{{\rm t},q}).\label{LikehoodsinglePath}
	\end{align}
	
	Then, by substituting $\delta_{q}$ in \eqref{stage2deltap_MLE} (i.e.,
	\eqref{Detailed_deltap_stage2}) into $L(\boldsymbol{{\bf \eta}}_{q};\{{\hat{{\bf Y}}}_{q}^{(k)}[n]\}_{n=1,\cdots,N})$
	in \eqref{LikehoodsinglePath}, we can have the simplified expression
	of $F(\bar{\boldsymbol{\eta}}_{q};\{{\hat{{\bf Y}}}_{q}^{(k)}[n]\}_{n=1,\cdots,N})$
	as
	\begin{align}
		& F(\bar{\boldsymbol{\eta}}_{q};\{{\hat{{\bf Y}}}_{q}^{(k)}[n]\}_{n=1,\cdots,N})\nonumber \\
		= & 2{\mathfrak{Re}}\left\{ \frac{{\bf a}_{{\rm B}}^{H}(\vartheta_{{\rm r},0})\left\{ \sum\limits _{n=1}^{N}(e^{j2\pi\tau_{q}\frac{(n-1)B}{N}}{\hat{{\bf Y}}}_{q}^{(k)}[n]\bm{{\Sigma}}_{q}^{H}{\bf X}^{H}[n])\right\} {\bf a}_{{\rm M}}(\vartheta_{{\rm t},q})}{N_{{\rm b}}{\bf a}_{{\rm M}}^{H}(\vartheta_{{\rm t},q})\left\{ \sum\limits _{n=1}^{N}({\bf X}[n]\bm{{\Sigma}}_{q}\bm{{\Sigma}}_{q}^{H}{\bf X}^{H}[n])\right\} {\bf a}_{{\rm M}}(\vartheta_{{\rm t},q})}\right.\nonumber \\
		& \cdot\left.{\bf a}_{{\rm M}}^{H}(\vartheta_{{\rm t},q})\left(\sum\limits _{n=1}^{N}e^{-j2\pi\tau_{q}\frac{(n-1)B}{N}}{\bf X}[n]\bm{{\Sigma}}_{q}(\hat{\bf {Y}}_{q}^{(k)}[n])^{H}\right){\bf a}_{{\rm B}}(\vartheta_{{\rm r},0})\right\} \nonumber \\
		& -\left\{ \left|\frac{{\bf a}_{{\rm B}}^{H}(\vartheta_{{\rm r},0})\left\{ \sum\limits _{n=1}^{N}(e^{j2\pi\tau_{q}\frac{(n-1)B}{N}}{\hat{{\bf Y}}}_{q}^{(k)}[n]\bm{{\Sigma}}_{q}^{H}{\bf X}^{H}[n])\right\} {\bf a}_{{\rm M}}(\vartheta_{{\rm t},q})}{N_{{\rm b}}{\bf a}_{{\rm M}}^{H}(\vartheta_{{\rm t},q})\left\{ \sum\limits _{n=1}^{N}({\bf X}[n]\bm{{\Sigma}}_{q}\bm{{\Sigma}}_{q}^{H}{\bf X}^{H}[n])\right\} {\bf a}_{{\rm M}}(\vartheta_{{\rm t},q})}\right|^{2}\right.\nonumber \\
		& \cdot\left.N_{{\rm b}}{\bf a}_{{\rm M}}^{H}(\vartheta_{{\rm t},q})\left(\sum\limits _{n=1}^{N}{\bf X}[n]\bm{{\Sigma}}_{q}\bm{{\Sigma}}_{q}^{H}{\bf X}^{H}[n]\right){\bf a}_{{\rm M}}(\vartheta_{{\rm t},q}))\right\} \nonumber \\
		= & 2{\mathfrak{Re}}\left\{ \frac{\left|{\bf a}_{{\rm M}}^{H}(\vartheta_{{\rm t},q})\left\{ \sum\limits _{n=1}^{N}(e^{-j2\pi\tau_{l}\frac{(n-1)B}{N}}{\bf X}[n]\bm{{\Sigma}}_{q}({\hat{{\bf Y}}}_{q}^{(k)}[n])^{H})\right\} {\bf a}_{{\rm B}}(\vartheta_{{\rm r},0})\right|^{2}}{N_{{\rm b}}{\bf a}_{{\rm M}}^{H}(\vartheta_{{\rm t},q})\left\{ \sum\limits _{n=1}^{N}({\bf X}[n]\bm{{\Sigma}}_{q}\bm{{\Sigma}}_{q}^{H}{\bf X}^{H}[n])\right\} {\bf a}_{{\rm M}}(\vartheta_{{\rm t},q})}\right\} \nonumber \\
		& -\frac{\left|{\bf a}_{{\rm B}}^{H}(\vartheta_{{\rm r},0})\left\{ \sum\limits _{n=1}^{N}(e^{j2\pi\tau_{q}\frac{(n-1)B}{N}}{\hat{{\bf Y}}}_{q}^{(k)}[n]\bm{{\Sigma}}_{q}^{H}{\bf X}^{H}[n])\right\} {\bf a}_{{\rm M}}(\vartheta_{{\rm t},q})\right|^{2}}{N_{{\rm b}}{\bf a}_{{\rm M}}^{H}(\vartheta_{{\rm t},q})\left\{ \sum\limits _{n=1}^{N}({\bf X}[n]\bm{{\Sigma}}_{q}\bm{{\Sigma}}_{q}^{H}{\bf X}^{H}[n])\right\} {\bf a}_{{\rm M}}(\vartheta_{{\rm t},q})}\nonumber \\
		= & \frac{2\left|{\bf a}_{{\rm M}}^{H}(\vartheta_{{\rm t},q})\left\{ \sum\limits _{n=1}^{N}(e^{-j2\pi\tau_{l}\frac{(n-1)B}{N}}{\bf X}[n]\bm{{\Sigma}}_{q}({\hat{{\bf Y}}}_{q}^{(k)}[n])^{H})\right\} {\bf a}_{{\rm B}}(\vartheta_{{\rm r},0})\right|^{2}}{N_{{\rm b}}{\bf a}_{{\rm M}}^{H}(\vartheta_{{\rm t},q})\left\{ \sum\limits _{n=1}^{N}({\bf X}[n]\bm{{\Sigma}}_{q}\bm{{\Sigma}}_{q}^{H}{\bf X}^{H}[n])\right\} {\bf a}_{{\rm M}}(\vartheta_{{\rm t},q})}\nonumber \\
		& -\frac{\left|{\bf a}_{{\rm B}}^{H}(\vartheta_{{\rm r},0})\left\{ \sum\limits _{n=1}^{N}(e^{j2\pi\tau_{q}\frac{(n-1)B}{N}}{\hat{{\bf Y}}}_{q}^{(k)}[n]\bm{{\Sigma}}_{q}^{H}{\bf X}^{H}[n])\right\} {\bf a}_{{\rm M}}(\vartheta_{{\rm t},q})\right|^{2}}{N_{{\rm b}}{\bf a}_{{\rm M}}^{H}(\vartheta_{{\rm t},q})\left\{ \sum\limits _{n=1}^{N}({\bf X}[n]\bm{{\Sigma}}_{q}\bm{{\Sigma}}_{q}^{H}{\bf X}^{H}[n])\right\} {\bf a}_{{\rm M}}(\vartheta_{{\rm t},q})}\nonumber \\
		= & \frac{\left|{\bf a}_{{\rm B}}^{H}(\vartheta_{{\rm r},0})\left\{ \sum\limits _{n=1}^{N}(e^{j2\pi\tau_{q}\frac{(n-1)B}{N}}{\hat{{\bf Y}}}_{q}^{(k)}[n]\bm{{\Sigma}}_{q}^{H}{\bf X}^{H}[n])\right\} {\bf a}_{{\rm M}}(\vartheta_{{\rm t},q})\right|^{2}}{N_{{\rm b}}{\bf a}_{{\rm M}}^{H}(\vartheta_{{\rm t},q})\left\{ \sum\limits _{n=1}^{N}({\bf X}[n]\bm{{\Sigma}}_{q}\bm{{\Sigma}}_{q}^{H}{\bf X}^{H}[n])\right\} {\bf a}_{{\rm M}}(\vartheta_{{\rm t},q})}.\label{stage2_LikeliHood_simp}
	\end{align}
	
\section{Closed-form Solutions for Scenarios with 2D Rotation of MS}
\label{Appendix_B_add}
\subsection{Closed-form Solutions for Coordinates of the Scatterer}
\label{Appendices_B_add.A}
For notation simplicity, we equivalently rewrite the $\hat{{\bf {e}}}_{q}$
defined after \eqref{MScordVlosVlos_2Drott} as $\hat{{\bf {e}}}_{q}=[\hat{e}_{x}^{q},\hat{e}_{y}^{q},\hat{e}_{z}^{q}]^{T}$,
define $\hat{d}_{mqr}\triangleq(\hat{\tau}_{q}c-\left\Vert {\bf r}-{\bf b}\right\Vert _{2})$
depicting the distance of the MS-scatterer$q$-RIS path, and utilize
$d\triangleq{d}_{sr}^{q}$, which is the distance from the $q$th
scatterer to the RIS. In \eqref{EqutnSetFord_sr_q}, the $\hat{{\bf {e}}}_{q}$
and $\hat{d}_{mqr}$ are known based on the estimated channel parameters,
and the ${\hat{{\bf m}}}$ is known due to \eqref{MScordVlosVlos_2Drott}.
Thus, the only unknown variable in \eqref{EqutnSetFord_sr_q} is essentially
$d={d}_{sr}^{q}$. As long as the $d={d}_{sr}^{q}$ is obtained, the
${\bf s}^{q}$ can be recovered according to \eqref{SqSubEq1}.

From \eqref{SqSubEq1}, we have $s_{x}^{q}=r_{x}-d\hat{e}_{x}^{q}$,
$s_{y}^{q}=r_{y}-d\hat{e}_{y}^{q}$, and $s_{z}^{q}=r_{z}-d\hat{e}_{z}^{q}$.
From \eqref{SqSubEq2}, we have $(\hat{d}_{mqr}-d)=\left\Vert {\bf s}^{q}\!-\!{\hat{{\bf m}}}\right\Vert _{2}$,
which is equivalent to 
\begin{align}
	(\hat{d}_{mqr}-d) & =\sqrt{(s_{x}^{q}-\hat{m}_{x})^{2}+(s_{y}^{q}-\hat{m}_{y})^{2}+(s_{z}^{q}-\hat{m}_{z})^{2}}.\label{SqSubEq2-Form2}
\end{align}

By substituting $s_{x}^{q}=r_{x}-d\hat{e}_{x}^{q}$, $s_{y}^{q}=r_{y}-d\hat{e}_{y}^{q}$,
and $s_{z}^{q}=r_{z}-d\hat{e}_{z}^{q}$ into \eqref{SqSubEq2-Form2},
we have  
	\begin{align}
	(\hat{d}_{mqr}-d) & =\sqrt{(r_{x}-d\hat{e}_{x}^{q}-\hat{m}_{x})^{2}+(r_{y}-d\hat{e}_{y}^{q}-\hat{m}_{y})^{2}+(r_{z}-d\hat{e}_{z}^{q}-\hat{m}_{z})^{2}},\nonumber \\
	(\hat{d}_{mqr}-d)^{2} & \stackrel{(a)}{=}\!\!(r_{x}\!\!-\!\hat{m}_{x})^{2}\!+\!(r_{y}\!-\!\hat{m}_{y})^{2}\!+\!(r_{z}\!-\!\hat{m}_{z})^{2}\!-\!2d[(r_{x}\!-\!\hat{m}_{x})\hat{e}_{x}^{q}\!+\!(r_{y}\!-\!\hat{m}_{y})\hat{e}_{y}^{q}\!+\!(r_{z}\!-\!\hat{m}_{z})\hat{e}_{z}^{q}]\!+\!d^{2},\nonumber \\
	\hat{d}_{mqr}^{2}-2d\hat{d}_{mqr}+d^{2} & =\hat{d}_{rm}^{2}-2d[(r_{x}-\hat{m}_{x})\hat{e}_{x}^{q}+(r_{y}-\hat{m}_{y})\hat{e}_{y}^{q}+(r_{z}-\hat{m}_{z})\hat{e}_{z}^{q}]+d^{2},\nonumber \\
	\hat{d}_{mqr}^{2}-2d\hat{d}_{mqr} & =\hat{d}_{rm}^{2}-2d[(r_{x}-\hat{m}_{x})\hat{e}_{x}^{q}+(r_{y}-\hat{m}_{y})\hat{e}_{y}^{q}+(r_{z}-\hat{m}_{z})\hat{e}_{z}^{q}],\nonumber \\
	d & =(\hat{d}_{mqr}^{2}-\hat{d}_{rm}^{2})/\{2[\hat{d}_{mqr}-(r_{x}-\hat{m}_{x})\hat{e}_{x}^{q}-(r_{y}-\hat{m}_{y})\hat{e}_{y}^{q}-(r_{z}-\hat{m}_{z})\hat{e}_{z}^{q}]\},\nonumber \\
	d & =(\hat{d}_{mqr}^{2}-\hat{d}_{rm}^{2})/\{2[\hat{d}_{mqr}-({\bf r}-\hat{{\bf m}})^{T}\hat{{\bf {e}}}_{q}]\},\label{SolvingFord}
\end{align}
where $(e_{x}^{q})^{2}+(e_{y}^{q})^{2}+(e_{z}^{q})^{2}=1$
is utilized in the step $(a)$ of \eqref{SolvingFord}. From \eqref{SolvingFord},
we can coarsely estimate the distance $d={d}_{sr}^{q}$. Then, by
using \eqref{SqSubEq1}, the ${\bf s}^{q}$ can be calculated from
the estimated $\hat{d}={\hat{d}}_{sr}^{q}$ and $\hat{{\bf e}}_{q}$.
\subsection{Closed-form Solutions for Two Rotation Angles of the ULA at MS}
\label{Appendices_B_add.B}
By using the geometric relationships of \eqref{AnglesMS-RIS-VLOS}
and \eqref{AnglesMS-RIS-NLOS} in the MS-RIS link, we can build the
system of equations as 
\begin{subequations}
	\label{EqtnsForRotArry} 
	\begin{align}
		\frac{({\bf r}-\hat{{\bf m}})^{T}}{\left\Vert {\bf r}-\hat{{\bf m}}\right\Vert _{2}}\mathbf{e}_{\textrm{rot}} & =\textrm{sin}\hat{\theta}_{{\rm t},0},\label{EqtnsForRotArry-Sub1}\\
		\frac{(\hat{{\bf s}}^{q}-\hat{{\bf m}})^{T}}{\left\Vert \hat{{\bf s}}^{q}-\hat{{\bf m}}\right\Vert _{2}}\mathbf{e}_{\textrm{rot}} & =\textrm{sin}\hat{\theta}_{{\rm t},q},\label{EqtnsForRotArry-Sub2}
	\end{align}
\end{subequations}
where $\mathbf{e}_{\textrm{rot}}=[\sin\beta\cos\alpha,-\sin\beta\sin\alpha,\cos\beta]^{T}$
is unknown, but can be solved from \eqref{EqtnsForRotArry}. Obviously,
the $({\bf r}-\hat{{\bf m}})^{T}/\left\Vert {\bf r}-\hat{{\bf m}}\right\Vert _{2}$
and $(\hat{{\bf s}}^{q}-\hat{{\bf m}})^{T}/\left\Vert \hat{{\bf s}}^{q}-\hat{{\bf m}}\right\Vert _{2}$
are both unit direction vectors in the real number field. For simplicity,
we assume $\mathbf{a}=({\bf r}-\hat{{\bf m}})/\left\Vert {\bf r}-\hat{{\bf m}}\right\Vert _{2}$,
$\mathbf{b}=(\hat{{\bf s}}^{q}-\hat{{\bf m}})/\left\Vert \hat{{\bf s}}^{q}-\hat{{\bf m}}\right\Vert _{2}$,
and $\mathbf{e}=\mathbf{e}_{\textrm{rot}}$. Then, the \eqref{EqtnsForRotArry}
can be intuitively written as a purely mathematical problem as 
\begin{subequations}
	\label{EqtnsForRotArry-Smplfy} 
	\begin{align}
		\mathbf{a}^{T}\mathbf{e} & =d_{1},\label{Smplfy1}\\
		\mathbf{b}^{T}\mathbf{e} & =d_{2},\label{Smplfy2}\\
		\mathbf{e}^{T}\mathbf{e} & =1,\label{Smplfy3}
	\end{align}
\end{subequations}
where $\mathbf{a}=[a_{1},a_{2},a_{3}]^{T}$ and $\mathbf{b}=[b_{1},b_{2},b_{3}]^{T}$
are given unit vectors ($\|\mathbf{a}\|=\|\mathbf{b}\|=1$) in the
real number field, and $d_{1}=\sin\hat{\theta}_{t,0}$, $d_{2}=\sin\hat{\theta}_{t,q}$
are known scalars. Then, we have to find the unit direction vector
$\mathbf{e}=[x,y,z]^{T}$ that satisfies \eqref{EqtnsForRotArry-Smplfy}.
As long as the $\mathbf{e}$ is obtained, the two rotational angles
$\alpha$ and $\beta$ can be obtained. The detailed solving process
for $\mathbf{e}$ are as follows.

\subsubsection{Geometric Interpretation and Solution Overview}

Equation \eqref{Smplfy1} defines a plane perpendicular to vector
$\mathbf{a}$ at a distance $d_{1}$ from the origin. Similarly, equation
\eqref{Smplfy2} defines another plane perpendicular to $\mathbf{b}$
at a distance $d_{2}$. The intersection of these two planes (obviously,
$\mathbf{a}$ and $\mathbf{b}$ cannot be parallel from the physical
meaning) is a line $L$. Equation \eqref{Smplfy3} defines the unit
sphere centered at the origin. The solution(s) $\mathbf{e}$ are the
intersection point(s) of the line $L$ and the unit sphere. In 3D
space, a line and a sphere can intersect at \textbf{0, 1 (tangent)},
or \textbf{2} points.

\subsubsection{Solving Procedure}

The process of solving the system of equations in \eqref{EqtnsForRotArry-Smplfy}
consists of the following steps.

\text{1. {General Solution to the Linear System}}

The linear system formed by equations \eqref{Smplfy1} and \eqref{Smplfy2}
is underdetermined. Its general solution is the sum of a particular
solution $\mathbf{e}_{\textrm{p}}$ and the homogeneous solution $\mathbf{e}_{\textrm{h}}$
as 
\begin{equation}
	\mathbf{e}=\mathbf{e}_{\textrm{p}}+t\cdot\mathbf{e}_{\textrm{h}},\label{eq:general}
\end{equation}
where $t$ is a scalar parameter, $\mathbf{e}_{\textrm{p}}$ satisfies
$\mathbf{A}\mathbf{e}_{\textrm{p}}=\mathbf{d}$, and $\mathbf{e}_{\textrm{h}}$
satisfies $\mathbf{A}\mathbf{e}_{\textrm{h}}=\mathbf{0}$ (i.e., lies
in the null space of $\mathbf{A}$), with 
\begin{equation}
	\mathbf{A}=\begin{bmatrix}\mathbf{a}^{T}\\
		\mathbf{b}^{T}
	\end{bmatrix}=\begin{bmatrix}a_{1} & a_{2} & a_{3}\\
		b_{1} & b_{2} & b_{3}
	\end{bmatrix},\mathbf{d}=\begin{bmatrix}d_{1}\\
		d_{2}
	\end{bmatrix}.
\end{equation}

\text{2. Particular Solution}

A standard method to find a particular solution is to use the Moore-Penrose
pseudoinverse, which gives the minimum norm solution as 
\begin{equation}
	\mathbf{e}_{\textrm{p}}=\mathbf{A}^{\dagger}\mathbf{d},\label{eq:pseudo}
\end{equation}
where $\mathbf{A}^{\dagger}=\mathbf{A}^{T}(\mathbf{A}\mathbf{A}^{T})^{-1}$.
Obviously, the Moore-Penrose pseudoinverse $\mathbf{A}^{\dagger}$
exists if the $\mathbf{A}\mathbf{A}^{T}$ has an inverse.

First, we compute the Gram matrix as 
\begin{equation}
	\mathbf{A}\mathbf{A}^{T}=\begin{bmatrix}\mathbf{a}^{T}\mathbf{a} & \mathbf{a}^{T}\mathbf{b}\\
		\mathbf{b}^{T}\mathbf{a} & \mathbf{b}^{T}\mathbf{b}
	\end{bmatrix}=\begin{bmatrix}1 & g\\
		g & 1
	\end{bmatrix},\label{GramMtrx}
\end{equation}
where $g\triangleq\mathbf{a}^{T}\mathbf{b}$ and $\left|g\right|\leq1$.
The inverse of the Gram matrix is 
\begin{equation}
	(\mathbf{A}\mathbf{A}^{T})^{-1}=\frac{1}{\det(\mathbf{A}\mathbf{A}^{T})}\begin{bmatrix}1 & -g\\
		-g & 1
	\end{bmatrix}=\frac{1}{1-g^{2}}\begin{bmatrix}1 & -g\\
		-g & 1
	\end{bmatrix}.\label{GramMtrx_Invrs}
\end{equation}

From \eqref{GramMtrx_Invrs}, it is seen that the the Moore-Penrose
pseudoinverse exist as long as $\left|g\right|\neq1$, i.e., the vectors
of $({\bf r}-\hat{{\bf m}})$ and $(\hat{{\bf s}}^{q}-\hat{{\bf m}})$
should not be parallel. By substituting \eqref{GramMtrx_Invrs} into
\eqref{eq:pseudo}, we have 
\begin{align}
	\mathbf{e}_{\textrm{p}} & =\mathbf{A}^{T}(\mathbf{A}\mathbf{A}^{T})^{-1}\mathbf{d}\nonumber \\
	& =\frac{1}{1-g^{2}}\begin{bmatrix}a_{1} & b_{1}\\
		a_{2} & b_{2}\\
		a_{3} & b_{3}
	\end{bmatrix}\begin{bmatrix}1 & -g\\
		-g & 1
	\end{bmatrix}\begin{bmatrix}d_{1}\\
		d_{2}
	\end{bmatrix}\nonumber \\
	& =\frac{1}{1-g^{2}}\begin{bmatrix}a_{1} & b_{1}\\
		a_{2} & b_{2}\\
		a_{3} & b_{3}
	\end{bmatrix}\begin{bmatrix}d_{1}-gd_{2}\\
		d_{2}-gd_{1}
	\end{bmatrix}\nonumber \\
	& =\frac{1}{1-g^{2}}\left[(d_{1}-gd_{2})\mathbf{a}+(d_{2}-gd_{1})\mathbf{b}\right].\label{ParticularSltn_Specf}
\end{align}

\text{3. Homogeneous Solution}

The homogeneous solution must satisfy $\mathbf{a}^{T}\mathbf{e}_{\textrm{h}}=0$
and $\mathbf{b}^{T}\mathbf{e}_{\textrm{h}}=0$. This means $\mathbf{e}_{\textrm{h}}$
is orthogonal to both $\mathbf{a}$ and $\mathbf{b}$. In 3D space,
the cross product $\mathbf{a}\times\mathbf{b}$ is such a vector.
Thus, we take 
\begin{equation}
	\mathbf{e}_{\textrm{h}}=\mathbf{a}\times\mathbf{b}.\label{eq:homogeneous}
\end{equation}

Specifically, the cross product of two vectors $\ensuremath{\mathbf{a}=[a_{1},a_{2},a_{3}]^{T}}$
and $\ensuremath{\mathbf{b}=[b_{1},b_{2},b_{3}]^{T}}$ is most formally
computed using the determinant method as 
\begin{equation}
	\mathbf{a}\times\mathbf{b}=\begin{vmatrix}\mathbf{i} & \mathbf{j} & \mathbf{k}\\
		a_{1} & a_{2} & a_{3}\\
		b_{1} & b_{2} & b_{3}
	\end{vmatrix},
\end{equation}
where $\mathbf{i}$, $\mathbf{j}$, and $\mathbf{k}$ are the unit
vectors in the $x$, $y$, and $z$ directions, respectively. To compute
this determinant, we expand along the first row, and arrive at 
\begin{align}
	\mathbf{a}\times\mathbf{b} & =\mathbf{i}\begin{vmatrix}a_{2} & a_{3}\\
		b_{2} & b_{3}
	\end{vmatrix}-\mathbf{j}\begin{vmatrix}a_{1} & a_{3}\\
		b_{1} & b_{3}
	\end{vmatrix}+\mathbf{k}\begin{vmatrix}a_{1} & a_{2}\\
		b_{1} & b_{2}
	\end{vmatrix}\nonumber \\
	& =\mathbf{i}(a_{2}b_{3}-a_{3}b_{2})-\mathbf{j}(a_{1}b_{3}-a_{3}b_{1})+\mathbf{k}(a_{1}b_{2}-a_{2}b_{1}),
\end{align}
where the negative sign before the $\mathbf{j}$ component should
be noted. Then the final result of $\mathbf{a}\times\mathbf{b}$ can
be written as 
\begin{equation}
	\mathbf{a}\times\mathbf{b}=[a_{2}b_{3}-a_{3}b_{2},a_{3}b_{1}-a_{1}b_{3},a_{1}b_{2}-a_{2}b_{1}]^{T}.
\end{equation}

\text{4. Applying the Unit Sphere Constraint}

By substituting \eqref{eq:homogeneous} into \eqref{eq:general},
we have the general solution written as 
\begin{equation}
	\mathbf{e}=\mathbf{e}_{\textrm{p}}+t(\mathbf{a}\times\mathbf{b}).\label{eq:final_general}
\end{equation}

By substituting \eqref{eq:final_general} into the unit sphere equation
\eqref{Smplfy3}, we have 
\begin{subequations}
	\begin{align}
		[\mathbf{e}_{\textrm{p}}+t(\mathbf{a}\times\mathbf{b})]^{T}[\mathbf{e}_{\textrm{p}}+t(\mathbf{a}\times\mathbf{b})] & =1,\\
		\mathbf{e}_{\textrm{p}}^{T}\mathbf{e}_{\textrm{p}}+2t[\mathbf{e}_{\textrm{p}}^{T}(\mathbf{a}\times\mathbf{b})]+t^{2}(\mathbf{a}\times\mathbf{b})^{T}(\mathbf{a}\times\mathbf{b}) & =1,\\
		\mathbf{e}_{\textrm{p}}^{T}\mathbf{e}_{\textrm{p}}+2t[\mathbf{e}_{\textrm{p}}^{T}\mathbf{n}]+t^{2}\mathbf{n}^{T}\mathbf{n} & =1,\label{eq:substituted}
	\end{align}
\end{subequations}
where $\mathbf{n}\triangleq\mathbf{a}\times\mathbf{b}$. Notice that
the $\mathbf{e}_{\textrm{p}}$ in \eqref{ParticularSltn_Specf} is
a linear combination of $\mathbf{a}$ and $\mathbf{b}$, thus lies
in the plane spanned by $\mathbf{a}$ and $\mathbf{b}$. Since $\mathbf{n}$
is perpendicular to this plane, we have $\mathbf{e}_{\textrm{p}}^{T}\mathbf{n}=0$.
Then, the \eqref{eq:substituted} simplifies to 
\begin{equation}
	\|\mathbf{e}_{\textrm{p}}\|^{2}+t^{2}\|\mathbf{n}\|^{2}=1.\label{eq:simplified}
\end{equation}

The squared magnitude of the cross product is 
\begin{equation}
	\begin{split}\|\mathbf{n}\|^{2} & =\|\mathbf{a}\times\mathbf{b}\|^{2}=\|\mathbf{a}\|^{2}\|\mathbf{b}\|^{2}\sin^{2}\theta\\
		& =(1-\cos^{2}\theta)=1-g^{2}.
	\end{split}
	\label{SqrMag-CrssPrd}
\end{equation}
where $\theta$ is the angle between $\mathbf{a}$ and $\mathbf{b}$,
and $g=\cos\theta=\mathbf{a}^{T}\mathbf{b}$. By substituting \eqref{SqrMag-CrssPrd}
into \eqref{eq:simplified}, we have 
\begin{equation}
	\|\mathbf{e}_{\textrm{p}}\|^{2}+t^{2}(1-g^{2})=1.\label{eq:simplified2}
\end{equation}
By solving the \eqref{eq:simplified2} for the parameter $t^{2}$,
we have 
\begin{equation}
	t^{2}=\frac{1-\|\mathbf{e}_{\textrm{p}}\|^{2}}{1-g^{2}}.\label{eq:t_squared}
\end{equation}

\subsubsection{Solution Cases}

\label{subsec:Solution-Cases-and}

The nature of the solution depends on the right-hand side of \eqref{eq:t_squared}
as 
\begin{itemize}
	\item \textbf{No solution:} If $1-\|\mathbf{e}_{\textrm{p}}\|^{2}<0$ (i.e.,
	$\|\mathbf{e}_{\textrm{p}}\|^{2}>1$). The line defined by the linear
	system does not intersect the sphere. 
	\item \textbf{Unique solution:} If $1-\|\mathbf{e}_{\textrm{p}}\|^{2}=0$
	(i.e., $\|\mathbf{e}_{\textrm{p}}\|^{2}=1$). The line is tangent
	to the sphere at $\mathbf{e}=\mathbf{e}_{\textrm{p}}$ (i.e., $t=0$). 
	\item \textbf{Two solutions:} If $1-\|\mathbf{e}_{\textrm{p}}\|^{2}>0$
	(i.e., $\|\mathbf{e}_{\textrm{p}}\|^{2}<1$). The line intersects
	the sphere at two points. 
\end{itemize}
The final solution for the unit direction vector $\mathbf{e}$, when
it exists, is given by 
\begin{align}
	\hat{\mathbf{e}}_{\textrm{rot}}=\hat{\mathbf{e}} & =\mathbf{e}_{\textrm{p}}\pm\sqrt{\frac{1-\|\mathbf{e}_{\textrm{p}}\|^{2}}{1-g^{2}}}\cdot(\mathbf{a}\times\mathbf{b}),\label{eq:final_solution}
\end{align}
where 
\begin{align}
	\mathbf{e}_{\textrm{p}} & =\frac{(d_{1}-gd_{2})\mathbf{a}+(d_{2}-gd_{1})\mathbf{b}}{1-g^{2}},g=\mathbf{a}^{T}\mathbf{b}.
\end{align}

\subsubsection{Recovering Rotation Angles from the Unit Direction Vector}

According to the definition of the unit direction vector $\mathbf{e}_{\textrm{rot}}=[\sin\beta\cos\alpha,-\sin\beta\sin\alpha,\cos\beta]^{T}$,
after obtaining $\hat{\mathbf{e}}_{\textrm{rot}}=[\hat{e}_{\textrm{rot}}^{x},\hat{e}_{\textrm{rot}}^{y},\hat{e}_{\textrm{rot}}^{z}]^{T}$
by \eqref{eq:final_solution}, we have 
\begin{align}
	\hat{\beta} & =\arccos\hat{e}_{\textrm{rot}}^{z},\\
	\hat{\alpha} & =-\arctan(\hat{e}_{\textrm{rot}}^{y}/\hat{e}_{\textrm{rot}}^{x})\:\;\textrm{or}\:\;\pi-\arctan(\hat{e}_{\textrm{rot}}^{y}/\hat{e}_{\textrm{rot}}^{x}).\label{alpha_Recvr}
\end{align}

By limiting $-\pi<-\hat{\alpha}<0$ as assumed in Section \ref{System Model},
we have $-\hat{\alpha}=\arctan(\hat{e}_{\textrm{rot}}^{y}/\hat{e}_{\textrm{rot}}^{x})$
if $-\pi/2<\arctan(\hat{e}_{\textrm{rot}}^{y}/\hat{e}_{\textrm{rot}}^{x})<0$,
and $-\hat{\alpha}=\arctan(\hat{e}_{\textrm{rot}}^{y}/\hat{e}_{\textrm{rot}}^{x})-\pi$
if $0<\arctan(\hat{e}_{\textrm{rot}}^{y}/\hat{e}_{\textrm{rot}}^{x})<\pi/2$.
Thus, the \eqref{alpha_Recvr} can be expressed rigorously as 
\begin{align}
	\hat{\alpha} & =\left\{ \begin{array}{c}
		-\arctan(\hat{e}_{\textrm{rot}}^{y}/\hat{e}_{\textrm{rot}}^{x}),\textrm{if}-\pi/2<\arctan(\hat{e}_{\textrm{rot}}^{y}/\hat{e}_{\textrm{rot}}^{x})<0,\\
		\pi-\arctan(\hat{e}_{\textrm{rot}}^{y}/\hat{e}_{\textrm{rot}}^{x}),\textrm{if}\,\,0\leq\arctan(\hat{e}_{\textrm{rot}}^{y}/\hat{e}_{\textrm{rot}}^{x})<\pi/2.
	\end{array}\right.
\end{align}

Then, the two dimensional rotation angles $\hat{\alpha}$ and $\hat{\beta}$
are estimated. 

\section{Derivation of the FIM and Transformation Matrix}

\label{Appendix_C}

\subsection{Derivation of the FIM}

\label{Appendices C.A}

We find that the key to obtain the FIM ${\bf J}_{\boldsymbol{\eta}}$
in \eqref{FIM_stage2_rewritten} is to calculate $\frac{\partial{\tilde{{\bf H}}}[n]}{\partial[\boldsymbol{\eta}]_{u}},u=1,2,\cdots,6(Q+1)$,
which are given as follows.
\begin{align}
	\frac{\partial{\tilde{{\bf H}}}[n]}{\partial[\boldsymbol{\eta}]_{6q+1}}=\frac{\partial{\tilde{{\bf H}}}[n]}{\partial\tau_{q}} & =\frac{\partial\sum\limits _{q=0}^{Q}\delta_{q}e^{-j2\pi\tau_{q}\frac{(n-1)B}{N}}(\bm{{\Sigma}}_{q}^{T}\otimes{\tilde{{\bf H}}}_{q})}{\partial\tau_{q}}\nonumber \\
	& =-j2\pi\frac{(n-1)B}{N}\delta_{q}e^{-j2\pi\tau_{q}\frac{(n-1)B}{N}}(\bm{{\Sigma}}_{q}^{T}\otimes{\tilde{{\bf H}}}_{q}).
\end{align}
\begin{align}
	\frac{\partial{\tilde{{\bf H}}}[n]}{\partial[\boldsymbol{\eta}]_{6q+2}}=\frac{\partial{\tilde{{\bf H}}}[n]}{\partial\delta_{q,\mathrm{R}}} & =\frac{\partial\sum\limits _{q=0}^{Q}\delta_{q}e^{-j2\pi\tau_{q}\frac{(n-1)B}{N}}(\bm{{\Sigma}}_{q}^{T}\otimes{\tilde{{\bf H}}}_{q})}{\partial\delta_{q,\mathrm{R}}}\nonumber \\
	& =[\frac{\partial(\delta_{q,\mathrm{R}}+j\delta_{q,\mathrm{I}})}{\partial\delta_{q,\mathrm{R}}}]e^{-j2\pi\tau_{q}\frac{(n-1)B}{N}}(\bm{{\Sigma}}_{q}^{T}\otimes{\tilde{{\bf H}}}_{q})\nonumber \\
	& =e^{-j2\pi\tau_{q}\frac{(n-1)B}{N}}(\bm{{\Sigma}}_{q}^{T}\otimes{\tilde{{\bf H}}}_{q}).
\end{align}
\begin{align}
	\frac{\partial{\tilde{{\bf H}}}[n]}{\partial[\boldsymbol{\eta}]_{6q+3}}=\frac{\partial{\tilde{{\bf H}}}[n]}{\partial\delta_{q,\mathrm{I}}} & =\frac{\partial\sum\limits _{q=0}^{Q}\delta_{q}e^{-j2\pi\tau_{q}\frac{(n-1)B}{N}}(\bm{{\Sigma}}_{q}^{T}\otimes{\tilde{{\bf H}}}_{q})}{\partial\delta_{q,\mathrm{I}}}\nonumber \\
	& =[\frac{\partial(\delta_{q,\mathrm{R}}+j\delta_{q,\mathrm{I}})}{\partial\delta_{q,\mathrm{I}}}]e^{-j2\pi\tau_{q}\frac{(n-1)B}{N}}(\bm{{\Sigma}}_{q}^{T}\otimes{\tilde{{\bf H}}}_{q})\nonumber \\
	& =je^{-j2\pi\tau_{q}\frac{(n-1)B}{N}}(\bm{{\Sigma}}_{q}^{T}\otimes{\tilde{{\bf H}}}_{q}).
\end{align}
\begin{align}
	\frac{\partial{\tilde{{\bf H}}}[n]}{\partial[\boldsymbol{\eta}]_{6q+4}}=\frac{\partial{\tilde{{\bf H}}}[n]}{\partial\theta_{{\rm t},q}} & =\frac{\partial\sum\limits _{q=0}^{Q}\delta_{q}e^{-j2\pi\tau_{q}\frac{(n-1)B}{N}}(\bm{{\Sigma}}_{q}^{T}\otimes{\tilde{{\bf H}}}_{q})}{\partial\theta_{{\rm t},q}}\nonumber \\
	& =\delta_{q}e^{-j2\pi\tau_{q}\frac{(n-1)B}{N}}\bm{{\Sigma}}_{q}^{T}\otimes\frac{\partial{\tilde{{\bf H}}}_{q}}{\partial\theta_{{\rm t},q}}\nonumber \\
	& =\delta_{q}e^{-j2\pi\tau_{q}\frac{(n-1)B}{N}}\left\{ \bm{{\Sigma}}_{q}^{T}\otimes[{\bf a}_{{\rm B}}(\vartheta_{{\rm r},0})\frac{\partial{\bf a}_{\mathrm{M}}^{H}(\vartheta_{{\rm t},q})}{\partial\theta_{{\rm t},q}}]\right\} \nonumber \\
	& =j2\pi\frac{d}{\lambda}\cos(\theta_{{\rm t},q})\delta_{q}e^{-j2\pi\tau_{q}\frac{(n-1)B}{N}}\left[\bm{{\Sigma}}_{q}^{T}\otimes({\tilde{{\bf H}}}_{q}{\bf D}_{N_{{\rm m}}})\right],
\end{align}
where $q=0,1,\cdots,Q$, $\frac{\partial{\bf a}_{{\rm M}}^{H}(\vartheta_{{\rm t},q})}{\partial\theta_{{\rm t},q}}=j2\pi\frac{d}{\lambda}\cos(\theta_{{\rm t},q}){\bf a}_{{\rm M}}^{H}(\vartheta_{{\rm t},q}){\bf D}_{N_{{\rm m}}}$,
and ${\bf D}_{N_{{\rm m}}}\triangleq{\rm diag}(0,1,\cdots,N_{{\rm m}}-1)$.
\begin{align}
	\frac{\partial{\tilde{{\bf H}}}[n]}{\partial[\boldsymbol{\eta}]_{6q+5}}=\frac{\partial{\tilde{{\bf H}}}[n]}{\partial\phi_{{\rm in},q}} & =\frac{\partial\sum\limits _{q=0}^{Q}\delta_{q}e^{-j2\pi\tau_{q}\frac{(n-1)B}{N}}(\bm{{\Sigma}}_{q}^{T}\otimes{\tilde{{\bf H}}}_{q})}{\partial\phi_{{\rm in},q}}\nonumber \\
	& =\delta_{q}e^{-j2\pi\tau_{q}\frac{(n-1)B}{N}}(\frac{\partial\bm{{\Sigma}}_{q}^{T}}{\partial\phi_{{\rm in},q}}\otimes{\tilde{{\bf H}}}_{q}),
\end{align}
where $q=0,1,\cdots,Q$. The $\frac{\partial\bm{{\Sigma}}_{q}^{T}}{\partial\phi_{{\rm in},q}}$
can be represented as follows:
\begin{align}
	\frac{\partial\bm{{\Sigma}}_{q}^{T}}{\partial\phi_{{\rm in},q}} & =\left\{ {\rm diag}[\frac{\partial{\bf a}_{{\rm R}}^{H}(\triangle\omega_{q}^{{\rm a}},\triangle\omega_{q}^{{\rm e}})}{\partial\phi_{{\rm in},q}}{\bf g}_{1},\cdots,\frac{\partial{\bf a}_{{\rm R}}^{H}(\triangle\omega_{q}^{{\rm a}},\triangle\omega_{q}^{{\rm e}})}{\partial\phi_{{\rm in},q}}{\bf g}_{T}]\right\} ^{T}.\label{TheDeriveDiag_phi}
\end{align}

The $\frac{\partial{\bf a}_{{\rm R}}^{H}(\triangle\omega_{q}^{{\rm a}},\triangle\omega_{q}^{{\rm e}})}{\partial\phi_{{\rm in},q}}$
in \eqref{TheDeriveDiag_phi} can be calculated by
\begin{align}
	\frac{\partial{\bf a}_{{\rm R}}^{H}(\triangle\omega_{q}^{{\rm a}},\triangle\omega_{q}^{{\rm e}})}{\partial\phi_{{\rm in},q}}= & \frac{\partial[{\bf f}(\triangle\omega_{q}^{{\rm e}},N_{{\rm e}})\otimes{\bf f}(\triangle\omega_{q}^{{\rm a}},N_{{\rm a}})]^{H}}{\partial\phi_{{\rm in},q}}\nonumber \\
	= & \frac{\partial{\bf f}^{H}(\triangle\omega_{q}^{{\rm e}},N_{{\rm e}})}{\partial\phi_{{\rm in},q}}\otimes{\bf f}^{H}(\triangle\omega_{q}^{{\rm a}},N_{{\rm a}})+{\bf f}^{H}(\triangle\omega_{q}^{{\rm e}},N_{{\rm e}})\otimes\frac{\partial{\bf f}^{H}(\triangle\omega_{q}^{{\rm a}},N_{{\rm a}})}{\partial\phi_{{\rm in},q}}\nonumber \\
	= & j2\pi\frac{d_{{\rm e}}}{\lambda}\sin\phi_{{\rm in},q}[{\bf f}^{H}(\triangle\omega_{q}^{{\rm e}},N_{{\rm e}}){\bf D}_{N_{{\rm e}}}]\otimes{\bf f}^{H}(\triangle\omega_{q}^{{\rm a}},N_{{\rm a}})\nonumber \\
	& -j2\pi\frac{d_{{\rm a}}}{\lambda}\sin\psi_{{\rm in},q}\cos\phi_{{\rm in},q}{\bf f}^{H}(\triangle\omega_{q}^{{\rm e}},N_{{\rm e}})\otimes[{\bf f}^{H}(\triangle\omega_{q}^{{\rm a}},N_{{\rm a}}){\bf D}_{N_{{\rm a}}}]\nonumber \\
	\overset{(k)}{=} & j2\pi\frac{d_{{\rm e}}}{\lambda}\sin\phi_{{\rm in},q}[{\bf f}^{H}(\triangle\omega_{q}^{{\rm e}},N_{{\rm e}})\otimes{\bf f}^{H}(\triangle\omega_{q}^{{\rm a}},N_{{\rm a}})]({\bf D}_{N_{{\rm e}}}\otimes{\bf I}_{N_{{\rm a}}})\nonumber \\
	& -j2\pi\frac{d_{{\rm a}}}{\lambda}\sin\psi_{{\rm in},q}\cos\phi_{{\rm in},q}[{\bf f}{}^{H}(\triangle\omega_{q}^{{\rm e}},N_{{\rm e}})\otimes{\bf f}^{H}(\triangle\omega_{q}^{{\rm a}},N_{{\rm a}})]({\bf I}_{N_{{\rm e}}}\otimes{\bf D}_{N_{{\rm a}}})\nonumber \\
	= & j2\pi{\bf a}_{{\rm R}}^{H}(\triangle\omega_{q}^{{\rm a}},\triangle\omega_{q}^{{\rm e}})[\frac{d_{{\rm e}}}{\lambda}\sin\phi_{{\rm in},q}({\bf D}_{N_{{\rm e}}}\otimes{\bf I}_{N_{{\rm a}}}){\!}-{\!}\frac{d_{{\rm a}}}{\lambda}\sin\psi_{{\rm in},q}\cos\phi_{{\rm in},q}({\bf I}_{N_{{\rm e}}}\otimes{\bf D}_{N_{{\rm a}}})]\nonumber \\
	\overset{(l)}{=} & {\bf a}_{{\rm R}}^{H}(\triangle\omega_{q}^{{\rm a}},\triangle\omega_{q}^{{\rm e}}){\bf D}_{\phi_{{\rm in},q}},\label{TheDeriveVctr_phi}
\end{align}
where the step $(k)$ is obtained due to $(\mathbf{A}\mathbf{B})\otimes(\mathbf{CD})=(\mathbf{A\otimes}\mathbf{C})(\mathbf{B}\mathbf{\otimes D})$.
In the step $(l)$ of \eqref{TheDeriveVctr_phi}, we define ${\bf D}_{\phi_{{\rm in},q}}$
as
\begin{align}
	{\bf D}_{\phi_{{\rm in},q}} & \triangleq j2\pi[\frac{d_{{\rm e}}}{\lambda}\sin\phi_{{\rm in},q}({\bf D}_{N_{{\rm e}}}\otimes{\bf I}_{N_{{\rm a}}})-\frac{d_{{\rm a}}}{\lambda}\sin\psi_{{\rm in},q}\cos\phi_{{\rm in},q}({\bf I}_{N_{{\rm e}}}\otimes{\bf D}_{N_{{\rm a}}})],
\end{align}
where ${\bf D}_{N_{{\rm e}}}\triangleq{\rm diag}(0,1,\cdots,N_{{\rm e}}-1)$
and ${\bf D}_{N_{{\rm a}}}\triangleq{\rm diag}(0,1,\cdots,N_{{\rm a}}-1)$.
\begin{align}
	\frac{\partial{\tilde{{\bf H}}}[n]}{\partial[\boldsymbol{\eta}]_{6q+6}}=\frac{\partial{\tilde{{\bf H}}}[n]}{\partial\psi_{{\rm in},q}} & =\frac{\partial\sum\limits _{q=0}^{Q}\delta_{q}e^{-j2\pi\tau_{q}\frac{(n-1)B}{N}}(\bm{{\Sigma}}_{q}^{T}\otimes{\tilde{{\bf H}}}_{q})}{\partial\psi_{{\rm in},q}}\nonumber \\
	& =\delta_{q}e^{-j2\pi\tau_{q}\frac{(n-1)B}{N}}(\frac{\partial\bm{{\Sigma}}_{q}^{T}}{\partial\psi_{{\rm in},q}}\otimes{\tilde{{\bf H}}}_{q}),
\end{align}
where $q=0,1,\cdots,Q$. The $\frac{\partial\bm{{\Sigma}}_{q}^{T}}{\partial\psi_{{\rm in},q}}$
can be represented as follows:
\begin{align}
	\frac{\partial\bm{{\Sigma}}_{q}^{T}}{\partial\psi_{{\rm in},q}} & =\left\{ {\rm diag}[\frac{\partial{\bf a}_{{\rm R}}^{H}(\triangle\omega_{q}^{{\rm a}},\triangle\omega_{q}^{{\rm e}})}{\partial\psi_{{\rm in},q}}{\bf g}_{1},\cdots,\frac{\partial{\bf a}_{{\rm R}}^{H}(\triangle\omega_{q}^{{\rm a}},\triangle\omega_{q}^{{\rm e}})}{\partial\psi_{{\rm in},q}}{\bf g}_{T}]\right\} ^{T}.\label{TheDeriveDiag_psi}
\end{align}

The $\frac{\partial{\bf a}_{{\rm R}}^{H}(\triangle\omega_{q}^{{\rm a}},\triangle\omega_{q}^{{\rm e}})}{\partial\psi_{{\rm in},q}}$
in \eqref{TheDeriveDiag_psi} can be calculated by
\begin{align}
	\frac{\partial{\bf a}_{{\rm R}}^{H}(\triangle\omega_{q}^{{\rm a}},\triangle\omega_{q}^{{\rm e}})}{\partial\psi_{{\rm in},q}}= & \frac{\partial[{\bf f}(\triangle\omega_{q}^{{\rm e}},N_{{\rm e}})\otimes{\bf f}(\triangle\omega_{q}^{{\rm a}},N_{{\rm a}})]^{H}}{\partial\psi_{{\rm in},q}}\nonumber \\
	= & {\bf f}^{H}(\triangle\omega_{q}^{{\rm e}},N_{{\rm e}})\otimes\frac{\partial{\bf f}^{H}(\triangle\omega_{q}^{{\rm a}},N_{{\rm a}})}{\partial\psi_{{\rm in},q}}\nonumber \\
	= & -j2\pi\frac{d_{{\rm a}}}{\lambda}\cos\psi_{{\rm in},q}\sin\phi_{{\rm in},q}{\bf f}^{H}(\triangle\omega_{q}^{{\rm e}},N_{{\rm e}})\otimes[{\bf f}^{H}(\triangle\omega_{q}^{{\rm a}},N_{{\rm a}}){\bf D}_{N_{{\rm a}}}]\nonumber \\
	\overset{(m)}{=} & -j2\pi\frac{d_{{\rm a}}}{\lambda}\cos\psi_{{\rm in},q}\sin\phi_{{\rm in},q}[{\bf f}^{H}(\triangle\omega_{q}^{{\rm e}},N_{{\rm e}})\otimes{\bf f}^{H}(\triangle\omega_{q}^{{\rm a}},N_{{\rm a}})]({\bf I}_{N_{{\rm e}}}\otimes{\bf D}_{N_{{\rm a}}})\nonumber \\
	= & -j2\pi\frac{d_{{\rm a}}}{\lambda}\cos\psi_{{\rm in},q}\sin\phi_{{\rm in},q}{\bf a}_{{\rm R}}^{H}(\triangle\omega_{q}^{{\rm a}},\triangle\omega_{q}^{{\rm e}})({\bf I}_{N_{{\rm e}}}\otimes{\bf D}_{N_{{\rm a}}})\nonumber \\
	\overset{(n)}{=} & {\bf a}_{{\rm R}}^{H}(\triangle\omega_{q}^{{\rm a}},\triangle\omega_{q}^{{\rm e}}){\bf D}_{\psi_{{\rm in},q}},\label{TheDeriveVctr_psi}
\end{align}
where the step $(m)$ is obtained by using $(\mathbf{A}\mathbf{B})\otimes(\mathbf{CD})=(\mathbf{A\otimes}\mathbf{C})(\mathbf{B}\mathbf{\otimes D})$.
In the step $(n)$ of \eqref{TheDeriveVctr_psi}, we define ${\bf D}_{\psi_{{\rm in},q}}$
as
\begin{align}
	{\bf D}_{\psi_{{\rm in},q}} & \triangleq-j2\pi\frac{d_{{\rm a}}}{\lambda}\cos\psi_{{\rm in},q}\sin\phi_{{\rm in},q}({\bf I}_{N_{{\rm e}}}\otimes{\bf D}_{N_{{\rm a}}}),
\end{align}
where ${\bf D}_{N_{{\rm a}}}\triangleq{\rm diag}(0,1,\cdots,N_{{\rm a}}-1)$.
	
\subsection{Derivation of the Transformation Matrix}
\label{Appendices C.B}
The transformation matrix $\mathbf{T}$ can be calculated according
to the definition of $\mathbf{T}\triangleq\frac{\partial\boldsymbol{\eta}^{T}}{\partial\tilde{\bm{{\eta}}}}\in\mathbb{R}^{(5Q+7)\times(6Q+6)}$.
The unknown channel parameters are are collected in the vector $\ensuremath{\bm{{\eta}}\in\mathbb{R}^{6(Q+1)}}$
as $\bm{{\eta}}=[\bm{{\eta}}_{0}^{T},\bm{{\eta}}_{1}^{T},\cdots,\bm{{\eta}}_{q}^{T},\cdots,\bm{{\eta}}_{Q}^{T}]^{T}$,
where $\ensuremath{\bm{{\eta}}_{q}=[\tau_{q},\bm{{h}}_{q}^{T},\bm{{\theta}}_{q}^{T}]^{T}}$,
$\ensuremath{\bm{{h}}_{q}=[\delta_{q,\textrm{R}},\delta_{q,\textrm{I}}]^{T}}$,
$\ensuremath{\delta_{q,\textrm{R}}=\mathfrak{Re}(\delta_{q})}$, $\ensuremath{\delta_{q,\textrm{I}}=\mathfrak{Im}(\delta_{q})}$
and $\ensuremath{\bm{{\theta}}_{q}=[\theta_{{\rm t},q},\phi_{{\rm in},q},\psi_{{\rm in},q}]^{T}}$.
The final parameters are collected in the vector $\ensuremath{\tilde{\bm{{\eta}}}\in\mathbb{R}^{2(Q+1)+3Q+5}}$
as $\tilde{\bm{{\eta}}}=[\bm{{h}}_{0}^{T},\cdots,\bm{{h}}_{q}^{T},\cdots,\bm{{h}}_{Q}^{T},\tilde{{\bf {m}}}^{T},({\bf s}^{1})^{T},\cdots,({\bf s}^{q})^{T},\cdots,({\bf s}^{Q})^{T}]^{T}$,
where $\ensuremath{\tilde{{\bf {m}}}=[{\bf m}^{T},\alpha,\beta]^{T}}$.

We give the derivatives of each channel parameter with respect to
different final parameters as follows.

(1) The derivatives of TOA $\tau_{q}$ with respect to different final
parameters are given as follows. 
\begin{equation}
	\begin{aligned}\frac{\partial\tau_{q_{1}}}{\partial\delta_{q_{2},{\rm R}}} & =\frac{\partial\tau_{q_{1}}}{\partial\delta_{q_{2},{\rm I}}}=\frac{\partial\tau_{q_{1}}}{\partial\alpha}=\frac{\partial\tau_{q_{1}}}{\partial \beta}=0, \forall q_{1},q_{2}=0,1,\cdots,Q,
	\end{aligned}
\end{equation}
\begin{align}
	\frac{\partial\tau_{q_{1}}}{\partial{\bf m}} & =\begin{cases}
		\frac{{\bf m}-{\bf r}}{c\cdot\left\Vert {\bf m}-{\bf r}\right\Vert _{2}}, & q_{1}{\rm =0}\\
		\frac{{\bf m}-{\bf s}^{q_{1}}}{c\cdot\left\Vert {\bf m}-{\bf s}^{q_{1}}\right\Vert _{2}}, & q_{1}{\rm =1,}\cdots,Q
	\end{cases},
\end{align}
\begin{align}
	\frac{\partial\tau_{q_{1}}}{\partial{\bf s}^{q_{2}}} & =\begin{cases}
		{\bf 0}, & q_{1}{\rm =0}\,\textrm{or}\,q_{1}\neq q_{2}\\
		\frac{{\bf s}^{q_{1}}-{\bf r}}{c\cdot\left\Vert {\bf s}^{q_{1}}-{\bf r}\right\Vert _{2}}-\frac{{\bf m}-{\bf s}^{q_{1}}}{c\cdot\left\Vert {\bf m}-{\bf s}^{q_{1}}\right\Vert _{2}}, & q_{1}=q_{2}={\rm 1,}\cdots,Q
	\end{cases}.
\end{align}
	
(2) The derivatives of the real part of channel gain $\delta_{q,\mathrm{R}}$
with respect to different final parameters are given as follows. 
\begin{align}
	\frac{\partial\delta_{q_{1},{\rm R}}}{\partial\tilde{\eta}} & =\begin{cases}
		1, & \tilde{\eta}=\delta_{q_{1},{\rm R}}\\
		0, & \tilde{\eta}\ne\delta_{q_{1},{\rm R}},\forall\tilde{\eta}\in\tilde{\bm{{\eta}}}
	\end{cases}.
\end{align}

(3) The derivatives of the imaginary part of channel gain $\delta_{q,\mathrm{I}}$
with respect to different final parameters are given as follows. 
\begin{align}
	\frac{\partial\delta_{q_{1},{\rm I}}}{\partial\tilde{\eta}} & =\begin{cases}
		1, & \tilde{\eta}=\delta_{q_{1},{\rm I}}\\
		0, & \tilde{\eta}\ne\delta_{q_{1},{\rm I}},\forall\tilde{\eta}\in\tilde{\bm{{\eta}}}
	\end{cases}.
\end{align}
	
(4) The derivatives of the AOD $\theta_{\mathrm{t},q}$ with respect
to different final parameters are given as follows. 
\begin{align}
	\frac{\partial\theta_{\mathrm{t},q_{1}}}{\partial\delta_{q_{2},{\rm R}}} & =\frac{\partial\theta_{\mathrm{t},q_{1}}}{\partial\delta_{q_{2},{\rm I}}}=0,\forall q_{1},q_{2}=0,1,\cdots,Q.
\end{align}

In the scenario with a 1D rotational angle of the MS, by defining
$\mathbf{a}=[\cos\alpha,-\sin\alpha,0]^{T}$, the expression for $\partial\theta_{\mathrm{t},q_{1}}/{\partial{\bf m}}$
is 
\begin{align}
	\frac{\partial\theta_{\mathrm{t},q_{1}}}{\partial{\bf m}} & =\begin{cases}
		\frac{\left[\mathbf{a}^{T}({\bf r}-{\bf m})\right]({\bf r}-{\bf m})-\mathbf{a}\left\Vert {\bf r}-{\bf m}\right\Vert _{2}^{2}}{\left\Vert {\bf r}-{\bf m}\right\Vert _{2}^{2}\sqrt{\left\Vert {\bf r}-{\bf m}\right\Vert _{2}^{2}-\left[\mathbf{a}^{T}({\bf r}-{\bf m})\right]^{2}}},q_{1}{\rm =0},\\
		\frac{\left[\mathbf{a}^{T}({\bf s}^{q_{1}}-{\bf m})\right]({\bf s}^{q_{1}}-{\bf m})-\mathbf{a}\left\Vert {\bf s}^{q_{1}}-{\bf m}\right\Vert _{2}^{2}}{\left\Vert {\bf s}^{q_{1}}-{\bf m}\right\Vert _{2}^{2}\sqrt{\left\Vert {\bf s}^{q_{1}}-{\bf m}\right\Vert _{2}^{2}-\left[\mathbf{a}^{T}({\bf s}^{q_{1}}-{\bf m})\right]^{2}}},q_{1}{\rm =1,}\cdots,Q.
	\end{cases}\label{thetaq_deriv_m}
\end{align}

The expression for $\partial\theta_{\mathrm{t},q_{1}}/\partial\alpha$
is given in
\begin{align}
	\frac{\partial\theta_{\mathrm{t},q_{1}}}{\partial\alpha} & =\begin{cases}
		\frac{-(r_{x}-m_{x})\sin\alpha-(r_{y}-m_{y})\cos\alpha}{\sqrt{\left\Vert {\bf r}-{\bf m}\right\Vert _{2}^{2}-\left[(r_{x}-m_{x})\cos\alpha-(r_{y}-m_{y})\sin\alpha\right]^{2}}}, & q_{1}{\rm =0}\\
		\frac{-(s_{x}^{q_{1}}-m_{x})\sin\alpha-(s_{y}^{q_{1}}-m_{y})\cos\alpha}{\sqrt{\left\Vert {\bf s}^{q_{1}}-{\bf m}\right\Vert _{2}^{2}-\left[(s_{x}^{q_{1}}-m_{x})\cos\alpha-(s_{y}^{q_{1}}-m_{y})\sin\alpha\right]^{2}}}, & q_{1}{\rm =1,}\cdots,Q
	\end{cases},\label{thetaq_deriv_arfa}
\end{align}

The expression for $\partial\theta_{\mathrm{t},q_{1}}/{\partial{\bf s}^{q_{2}}}$
is given by 
\begin{align}
	\frac{\partial\theta_{\mathrm{t},q_{1}}}{\partial{\bf s}^{q_{2}}} & =\begin{cases}
		\mathbf{0},q_{1}=0{\ }\textrm{or}{\ }q_{1}\ne q_{2},\\
		\frac{\mathbf{a}\left\Vert {\bf s}^{q_{1}}-{\bf m}\right\Vert _{2}^{2}-\left[\mathbf{a}^{T}({\bf s}^{q_{1}}-{\bf m})\right]({\bf s}^{q_{1}}-{\bf m})}{\left\Vert {\bf s}^{q_{1}}-{\bf m}\right\Vert _{2}^{2}\sqrt{\left\Vert {\bf s}^{q_{1}}-{\bf m}\right\Vert _{2}^{2}-\left[\mathbf{a}^{T}({\bf s}^{q_{1}}-{\bf m})\right]^{2}}},q_{1}=q_{2}\ne0.
	\end{cases}\label{thetaq_deriv_sq}
\end{align}

In the scenario with 2D rotational angles of the MS, by defining $\mathbf{e}_{\textrm{rot}}=[\sin\beta\cos\alpha,-\sin\beta\sin\alpha,\cos\beta]^{T}$,
	the new version of $\partial\theta_{\mathrm{t},q_{1}}/\partial{\bf m}$
	is 
\begin{align}
	\frac{\partial\theta_{\mathrm{t},q_{1}}}{\partial{\bf m}} & =\begin{cases}
		\frac{\left[\mathbf{e}_{\textrm{rot}}^{T}({\bf r}-{\bf m})\right]({\bf r}-{\bf m})-\mathbf{e}_{\textrm{rot}}\left\Vert {\bf r}-{\bf m}\right\Vert _{2}^{2}}{\left\Vert {\bf r}-{\bf m}\right\Vert _{2}^{2}\sqrt{\left\Vert {\bf r}-{\bf m}\right\Vert _{2}^{2}-\left[\mathbf{e}_{\textrm{rot}}^{T}({\bf r}-{\bf m})\right]^{2}}},q_{1}{\rm =0},\\
		\frac{\left[\mathbf{e}_{\textrm{rot}}^{T}({\bf s}^{q_{1}}-{\bf m})\right]({\bf s}^{q_{1}}-{\bf m})-\mathbf{e}_{\textrm{rot}}\left\Vert {\bf s}^{q_{1}}-{\bf m}\right\Vert _{2}^{2}}{\left\Vert {\bf s}^{q_{1}}-{\bf m}\right\Vert _{2}^{2}\sqrt{\left\Vert {\bf s}^{q_{1}}-{\bf m}\right\Vert _{2}^{2}-\left[\mathbf{e}_{\textrm{rot}}^{T}({\bf s}^{q_{1}}-{\bf m})\right]^{2}}},q_{1}{\rm =1,}\cdots,Q.
	\end{cases}\label{thetaq_deriv_m_revised}
\end{align}

The new version for $\partial\theta_{\mathrm{t},q_{1}}/\partial\alpha$
	in the 2D rotation scenario is given in
	\begin{align}
		\frac{\partial\theta_{\mathrm{t},q_{1}}}{\partial\alpha} & =\begin{cases}
			\frac{-(r_{x}-m_{x})\sin\beta\sin\alpha-(r_{y}-m_{y})\sin\beta\cos\alpha}{\sqrt{\left\Vert {\bf r}-{\bf m}\right\Vert _{2}^{2}-\left[\mathbf{e}_{\textrm{rot}}^{T}({\bf r}-{\bf m})\right]^{2}}}, & q_{1}{\rm =0}\\
			\frac{-(s_{x}^{q_{1}}-m_{x})\sin\beta\sin\alpha-(s_{y}^{q_{1}}-m_{y})\sin\beta\cos\alpha}{\sqrt{\left\Vert {\bf s}^{q_{1}}-{\bf m}\right\Vert _{2}^{2}-\left[\mathbf{e}_{\textrm{rot}}^{T}({\bf s}^{q_{1}}-{\bf m})\right]^{2}}}, & q_{1}{\rm =1,}\cdots,Q
		\end{cases},\label{thetaq_deriv_arfa_revised}
	\end{align}
	
The expression for the newly-added $\partial\theta_{\mathrm{t},q_{1}}/\partial\beta$
	in the 2D rotation scenario is given in
	\begin{align}
		\frac{\partial\theta_{\mathrm{t},q_{1}}}{\partial\beta} & =\begin{cases}
			\frac{(r_{x}-m_{x})\cos\beta\cos\alpha-(r_{y}-m_{y})\cos\beta\sin\alpha-(r_{z}-m_{z})\sin\beta}{\sqrt{\left\Vert {\bf r}-{\bf m}\right\Vert _{2}^{2}-[\mathbf{e}_{\textrm{rot}}^{T}({\bf r}-{\bf m})]^{2}}}, & q_{1}{\rm =0}\\
			\frac{(s_{x}^{q_{1}}-m_{x})\cos\beta\cos\alpha-(s_{y}^{q_{1}}-m_{y})\cos\beta\sin\alpha-(s_{z}^{q_{1}}-m_{z})\sin\beta}{\sqrt{\left\Vert {\bf s}^{q_{1}}-{\bf m}\right\Vert _{2}^{2}-[\mathbf{e}_{\textrm{rot}}^{T}({\bf s}^{q_{1}}-{\bf m})]^{2}}}, & q_{1}{\rm =1,}\cdots,Q
		\end{cases},\label{thetaq_deriv_beta_added}
	\end{align}

The new version for $\partial\theta_{\mathrm{t},q_{1}}/{\partial{\bf s}^{q_{2}}}$
	in the 2D rotation scenario is
\begin{align}
	\frac{\partial\theta_{\mathrm{t},q_{1}}}{\partial{\bf s}^{q_{2}}} & =\begin{cases}
		\mathbf{0},q_{1}=0{\ }\textrm{or}{\ }q_{1}\ne q_{2},\\
		\frac{\mathbf{e}_{\textrm{rot}}\left\Vert {\bf s}^{q_{1}}-{\bf m}\right\Vert _{2}^{2}-\left[\mathbf{e}_{\textrm{rot}}^{T}({\bf s}^{q_{1}}-{\bf m})\right]({\bf s}^{q_{1}}-{\bf m})}{\left\Vert {\bf s}^{q_{1}}-{\bf m}\right\Vert _{2}^{2}\sqrt{\left\Vert {\bf s}^{q_{1}}-{\bf m}\right\Vert _{2}^{2}-\left[\mathbf{e}_{\textrm{rot}}^{T}({\bf s}^{q_{1}}-{\bf m})\right]^{2}}},q_{1}=q_{2}\ne0.
	\end{cases}\label{thetaq_deriv_sq_revised}
\end{align}
	
(5) The derivatives of the elevation angle $\phi_{\mathrm{in},q}$
at the RIS with respect to different final parameters are given as
follows: 
\begin{equation}
	\begin{aligned}\frac{\partial\phi_{\mathrm{in},q_{1}}}{\partial\delta_{q_{2},\mathbb{R}}} & =\frac{\partial\phi_{\mathrm{in},q_{1}}}{\partial\delta_{q_{2},\mathbb{I}}}=\frac{\partial\phi_{\mathrm{in},q_{1}}}{\partial\alpha}=\frac{\partial\phi_{\mathrm{in},q_{1}}}{\partial\beta}=0,\forall\ q_{1},q_{2}=0,1,\dots,Q,
	\end{aligned}
\end{equation}
\begin{align}
	\frac{\partial\phi_{\mathrm{in},q_{1}}}{\partial m_{x}} & =\begin{cases}
		\frac{-(r_{x}-m_{x})(r_{z}-m_{z})}{\left\Vert {\bf r}-{\bf m}\right\Vert _{2}^{2}\sqrt{\left\Vert {\bf r}-{\bf m}\right\Vert _{2}^{2}-(r_{z}-m_{z})^{2}}}, & q_{1}{\rm =0}\\
		0, & q_{1}{\rm =1,}\cdots,Q
	\end{cases},
\end{align}
\begin{align}
	\frac{\partial\phi_{\mathrm{in},q_{1}}}{\partial m_{y}} & =\begin{cases}
		\frac{-(r_{y}-m_{y})(r_{z}-m_{z})}{\left\Vert {\bf r}-{\bf m}\right\Vert _{2}^{2}\sqrt{\left\Vert {\bf r}-{\bf m}\right\Vert _{2}^{2}-(r_{z}-m_{z})^{2}}}, & q_{1}{\rm =0}\\
		0, & q_{1}{\rm =1,}\cdots,Q
	\end{cases},
\end{align}
\begin{align}
	\frac{\partial\phi_{\mathrm{in},q_{1}}}{\partial m_{z}} & =\begin{cases}
		\frac{\sqrt{\left\Vert {\bf r}-{\bf m}\right\Vert _{2}^{2}-(r_{z}-m_{z})^{2}}}{\left\Vert {\bf r}-{\bf m}\right\Vert _{2}^{2}}, & q_{1}{\rm =0}\\
		0, & q_{1}{\rm =1,}\cdots,Q
	\end{cases},
\end{align}
\begin{align}
	\frac{\partial\phi_{\mathrm{in},q_{1}}}{\partial s_{x}^{q_{2}}} & =\begin{cases}
		0,q_{1}=0{\ }\textrm{or}{\ }q_{1}\ne q_{2}\\
		\frac{-(r_{x}-s_{x}^{q_{2}})(r_{z}-s_{z}^{q_{2}})}{\left\Vert {\bf r}-{\bf s}^{q_{2}}\right\Vert _{2}^{2}\sqrt{\left\Vert {\bf r}-{\bf s}^{q_{2}}\right\Vert _{2}^{2}-(r_{z}-s_{z}^{q_{2}})^{2}}},q_{1}=q_{2}\ne0
	\end{cases},
\end{align}
\begin{align}
	\frac{\partial\phi_{\mathrm{in},q_{1}}}{\partial s_{y}^{q_{2}}} & =\begin{cases}
		0,q_{1}=0{\ }\textrm{or}{\ }q_{1}\ne q_{2}\\
		\frac{-(r_{y}-s_{y}^{q_{2}})(r_{z}-s_{z}^{q_{2}})}{\left\Vert {\bf r}-{\bf s}^{q_{2}}\right\Vert _{2}^{2}\sqrt{\left\Vert {\bf r}-{\bf s}^{q_{2}}\right\Vert _{2}^{2}-(r_{z}-s_{z}^{q_{2}})^{2}}}q_{1}=q_{2}\ne0
	\end{cases},
\end{align}
\begin{align}
	\frac{\partial\phi_{\mathrm{in},q_{1}}}{\partial s_{z}^{q_{2}}} & =\begin{cases}
		0, & q_{1}=0{\ }\textrm{or}{\ }q_{1}\ne q_{2}\\
		\frac{\sqrt{\left\Vert {\bf r}-{\bf s}^{q_{2}}\right\Vert _{2}^{2}-(r_{z}-s_{z}^{q_{2}})^{2}}}{\left\Vert {\bf r}-{\bf s}^{q_{2}}\right\Vert _{2}^{2}} & q_{1}=q_{2}\ne0
	\end{cases}.
\end{align}
	
(6) The derivatives of the azimuth angle $\psi_{\mathrm{in},q}$ at
the RIS with respect to different final parameters are given as follows:
\begin{equation}
	\begin{aligned}\frac{\partial\psi_{\mathrm{in},q_{1}}}{\partial\tilde{\eta}}=0, & \forall\tilde{\eta}\in\{\delta_{q_{2},{\rm R}},\delta_{q_{2},{\rm I}},\alpha,\beta,m_{z},s_{z}^{q_{2}}\},\forall q_{1},q_{2}=0,1,\cdots,Q,
	\end{aligned}
\end{equation}
\begin{align}
	\frac{\partial\psi_{\mathrm{in},q_{1}}}{\partial m_{x}} & =\begin{cases}
		\frac{-(r_{x}-m_{x})(r_{y}-m_{y})}{\left[(r_{x}-m_{x})^{2}+(r_{y}-m_{y})^{2}\right]\sqrt{(r_{x}-m_{x})^{2}}},q_{1}{\rm =0}\\
		0,q_{1}{\rm =1,}\cdots,Q
	\end{cases}
\end{align}
\begin{align}
	\frac{\partial\psi_{\mathrm{in},q_{1}}}{\partial m_{y}} & =\begin{cases}
		\frac{(r_{x}-m_{x})^{2}}{\left[(r_{x}-m_{x})^{2}+(r_{y}-m_{y})^{2}\right]\sqrt{(r_{x}-m_{x})^{2}}},q_{1}{\rm =0}\\
		0,q_{1}{\rm =1,}\cdots,Q
	\end{cases}
\end{align}
\begin{align}
	\frac{\partial\psi_{\mathrm{in},q_{1}}}{\partial s_{x}^{q_{2}}}=\begin{cases}
		0,q_{1}=0{\ }\textrm{or}{\ }q_{1}\ne q_{2}\\
		\frac{-(r_{x}-s_{x}^{q_{2}})(r_{y}-s_{y}^{q_{2}})}{\left[(r_{x}-s_{x}^{q_{2}})^{2}+(r_{y}-s_{y}^{q_{2}})^{2}\right]\sqrt{(r_{x}-s_{x}^{q_{2}})^{2}}},q_{1}=q_{2}\ne0
	\end{cases}
\end{align}
\begin{align}
	\frac{\partial\psi_{\mathrm{in},q_{1}}}{\partial s_{y}^{q_{2}}} & =\begin{cases}
		0,q_{1}=0{\ }\textrm{or}{\ }q_{1}\ne q_{2}\\
		\frac{(r_{x}-s_{x}^{q_{2}})^{2}}{\left[(r_{x}-s_{x}^{q_{2}})^{2}+(r_{y}-s_{y}^{q_{2}})^{2}\right]\sqrt{(r_{x}-s_{x}^{q_{2}})^{2}}},q_{1}=q_{2}\ne0
	\end{cases}
\end{align}
	
\end{appendices}
	
	\bibliographystyle{IEEEtran}
	\bibliography{IEEEabrv,References_temp_Revise_Arxiv}
	
	
	
\end{document}